\documentclass[twocolumn]{scrartcl}

\usepackage[english]{babel}
\usepackage{times}
\usepackage[utf8]{inputenc}
\usepackage[T1]{fontenc}

\usepackage{amsmath, amsthm, bm}
\usepackage{amsfonts, amssymb}
\usepackage{graphicx}
\usepackage{algcompatible}
\usepackage{newfloat}
\usepackage[caption=false]{subfig}

\newcommand{\G}{\mathcal{G}}
\newcommand{\Hg}{\mathcal{H}}
\newcommand{\E}{\mathcal{E}}
\newcommand{\V}{\mathcal{V}}
\usepackage[left=2cm,right=2cm,top=2cm,bottom=2cm]{geometry}

\DeclareGraphicsExtensions{.jpg,.mps,.pdf,.png}
\graphicspath{{tex/images/}, {images/}, {.}}

\algnewcommand\algorithmicreturn{\textbf{return }}
\algnewcommand\RETURN{\State \algorithmicreturn}

\DeclareFloatingEnvironment[ fileext=loa, listname=List of Algorithms,
name=ALGORITHM, placement=tbhp, ]{algorithm}

\title{From graphs to signals and back: Identification of network structures
  using spectral analysis}

\author{Ronan Hamon \and Pierre Borgnat \and Patrick Flandrin \and C\'eline
  Robardet}

\begin{document}
\maketitle

\begin{abstract}
  The structure of networks describing interactions between entities gives
  significant insights about how these systems work. Recently, an approach has
  been proposed to transform a graph into a collection of signals, using a
  multidimensional scaling technique on a distance matrix representing relations
  between vertices of the graph as points in a Euclidean space: coordinates are
  interpreted as components, or signals, indexed by the vertices. In this
  article, we propose several extensions to this approach: We first extend the
  current methodology, enabling us to highlight connections between properties of
  the collection of signals and graph structures, such as communities, regularity
  or randomness, as well as combinations of those. A robust inverse
  transformation method is next described, taking into account possible changes
  in the signals compared to original ones. This technique uses, in addition to
  the relationships between the points in the Euclidean space, the energy of each
  signal, coding the different scales of the graph structure. These contributions
  open up new perspectives by enabling processing of graphs through the
  processing of the corresponding collection of signals. A technique of denoising
  of a graph by filtering of the corresponding signals is then described,
  suggesting considerable potential of the approach.
\end{abstract}

\section{Introduction}

The development of data processing methods is crucial with the recent explosion
of data made possible by technological devices. Many systems, previously
inaccessible by traditional methods of quantitative analysis by lack of
information, are now described by huge amounts of data, so that the limiting
factor is now the absence of tools to give them sense. Among these systems, many
can be represented as networks i.e., a set of relationships between entities.
Network theory \cite{Newman2010} has been developed to supply toolboxes, such as
for detecting communities \cite{Fortunato2010}, in order to understand the
underlying properties of these systems, with many successes. More recently,
connections between signal processing and networks theory have tremendously
increased. The field of signal processing over networks has been extensively
studied in recent years \cite{Shuman2013} with the objective to transpose
concepts developed in classical signal processing, such as Fourier transform or
wavelets, in the graph domain. These works, 
based on spectral graph theory \cite{Chung1997}, have led to a growing set of
significant results, among them filtering of graph signals (i.e., signals
defined over a graph) \cite{Shuman2013, Sandryhaila2014}, spectral wavelets
\cite{Hammond2011}, wavelet filterbanks \cite{Narang2012, Shuman2013,
  Sakiyama2014, Nguyen2015} vertex-frequency analysis \cite{Shuman2015} of
graphs signals, multiscale community mining using graph wavelets
\cite{Tremblay2014a}, or sampling for graph signals \cite{Anis2014, Chen2015a,
  Marques2015, Gadde2015a, Tsitsvero2015a} added to considerations about
uncertainty principle on graphs \cite{Agaskar2013, Pasdeloup2015,
  Tsitsvero2015}.

For the study of networks themselves, another approach linked to signal
processing has been considered: it consists in a duality between graphs and
signals, based on the transformation of graphs into signals and reciprocally, in
order to take advantage of both signal processing and graph theory. With the
development of the network science \cite{Newman2010}, some works have proposed
to map time series into graph objects, aiming at using the wide range of
measures and tools defined on networks to highlight properties in the time
domain; this approach has been successfully used in the analysis of nonlinear
time series~\cite{Campanharo2011, Nunez2012}, in the characterization of
nonlinear dynamics~\cite{Zhang2006, Shimada2008, Donner2011} or in the
identification of invariant structures~\cite{Donges2012}, with significant
results in applications such as heart rate~\cite{Campanharo2011} or
earthquake~\cite{Aguilar-SanJuan2013} analyses.

Conversely, mapping a graph into time series has been less intensively studied.
Recently, Weng et al.~\cite{Weng2014} proposed to explore the structure of
scale-free networks using finite-memory random walks, where the values of time
series at time $t$ are the degree of the vertex visited by the walker at step
$t$. The resulting time series, obtained from different real-world networks,
exhibit correlations, linked with the scale-free property of these networks.
With a similar approach, Campanharo et al.~\cite{Campanharo2011} proposed a
random walk based algorithm to map graphs into time series by associating a
specific value in the time domain with vertices, with the particularity that the
graphs are themselves derived from time series.  Girault et
al.~\cite{Girault2014} extended this approach in the case where the graph is the
object of interest, by using semi-supervised learning to map vertices to signal
magnitudes such that the resulting time series are smooth. Using an alternative
approach, Haraguchi et al.~\cite{Haraguchi2009} and later Shimada et
al.~\cite{Shimada2012} proposed a deterministic method based on classical
multidimensional scaling (CMDS) to represent the vertices of the graph as a set
of points in a Euclidean space, where the relations described by the edges are
represented by distances between points.

An attractive feature of this latter approach, in comparison with random walks, 
is that it is fully deterministic since for a given graph, its representation in 
the signal domain remains the same. A second interesting point is that all 
information is included in the signals: from them, it is possible, under 
some assumptions, to compute exactly the original graph. 
Our contribution in this article elaborates on preliminary contributions along
this line in \cite{Hamon2013c}, and extends the method proposed in
\cite{Shimada2012} on several points: in Section~\ref{sec:transformation}, the
methodology is examined more deeply, and application of the method to a wider
variety of graph structures is studied through illustrations. A robust inverse
transformation is then proposed in Section~\ref{sec:inverse_transformation} to
transform back a collection of signals to a graph, with a focus on the case
where the collection of signals is degraded and is not anymore the direct result
of the transformation of a graph.  Finally, we develop a comprehensive framework
in Section~\ref{sec:spectral_analysis} to analyze graphs using tools from signal
processing, namely spectral analysis, with an illustration to the denoising of a
graph.  Note that the present work focuses on static graphs; in previous works,
some applications of the present framework have been proposed to study and
extract the topology of dynamic graphs \cite{Hamon2013a, Hamon2013b,
  Hamon2014a}.

\paragraph*{Notations}

Throughout the article, the following notations are adopted. Let $\G = (\V, \E)$
be a simple undirected and unweighted graph, where $\V$ is the set of vertices
of size $n$ and $\E$ the set of edges of size $m$. We note $\bm{A}$ its
adjacency matrix, whose element $a_{ij}$ is equal to $1$ if $(i,j) \in \E$, $0$
otherwise, and for $i, j \in \V$. The terms ``signals'' or ``components'' are
indistinctly used in the following, while avoiding the term time series, to
prevent the confusion with the case of dynamic graphs.

\section{From graph to signals}
\label{sec:transformation}

\subsection{Transformation using multidimensional scaling}

Shimada et al.~\cite{Shimada2012} proposed a method to transform a graph with
$n$ vertices into a collection of signals of $n$ points indexed by the vertices
of the graph by using multidimensional scaling (MDS) \cite{Borg2005}.

MDS is a set of mathematical techniques used to represent dissimilarities among
pairs of objects as distances between points in a multidimensional space whose
dimension is low. Classical MDS (CMDS) is a particular case of metric MDS where
the dissimilarities are assumed to be Euclidean distances. The matrix $\bm{X}$
of coordinates in the low-dimensional space can be computed analytically:
Starting with a distance matrix $\bm{\Delta} = (\delta_{ij})_{i,j=1,..,n}$, we
first compute a double centering of the matrix whose terms are squared:
$\bm{B} = -\frac{1}{2}\bm{J} \bm{\Delta}^{(2)}\bm{J}$ with
$\bm{\Delta}^{(2)} = \bm{\Delta} \circ \bm{\Delta}$ and
$\bm{J} = \bm{I}_n - \frac{1}{n}\bm{1}_n\bm{1}_n^T$ where $\bm{I}_n$ is the
identity matrix , $\bm{1}_n\bm{1}_n^T$ a $n\times n$ matrix of ones, and $\circ$
the Hadamard product. The CMDS solution is given by
$\bm{X} = \bm{Q}_+\bm{\Lambda}_+^{\frac{1}{2}}$ with $\bm{\Lambda_+}$ a diagonal
matrix whose terms are the strictly positive eigenvalues of the matrix $\bm{B}$
sorted in an increasing order and $\bm{Q_+}$ is the matrix of the corresponding
eigenvectors. The resulting signals are the components (or columns) of the
matrix $\bm{X}$. The $j$-th signal is noted $\bm{X}^{(j)}$. An alternative
approach to find the matrix $\bm{X}$ consists in solving the following
optimization problem:
$\min_{\bm{X}} {L(\bm{X}}) = \sum_{i<j}^n (\delta_{ij} - d_{ij}(\bm{X}))^2$
where $d_{ij}(\bm{X})$ is the Euclidean distance between the points $i$ and
$j$. An algorithm called \texttt{SMACOF} \cite{Borg2005} has been developed to
solve such problem. It is worth to note that the solution is not unique, as any
rotation, reflection or translation of points in the Euclidean space will
preserve the distances.

In \cite{Shimada2012}, CMDS is used to transform a graph into signals by
projecting vertices of the graph in a Euclidean space, such that distances
between these points correspond to relations in the graph. A distance matrix
between the vertices of the graph $\bm{\Delta}$ describes these relations:
Shimada et al. proposed the following definition from the adjacency matrix
$\bm{A}$ for the distance matrix:
\begin{align}
  \label{eq:delta}
  \bm{\Delta} = \bm{A} + w(\bm{1}_n\bm{1}_n^T - \bm{I}_n - \bm{A})
\end{align}
where $w$ is an arbitrary weight strictly greater than $1$. This definition does
not include a notion of proximity between vertices, beyond direct linkage: if
the two vertices are connected, the distance is equal to $1$, otherwise it is
equal to $w$. The remoteness between two vertices, which can be measured using
for instance the length of the shortest path between the two vertices, is not
taken into account in this definition: two pairs of unlinked vertices will have
a distance equal to $w$, whether they are close or not in the graph. One of the
advantage of this distance matrix, unlike a distance matrix based for instance
on the length of the shortest path between the vertices, is that it is Euclidean
(under some assumptions on the value of $w$, as discussed later in the paper),
and then there exists an exact solution. Furthermore, the information about the
proximity in the graph is no more induced from the graph, but is automatically
retrieved by the algorithm: if the considered Euclidean space is
low-dimensional, the distances cannot be well-preserved and only the distances
representing distant vertices will be respected. This enables us to look at,
component by component, how this approximation is performed, i.e., which
vertices are roughly considered as neighbors.

For a given distance matrix, it does not exist necessarily a configuration of
points $\bm{X}$ such that distances between these points are equal to distances
defined in the distance matrix. In the case where the distance matrix
$\bm{\Delta}$ is defined by Equation~\ref{eq:delta}, the structure of the graph
as well as the parameter $w$ has an influence on the existence of a solution
$\bm{X}$. This influence has been barely studied in \cite{Haraguchi2009}
and~\cite{Shimada2012}. In~\cite{Haraguchi2009}, the authors compare for
different values of $w$ the quality of the reconstruction on several graphs
generated using the Watts-Strogatz model and two real-world networks, by the
comparison of their edges sets. They conclude that the value $w$ should reside
between $1$ (excluded) and $1.01$. The same kind of approach is followed
in~\cite{Shimada2012}, but restricted to Watts-Strogatz model where the
probability of rewiring varies from $0.01$ to $1$. Their conclusion is that for
$n=400$, $w$ should be comprised between $1$ (excluded) and $1.14$, and that
this upper bound depends on the value of $n$ and should be as close to unity as
possible, without substantial argument for that. We propose in the following an
upper bound for the value of $w$ to guarantee that the matrix $\bm{\Delta}$ is
Euclidean, i.e., there exists a configuration $\bm{X}$ such that the Euclidean
distances between points are equal to $\bm{\Delta}$. The calculation of this
upper bound relies on the positive-semidefiniteness of the resulting matrix
$\bm{B}$, which is the covariance matrix of $\bm{X}$: $\bm{B} = \bm{X}\bm{X}^T$.
The study of the eigenvalues of $\bm{B}$ according to $w$ and the adjacency
matrix $\bm{A}$, detailed in \ref{apx:choice_w}, gives an upper bound for $w$:
$w \leq \sqrt{\frac{n}{n-2}}$ with $n$ the number of vertices. This result
agrees with the partial results obtained in \cite{Haraguchi2009}
and~\cite{Shimada2012}: $w$ should be close to $1$, all the greater given the
number of vertices $n$. In addition, in practice setting $w$ below this upper
bound is not required for most graphs, although it ensures that the matrix
$\bm{\Delta}$ will stay Euclidean, even with a peculiar topology of the graph.

Alternative distance matrix could be used instead of the one proposed in 
\cite{Shimada2012}. As described previously, a natural measure of closeness 
between two vertices in a graph is the length of the shortest path between the 
vertices \cite{Tenenbaum2000}. Distances also based on similarity between 
adjacent vertex sets are worth considering. These two alternative 
dissimilarities have the advantage to be easily extended to the case where the 
graph is weighted i.e., each edge has a weight giving its importance. There also 
exist alternative methods to CMDS to transform graph into points in a Euclidean 
space. For instance, the method of Laplacian eigenmaps, proposed by Belkin et 
al. \cite{Belkin2003}, have been used in a context of dimensionality reduction 
and data representation, and is based on a diagonalization of the Laplacian 
matrix of the graph. Many parallels can be done between the two approaches, even 
if they differ on the final representation of vertices in the Euclidean space 
\cite{Rossi2006}. Such a comparison is beyond the scope of this article and will 
not be discussed further here.

\subsection{Graph models and theoretical results}
\label{subsec:gs_theory}

\begin{figure*}

  \subfloat[\label{subfig:gs_rl-60-2}$2$-ring lattice with $60$ vertices
  (\textbf{2RL-60-10})]
  {
    \includegraphics[width=0.24\textwidth]{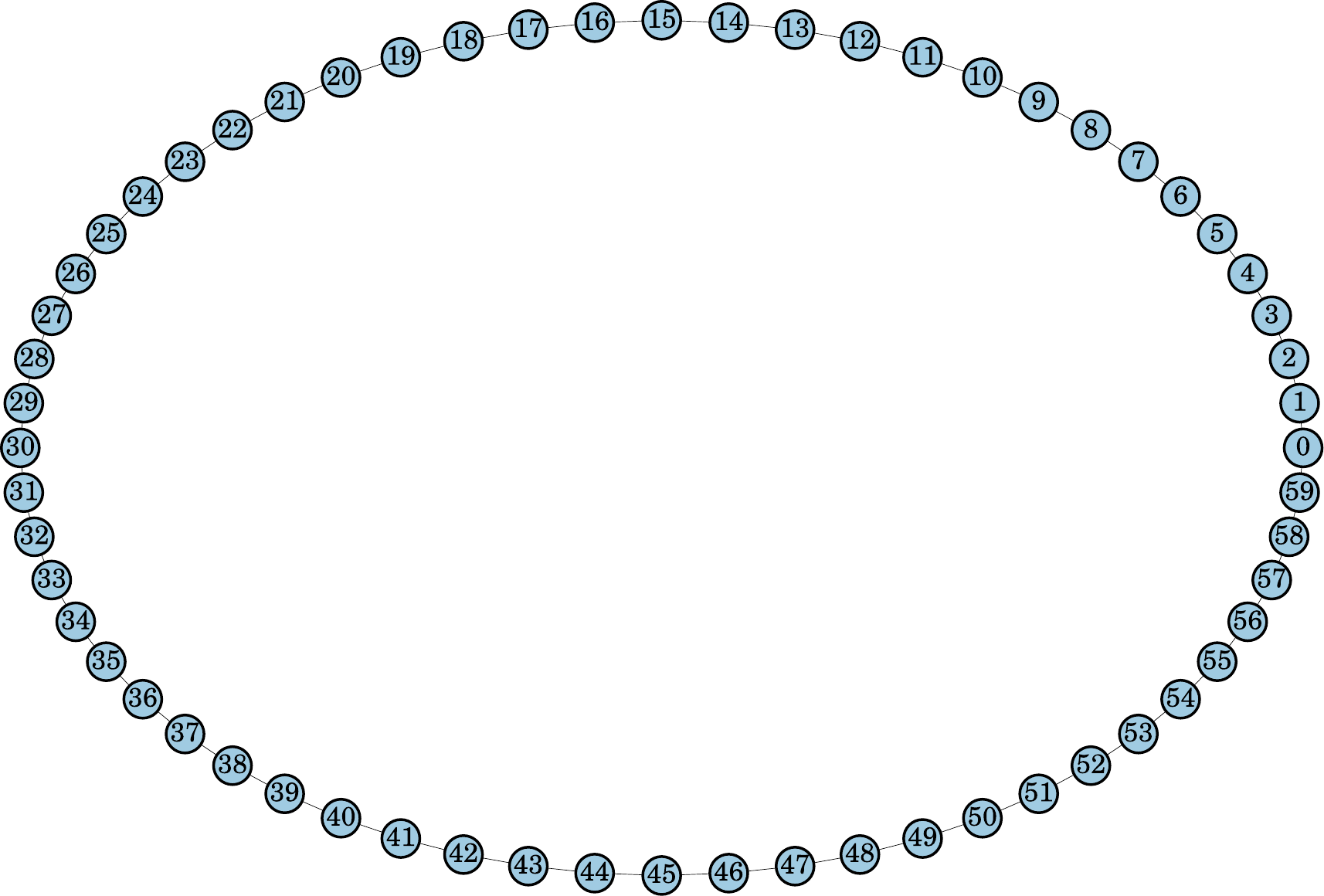}
    \includegraphics[width=0.24\textwidth]{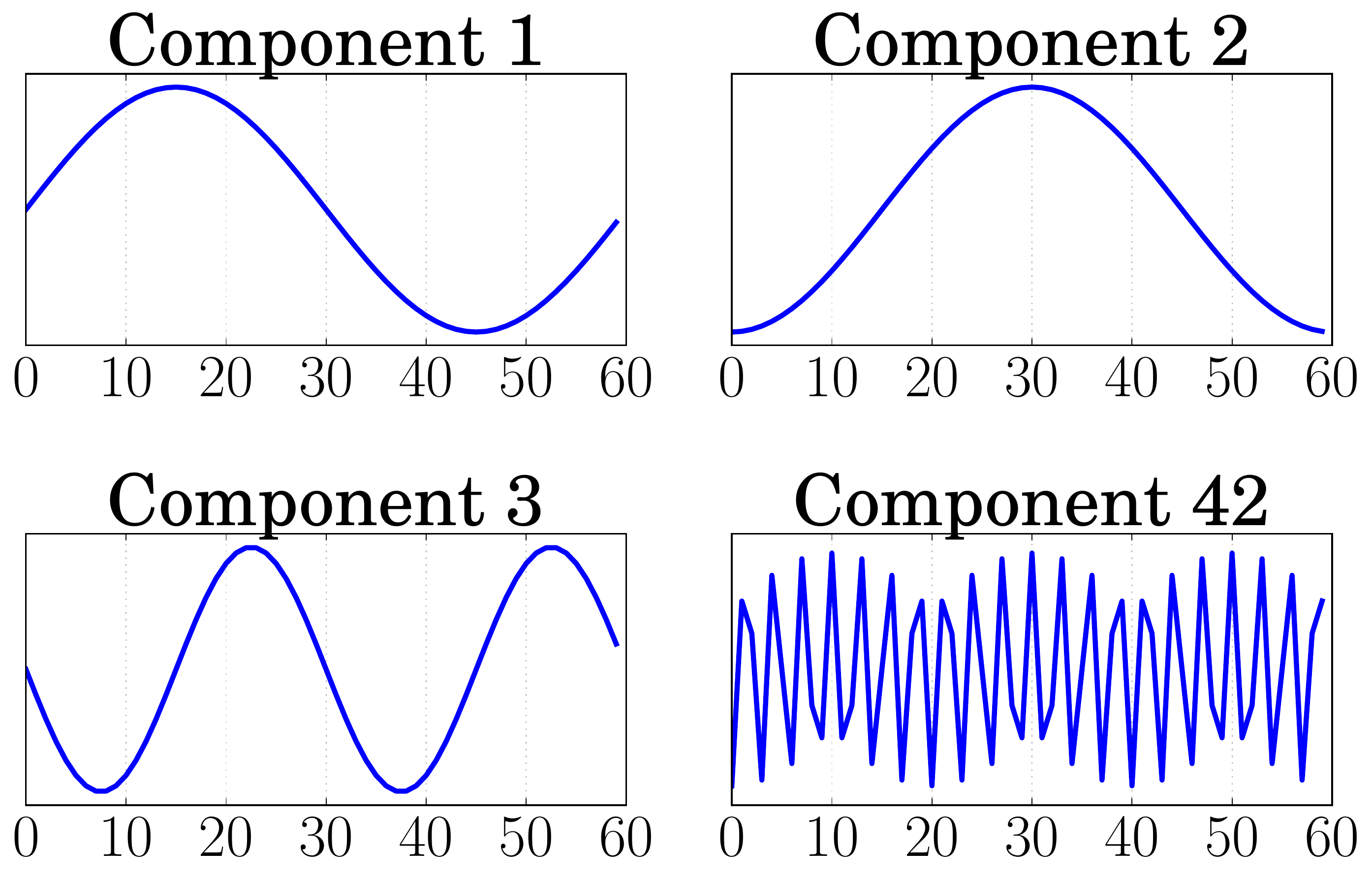}
  }
  \hspace{0.05\textwidth}
  \subfloat[\label{subfig:gs_rl-60-10}$10$-ring lattice with $60$ vertices
  (\textbf{RL-60-2})]
  {
    \includegraphics[width=0.24\textwidth]{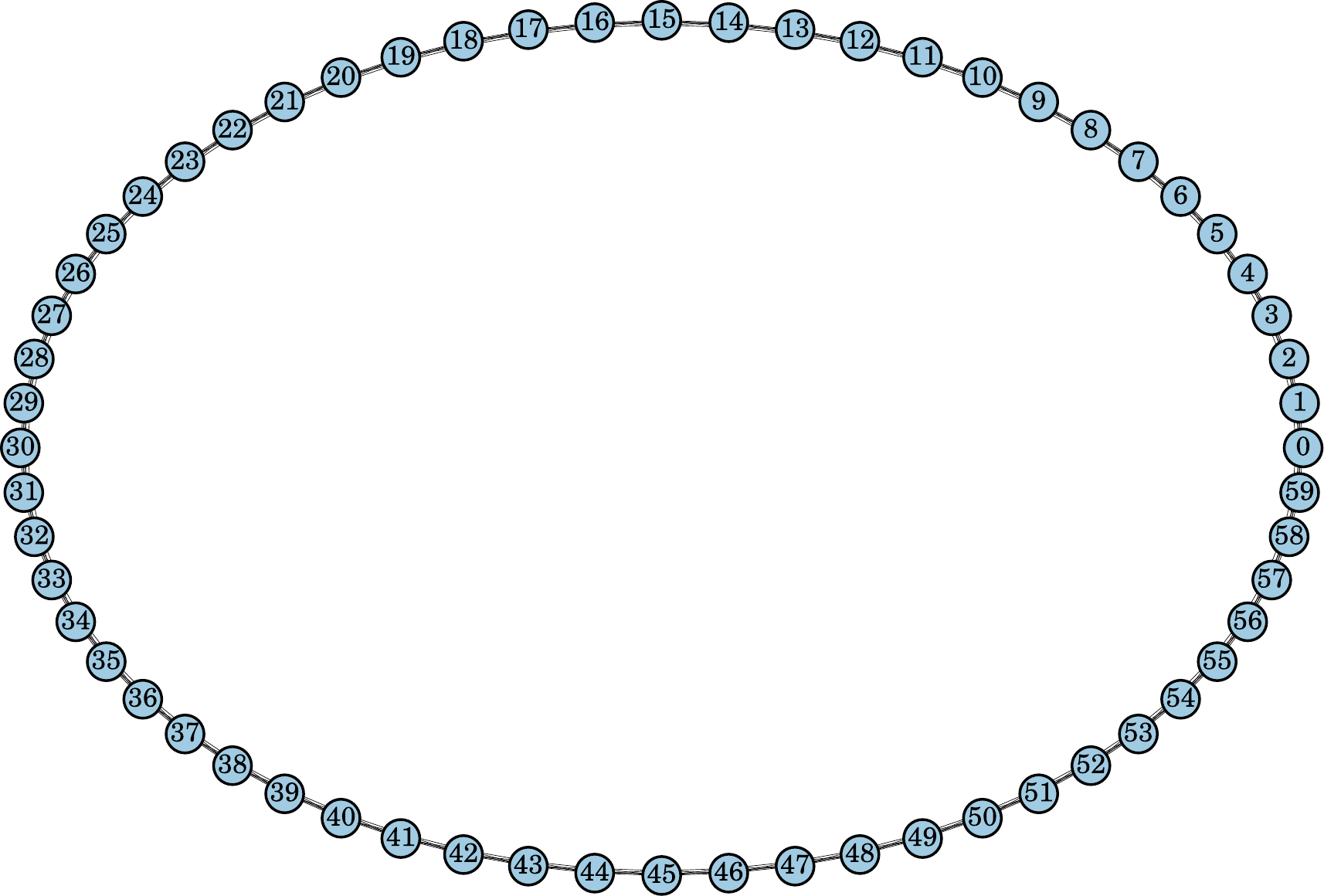}
    \includegraphics[width=0.24\textwidth]{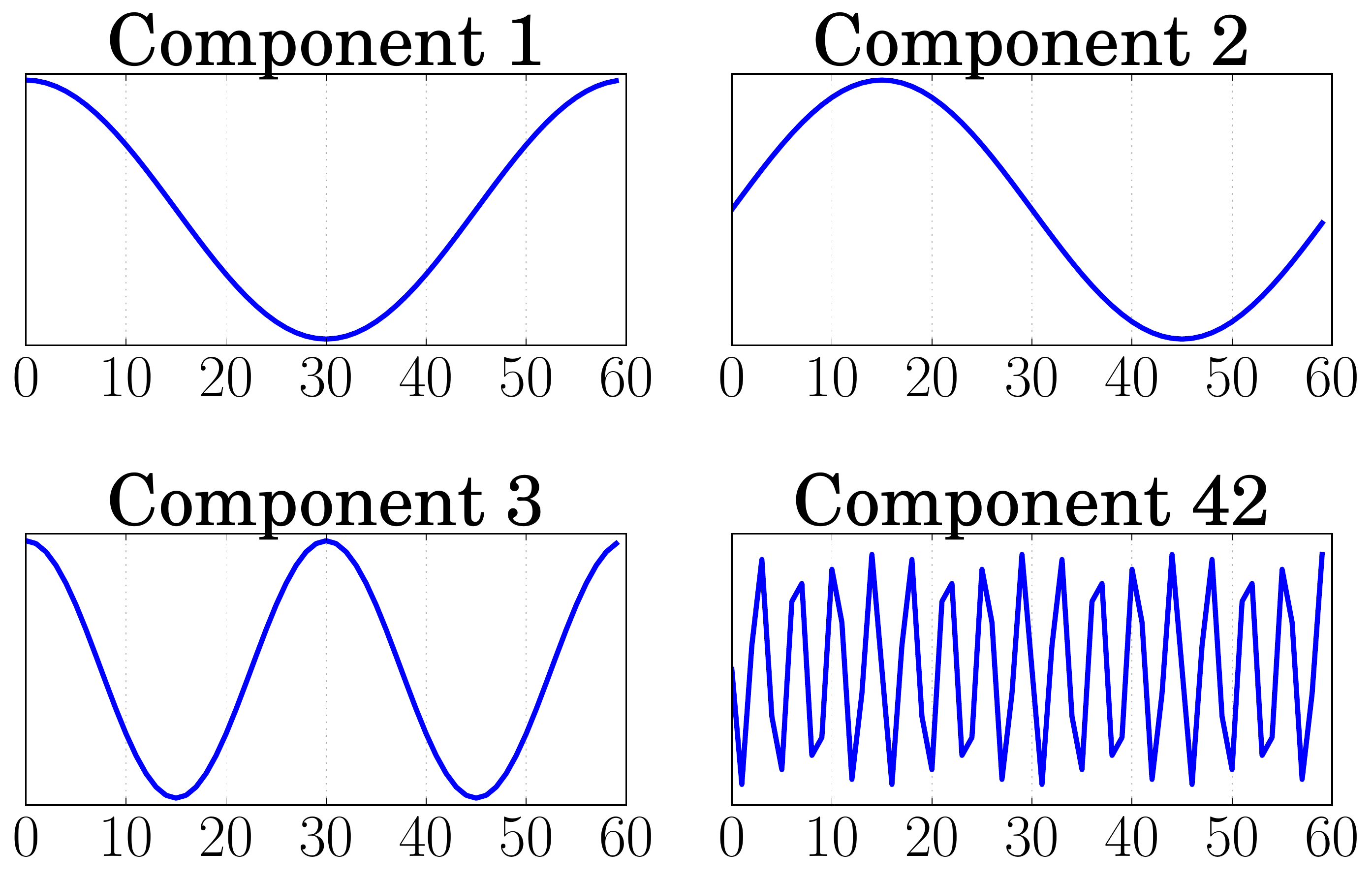}
  }
  
  \subfloat[\label{subfig:gs_ws-60-2-1}Watts-Strogatz model with $60$ vertices,
  $k=2$ and $p=0.1$ (\textbf{WS60-2-.1})]{
    \includegraphics[width=0.24\textwidth]{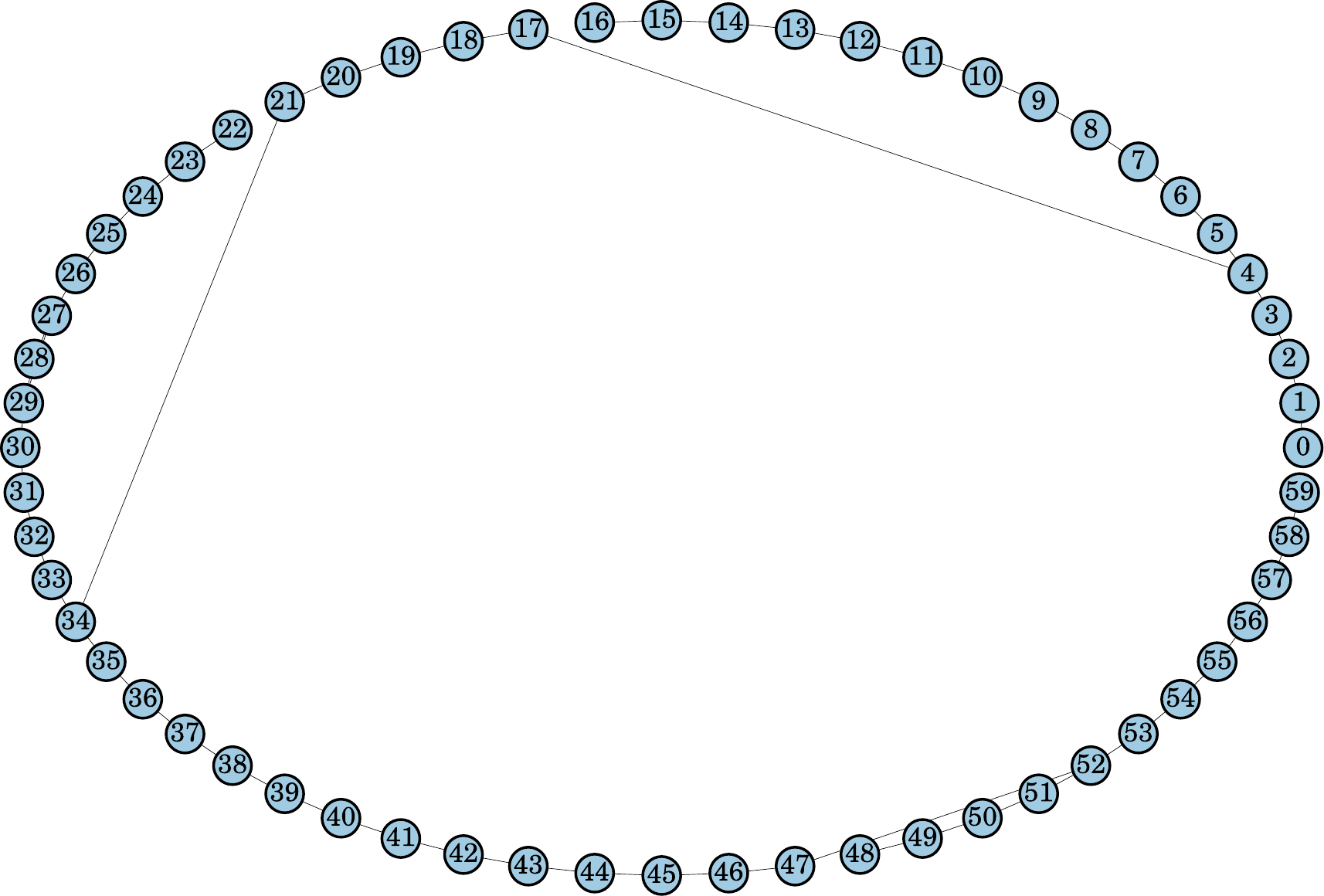}
    \includegraphics[width=0.24\textwidth]{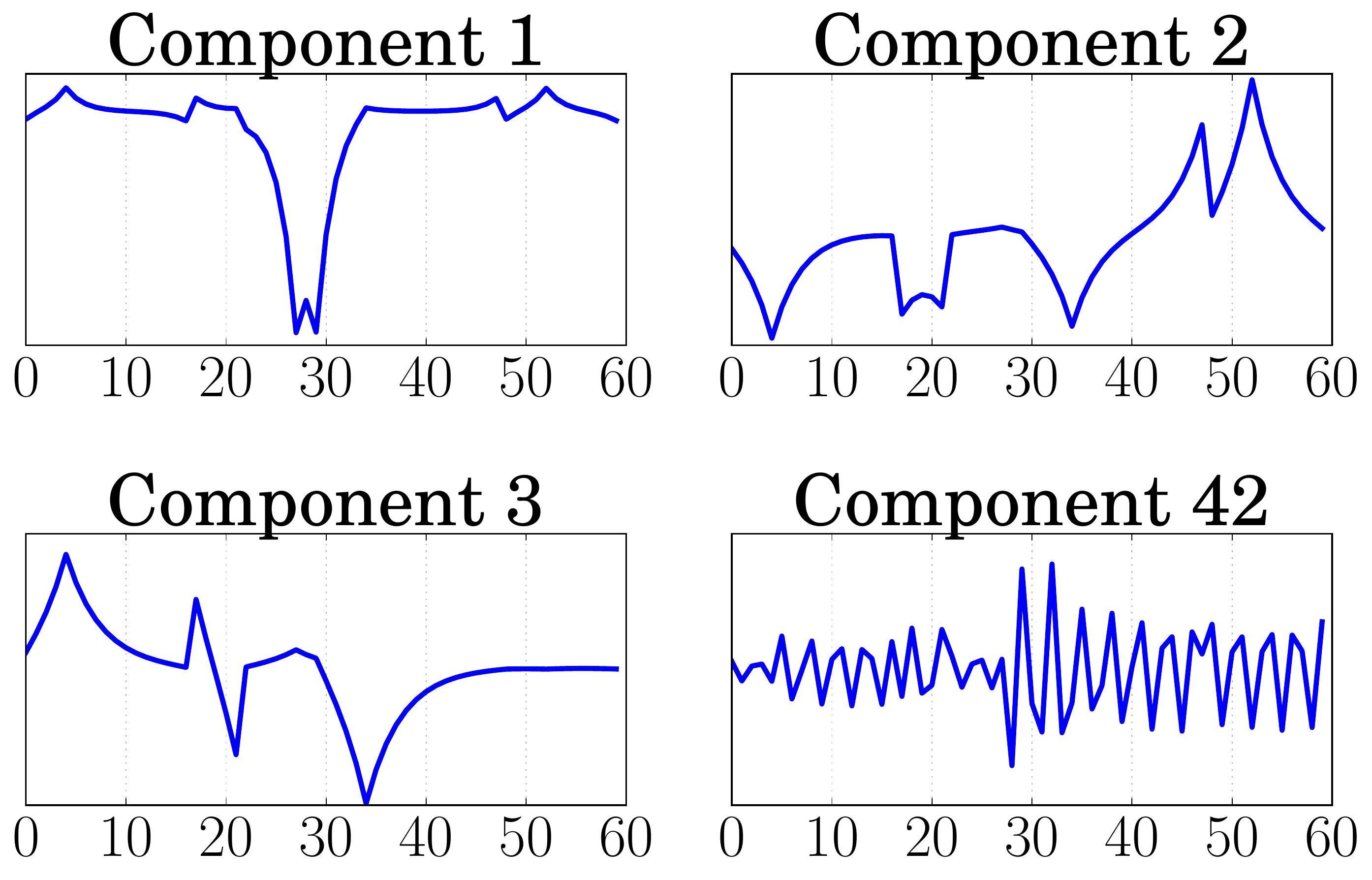}
  } 
  \hspace{0.05\textwidth} 
  \subfloat[\label{subfig:gs_ws-60-10-1}Watts-Strogatz model with $60$ vertices,
  $k=10$ and $p=0.1$ (\textbf{WS60-10-.1})]{
    \includegraphics[width=0.24\textwidth]{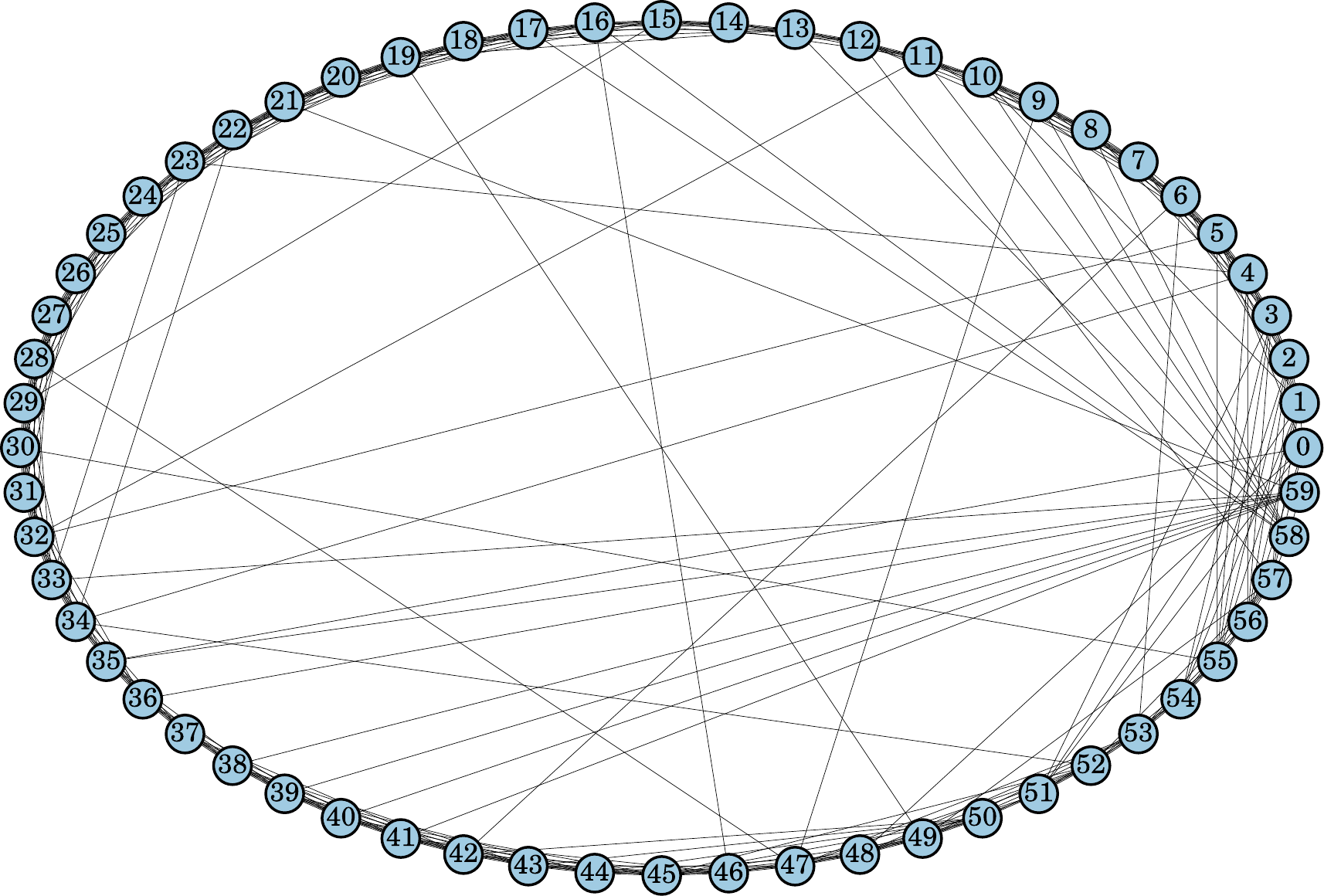}
    \includegraphics[width=0.24\textwidth]{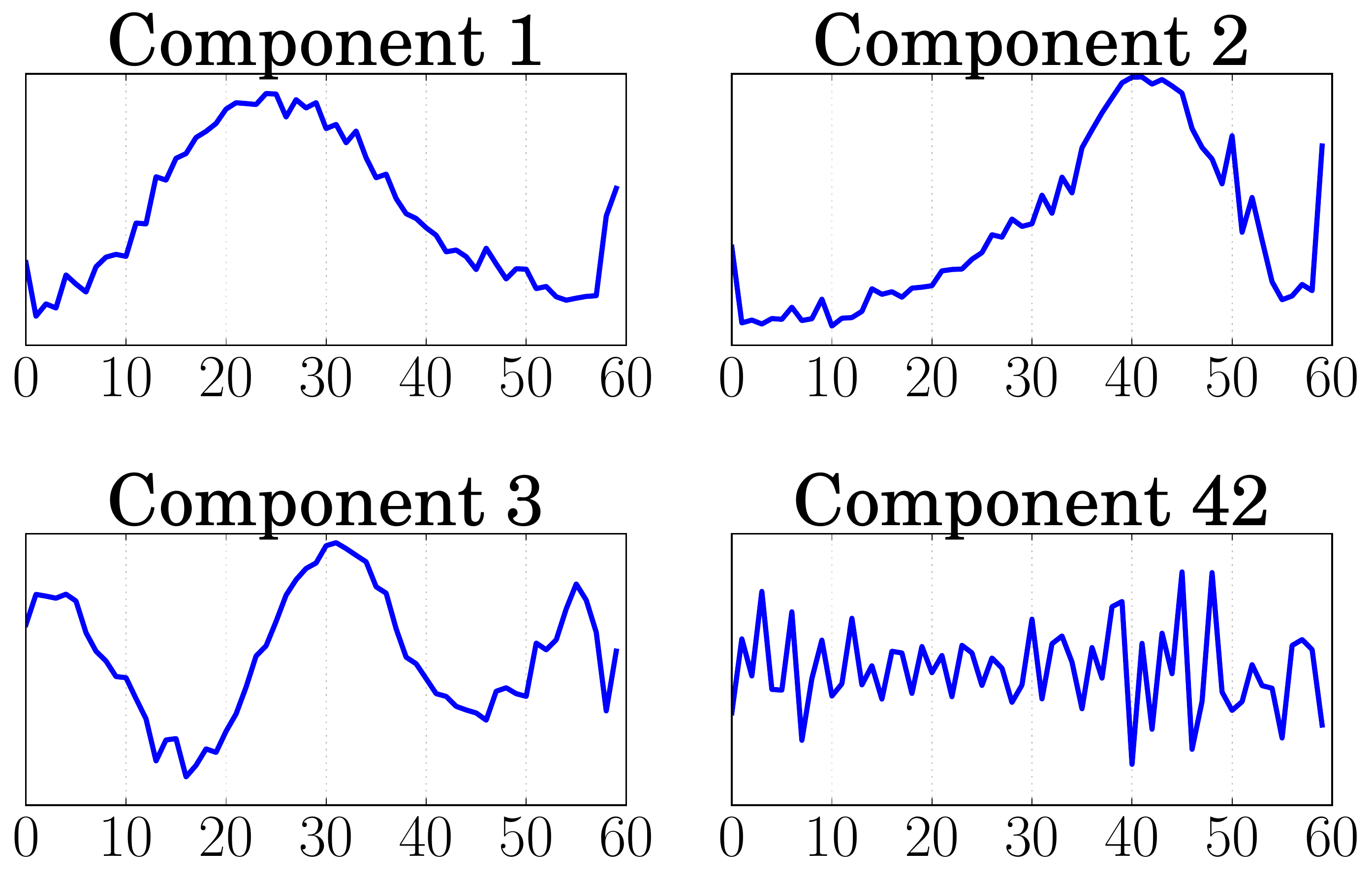}
  }
  
  \subfloat[\label{subfig:gs_sbm-60-2-7-10}Stochastic block model with 2
  communities, $60$ vertices, $p_w=0.7$ and $p_b=0.1$ (\textbf{SBM60-2})]
  {
    \includegraphics[width=0.24\textwidth]{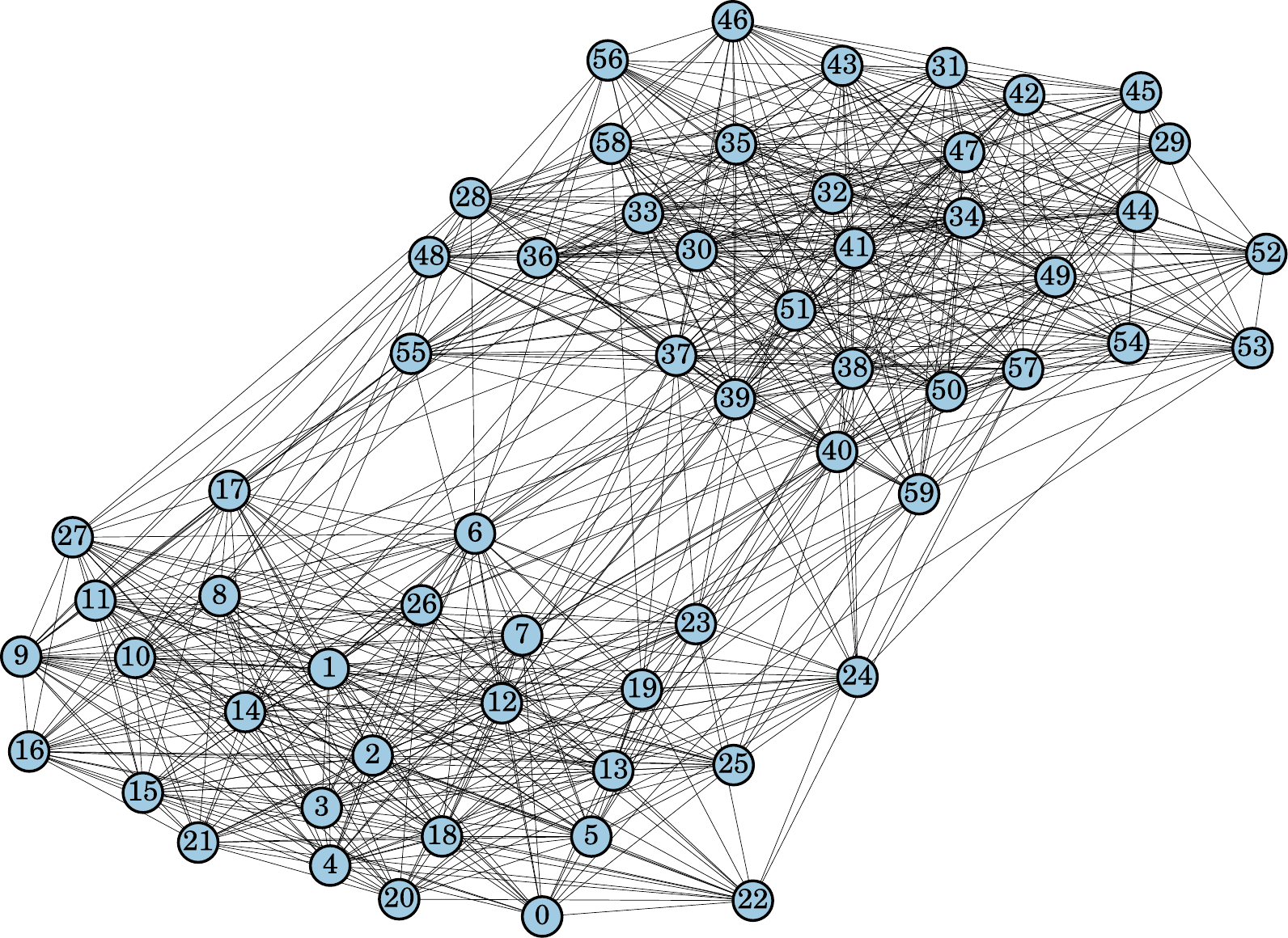}
    \includegraphics[width=0.24\textwidth]{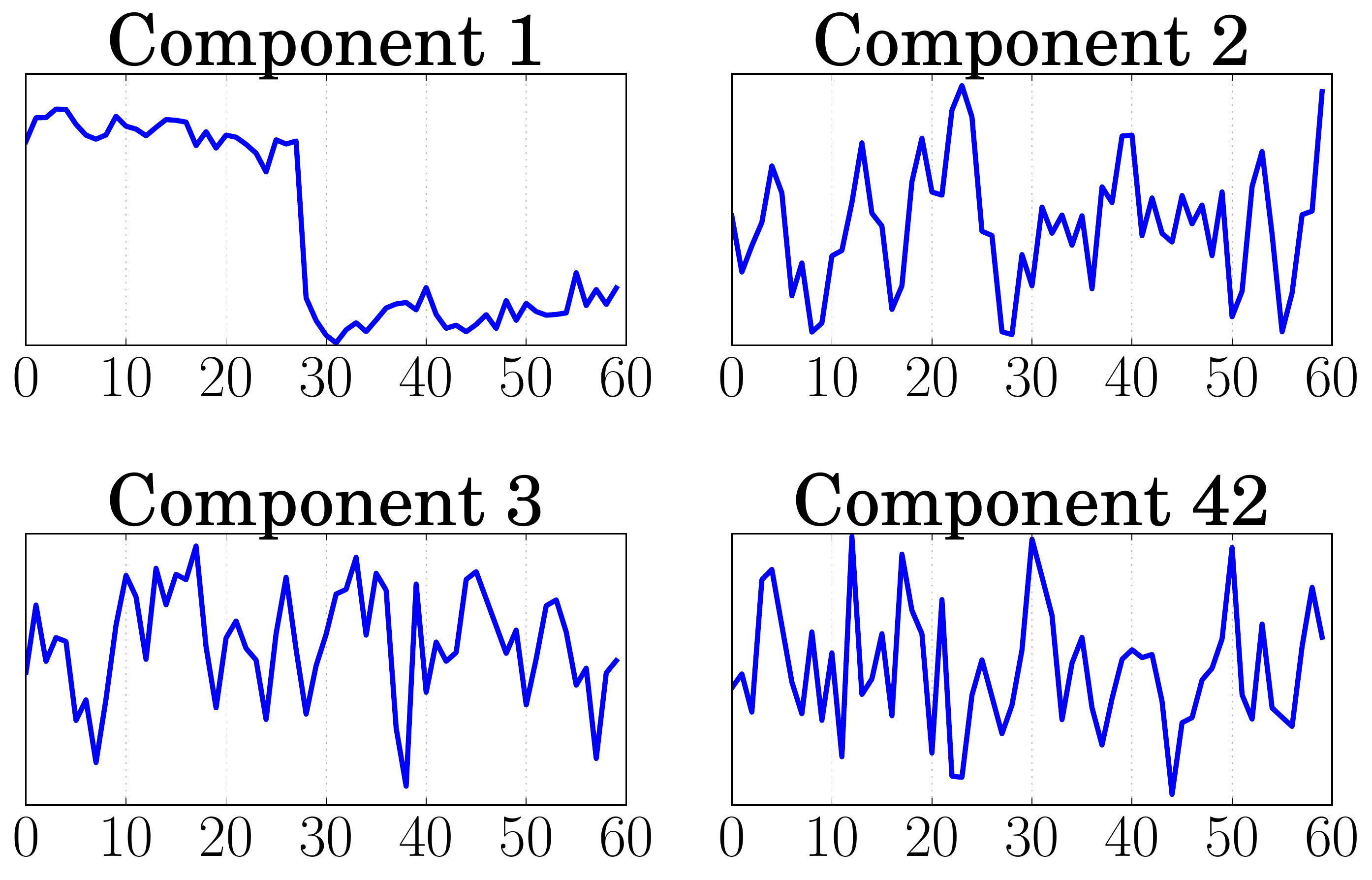}
  } 
  \hspace{0.05\textwidth}
  \subfloat[\label{subfig:gs_sbm-60-4-9-1}Stochastic block model with 4 
  communities, $60$ vertices, $p_w=0.9$ and $p_b=0.01$ (\textbf{SBM60-4})]
  {
    \includegraphics[width=0.24\textwidth]{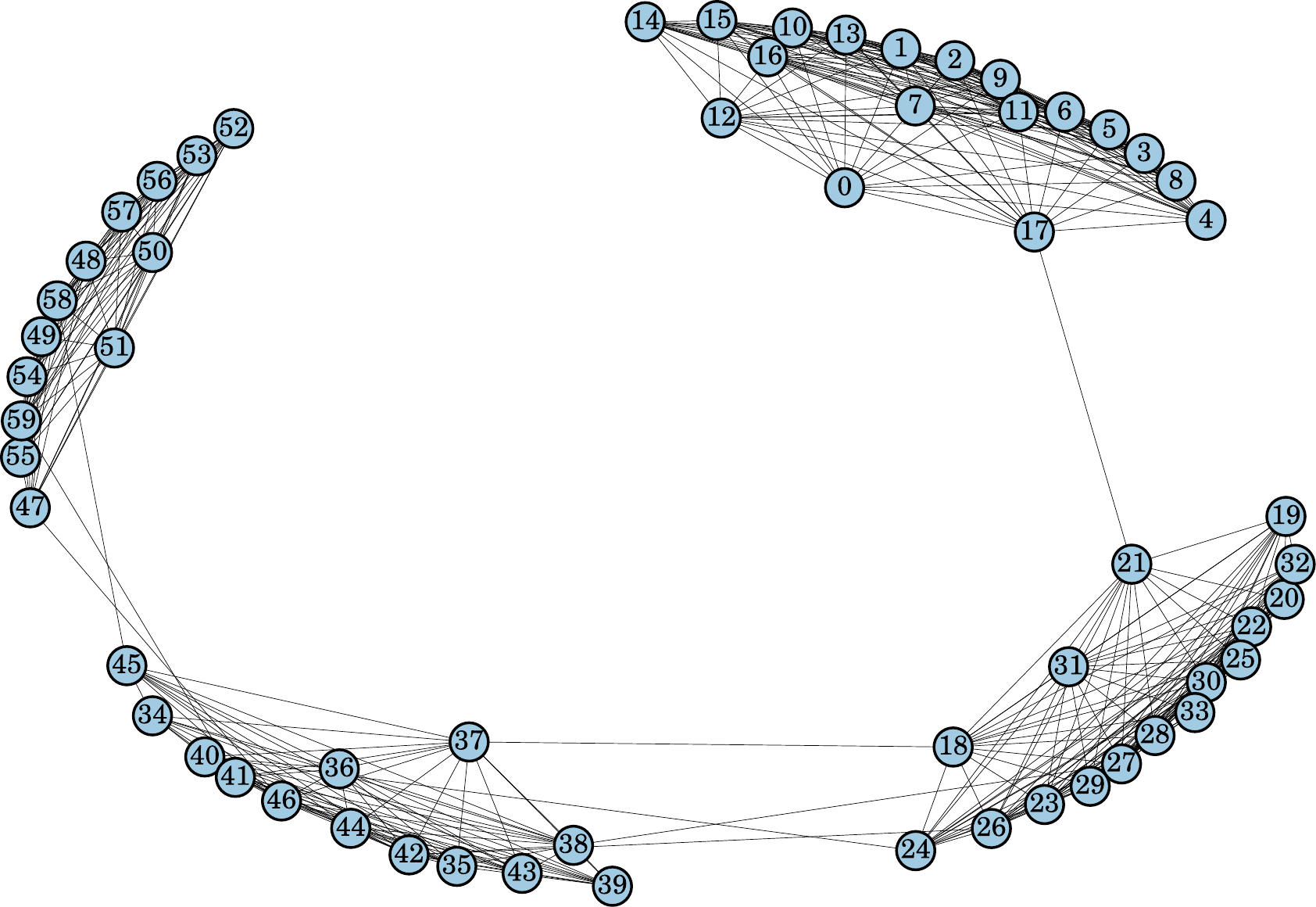}
    \includegraphics[width=0.24\textwidth]{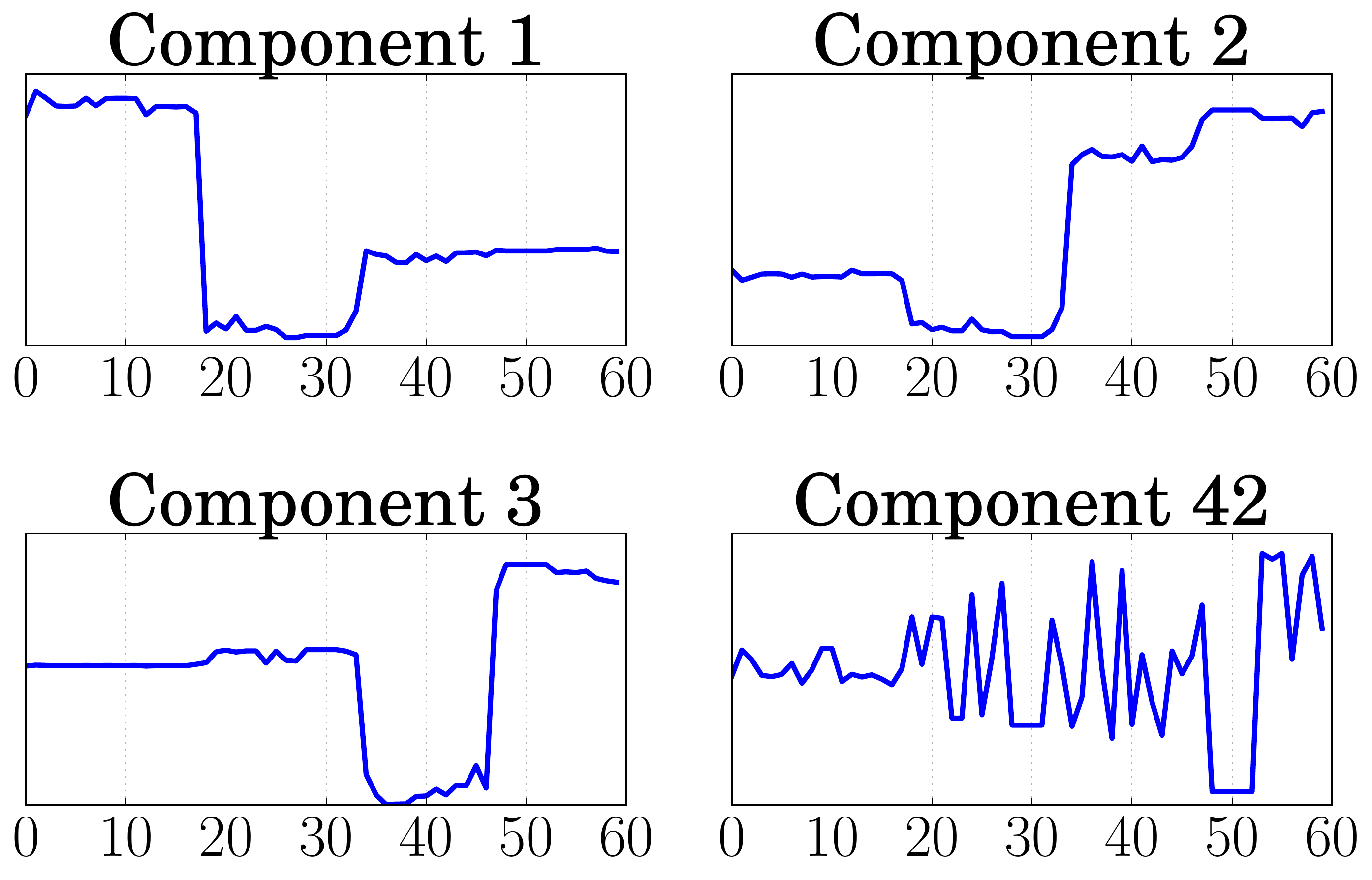}
  }
  
  \subfloat[\label{subfig:gs_er-60-4}Erd\"os-R\'enyi model with $60$ vertices
  and $p=0.4$ (\textbf{ER60-.4})]
  {
    \includegraphics[width=0.24\textwidth]{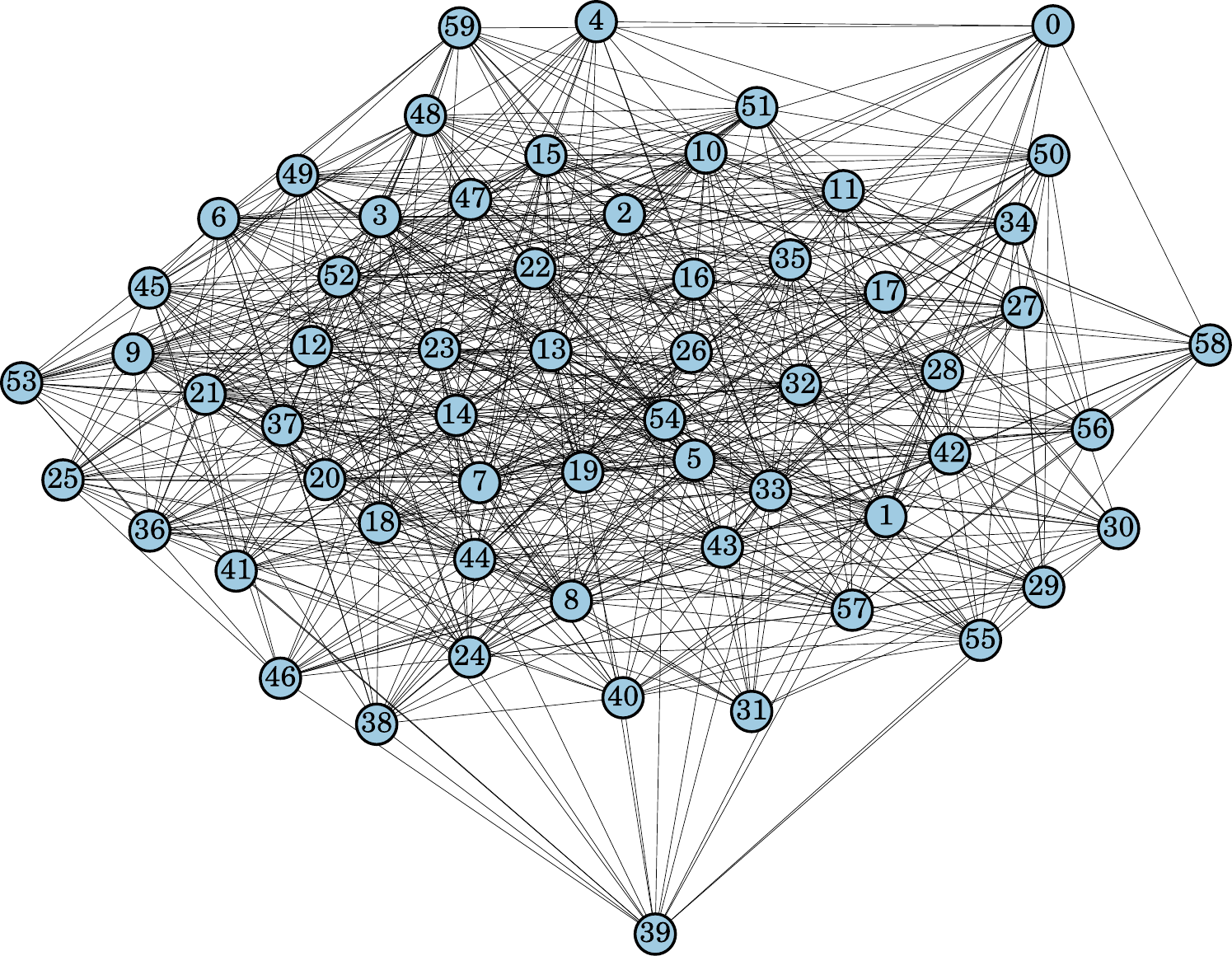}
    \includegraphics[width=0.24\textwidth]{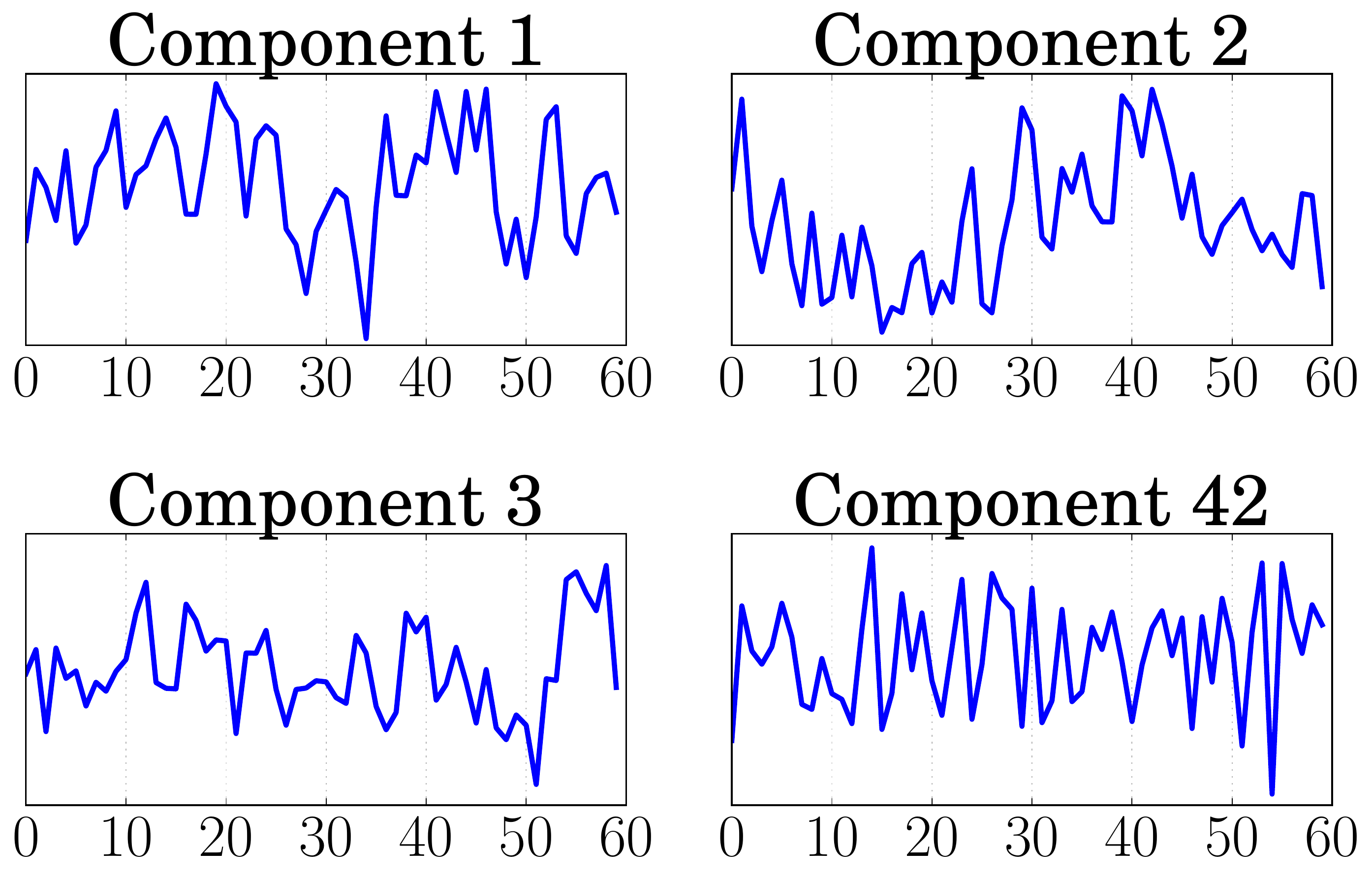}
  }
  \hspace{0.05\textwidth}
  \subfloat[\label{subfig:gs_bar-60-4}Barbell model with two cliques of $20$
  vertices and a path between them of $20$ vertices (\textbf{BAR60})]
  {
    \includegraphics[width=0.24\textwidth]{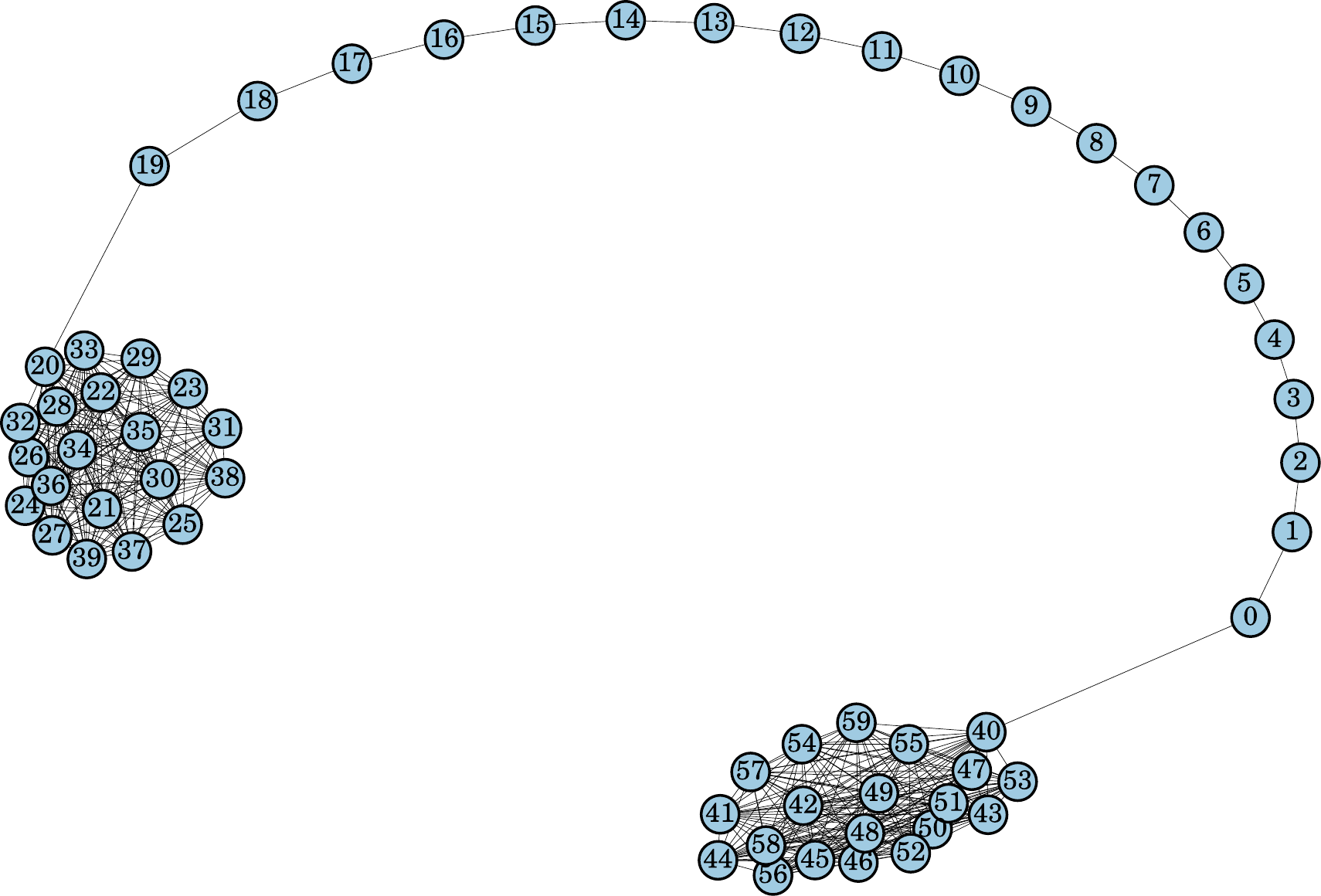}
    \includegraphics[width=0.24\textwidth]{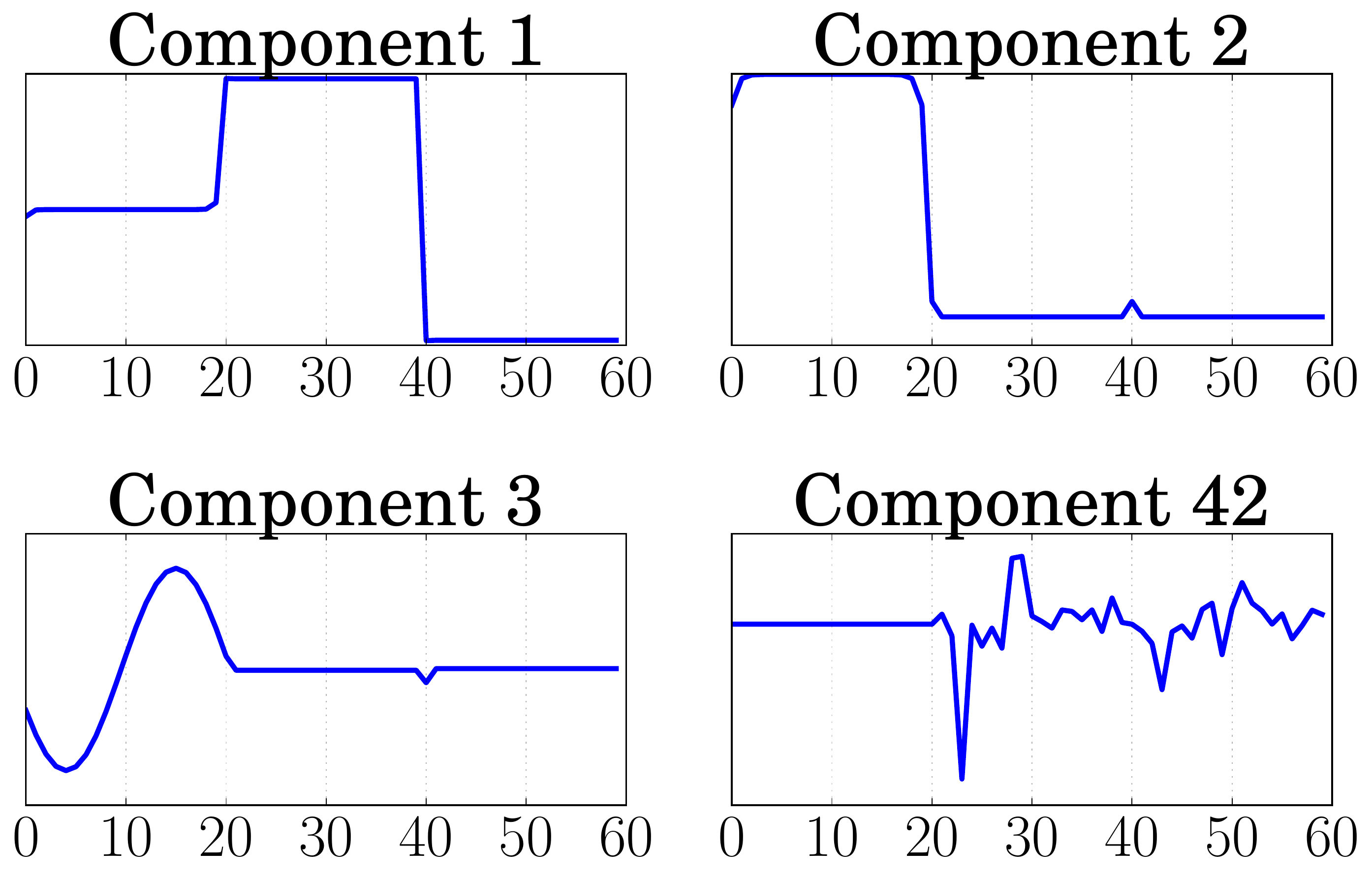}
  }
    
  \caption{\label{fig:GS_illustrations}Illustrations on several instances of
    graph models of the transformation of a graph into a collection of
    signals. All graphs have $n=60$ vertices. For each sub-figure, the left plot
    shows a two-dimensional representation of the graph, while the right plot
    displays the first three components, with high energies, and the arbitrarily
    chosen component 42, a low-energy component.}

\end{figure*}

This section introduces models that will be used to generate illustrations of 
the method throughout the article, as well as theoretical results associated to 
these models.

\paragraph*{$k$-ring lattices}

A $k$-ring lattice is a graph where each vertex $i$ is connected to the vertices 
$\{i-\frac{k}{2}, i-\frac{k}{2} + 1, \hdots, i-1, i+1, \hdots, i+\frac{k}{2}-1, 
i+\frac{k}{2}\}$, for $k \in \{2, 4, \hdots, n\}$. As Shimada et al. discussed 
in \cite{Shimada2012}, it is immediate to find expected eigenvalues and 
eigenvectors in this case using circulant matrix theory \cite{Gray2005}.

The computation of eigenvalues of $\bm{B}$ leads to
\begin{align} 
  \lambda_q = \frac{\alpha}{2n} \sum_{j=0}^{n-1} {\zeta^{jq}} - \frac{1}{2} \left( 
    \sum_{j=1}^{\frac{k}{2}} \zeta^{jq} + \sum_{j=n - \frac{k}{2}}^{n-1} \zeta^{jq} 
    + w^2\sum_{j=\frac{k}{2}+1}^{n-\frac{k}{2}-1} \zeta^{jq} \right)
\end{align}
where $\alpha = k + (n-1-k)w^2$ and $\zeta = e^{\frac{2i\pi}{n}}$ is the $n$th
root of the unity.  The eigenvectors are given by the columns of the
Fourier matrix. Details of the calculation are given in \ref{anx:kring}.

When the eigenvalues are ordered by the value of $q$, the eigenvectors are 
considered in increasing order of frequencies. The components are however sorted 
according to the energy of eigenvalues $\lambda_k$ when applying CMDS, which 
correspond to the sorting according to the value of $q$ only when $k=2$. In this 
case, the resulting signals are then harmonic oscillations whose frequencies 
increase as lower-energy components are considered. When $k$ increases, the 
components are no longer sorted by frequencies.

\paragraph*{Watts-Strogatz model}

The Watts-Strogatz model \cite{Watts1998} has been developed to build graphs 
with small-world property, where the average length of the shortest paths 
between vertices is low compared to the number of vertices, while the clustering 
coefficient is high. This property has been highlighted in many systems, such as 
for instance social networks. From a $k$-ring lattice, each edge $(u,v)$ is 
rewired with probability $p$ to become $(u,w)$, where $w$ is uniformly chosen 
among other vertices. Shimada et al. give a second-order approximation of the 
expected eigenvalues and eigenvectors according to the probability $p$, using 
perturbation theory. They showed that the correlation between the approximation 
and the actual signals is high when $p$ is low, and decreases when $p$ 
increases, which is consistent with intuition.

\paragraph*{Erd\"os-R\'enyi model}

A random graph is a graph whose set $\V$ of vertices is fixed and links 
between these vertices are drawn randomly \cite{Durrett2007}.  Random matrix 
theory suggests that eigenvectors are random vectors in $\mathbb{R}^n$, 
naturally conjectured to be distributed as i.i.d. Gaussian vectors 
\cite{Dekel2011}. As for eigenvalues, the Wigner results \cite{VanMieghem2011} 
show that the maximal eigenvalue is comprised between 
$np(1-\frac{1}{\sqrt{n}})$ and $np(1+\frac{1}{\sqrt{n}})$ while the remaining 
eigenvalues follow a semicircle law on the interval $[ -\frac{1}{\sqrt{n}}, 
\frac{1}{\sqrt{n}}]$, giving the energies of signals.

\paragraph*{Stochastic block model}

A simple stochastic block model \cite{Karrer2011} is used to generate a graph 
with communities. Each of the $n$ vertices is assigned to one of the $K$ 
communities, and edges between each pair of vertices is randomly and 
independently drawn according to probabilities depending on the group of 
vertices: if the two vertices belong to the same community, the probability, 
noted $p_w$, is close to $1$ while otherwise, the probability between vertices 
of different groups, noted $p_b$, is lower. The settings of $p_w$ and $p_b$ 
lead 
to customized density of edges within and between communities. In 
\cite{Meila2001}, intuitions about the shape of the signals are given, in an 
application of segmentation of images. They suggest that the eigenvectors of 
block matrices are piecewise constant with respect to the communities.

\paragraph*{Barbell model}

The barbell model \cite{Aldous2002} is defined as two cliques, each containing 
$n_c$ vertices, linked by a path with $n_p$ vertices. Based on the results 
defined previously, the shape of the eigenvectors is expected to be close to 
the 
ones in the case of stochastic block model, since each clique is community, as 
well as to the ones in the $2$-regular lattices, which is encountered in the 
path part of the graph.

\subsection{Illustrations}
\label{subsec:gs_illustrations}

Figure~\ref{fig:GS_illustrations} shows illustrations of the transformation of 
a graph into a collection of signals, on several instances of graph models 
introduced in the previous section, all set with $n=60$ vertices. For each 
sub-figure, the left plot shows a two-dimensional representation of the studied 
graph, while the right plot displays the first three components, with high 
energies, and the arbitrarily chosen component 42, a low-energy component.

In all cases, the obtained signals are consistent with the expected results 
described in the previous section. For the $k$-ring lattices 
(Figure~\ref{subfig:gs_rl-60-2} for $k=2$ and Figure~\ref{subfig:gs_rl-60-10} 
for $k=10$), the signals are harmonic oscillations, grouped in pairs of signals 
with the same frequency and a difference of phase equal to $\frac{\pi}{2}$. For 
the high-energy components, the value of $k$ does not have any influence. 
However, the frequency of signals differs, as expected, for lower-energy 
components, as the sorting of components is guided by the eigenvalues. 

When noise is added, here by considering the Watts-Strogatz model with a 
probability of rewiring equal to $p=0.1$ (Figure~\ref{subfig:gs_ws-60-2-1} for 
$k=2$ and Figure~\ref{subfig:gs_ws-60-10-1} for $k=10$), the signals are also 
noisy, as described by Shimada et al.~\cite{Shimada2012}. According to the 
value of $k$, the consequences of the noise on the shape of the resulting 
signals differ: for $k=10$, more edges are rewired, but the structure is barely 
affected since each vertex in linked with its $10$ nearest neighbors: The ring 
is then preserved, leading to noisy harmonic oscillations. However, when 
$k=2$, 
one single rewiring causes the disappearance of the ring: the structure of the 
graph is then different from a regular lattice, leading to high changes in 
signals.

Figures~\ref{subfig:gs_sbm-60-2-7-10} and \ref{subfig:gs_sbm-60-4-9-1} show two 
instances of the stochastic block model, the first one with two communities with 
$p_w=0.7$ and $p_b=0.1$ and the second one with four communities and $p_w=0.9$ 
and $p_b=0.01$. An interesting observation is that the high-energy components 
display the structure of communities, with noisy plateaus corresponding to the 
dense parts of the graph. It is worth noting that the number of relevant 
high-energy components is equal to the number of communities in the graph minus 
one, as one component is sufficient to discriminate two communities. As for the 
low-energy components, they are noisy signals, corresponding to the structure 
inside communities. These random structures correspond to a random graph of type 
Erd\"os-R\'enyi, an instance of which is represented in 
Figure~\ref{subfig:gs_er-60-4}: the resulting signals do not exhibit any 
structure, only looking as  white noise. 

Finally, Figure~\ref{subfig:gs_bar-60-4} displays an instance of the Barbell 
model, which combines features obtained for the previous instances: plateaus for 
the first high-energy components, as highlighted for the stochastic block model, 
describing the cliques, harmonic oscillation of the third component (only the 
part which corresponds to a path), found for the $k$-ring lattice, and finally 
noisy signal for low-energy components, as seen for the Erd\"os-R\'enyi model. 

These illustrations show the connection between graph structure and the 
resulting signals after transformation, that will be used in 
Section~\ref{sec:spectral_analysis} to study the topology of the graph using 
spectral analysis, and to perform standard operations, such as filtering, on 
graphs. Before that, a study of the inverse transformation is proposed, as it 
will enable us to represent a collection of signals in the graph domain.


\section{Inverse transformation: From signals to graph}
\label{sec:inverse_transformation}

\subsection{Statement of the problem}

Transformation from signals, or time series, to graph has been studied in many
applications, as in \cite{Campanharo2011} or \cite{Nunez2012}, in order to use
network theory as a tool for the understanding of time series. These methods are
nevertheless of no use in our case, because the signals are themselves a
representation of a specific graph, hence it can be seen as a restoration
problem.  Ignoring this would lead to inconsistent results between the
represented and the reconstructed graphs. Hence, the inverse transformation
shall take into account this knowledge to preserve the original topology of the
underlying graph. By construction of the collection of signals $\bm{X}$, the
perfect retrieval of the underlying graph is easily reachable, by considering
the distances between each point: As built using CMDS, these distances represent
the adjacency matrix. However, when $\bm{X}$ is degraded or modified, for
instance by filtering (see Section~\ref{subsec:filtering}), the distances are no
longer directly the ones computed between vertices, even if they stay in the
neighborhood of these distances. It is nevertheless not possible to directly
retrieve the comprehensive set of links of the graph. This case is yet worth
considering if processing of signals is performed in a goal of analysis:
altering the signals will also alter the distances, preventing the direct
inverse transformation.

\begin{figure*}

  \subfloat[\label{subfig:it_illus_58_0}]{
  \includegraphics[width=0.24\textwidth]{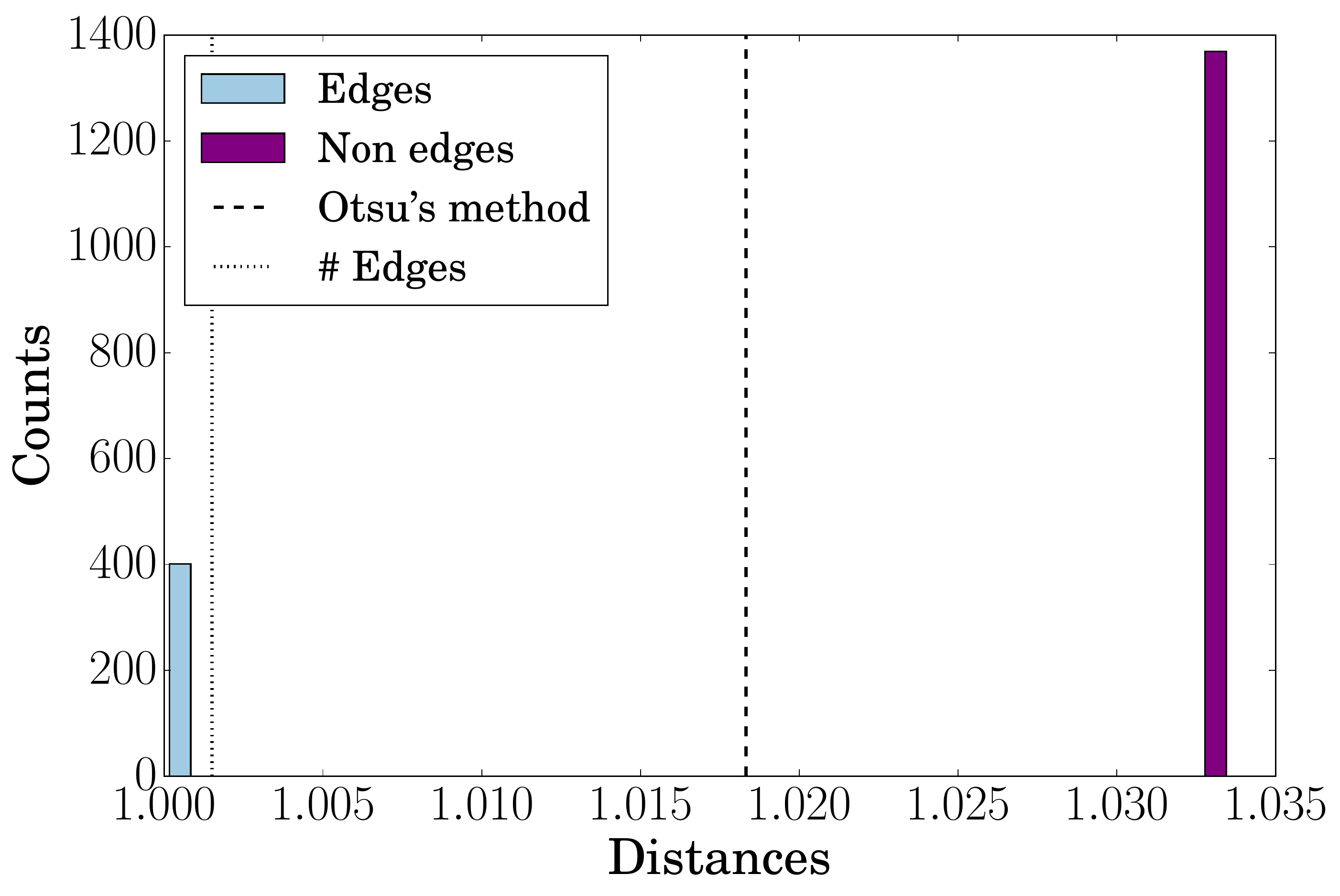}
}
  \subfloat[\label{subfig:it_illus_57_0}]{
  \includegraphics[width=0.24\textwidth]{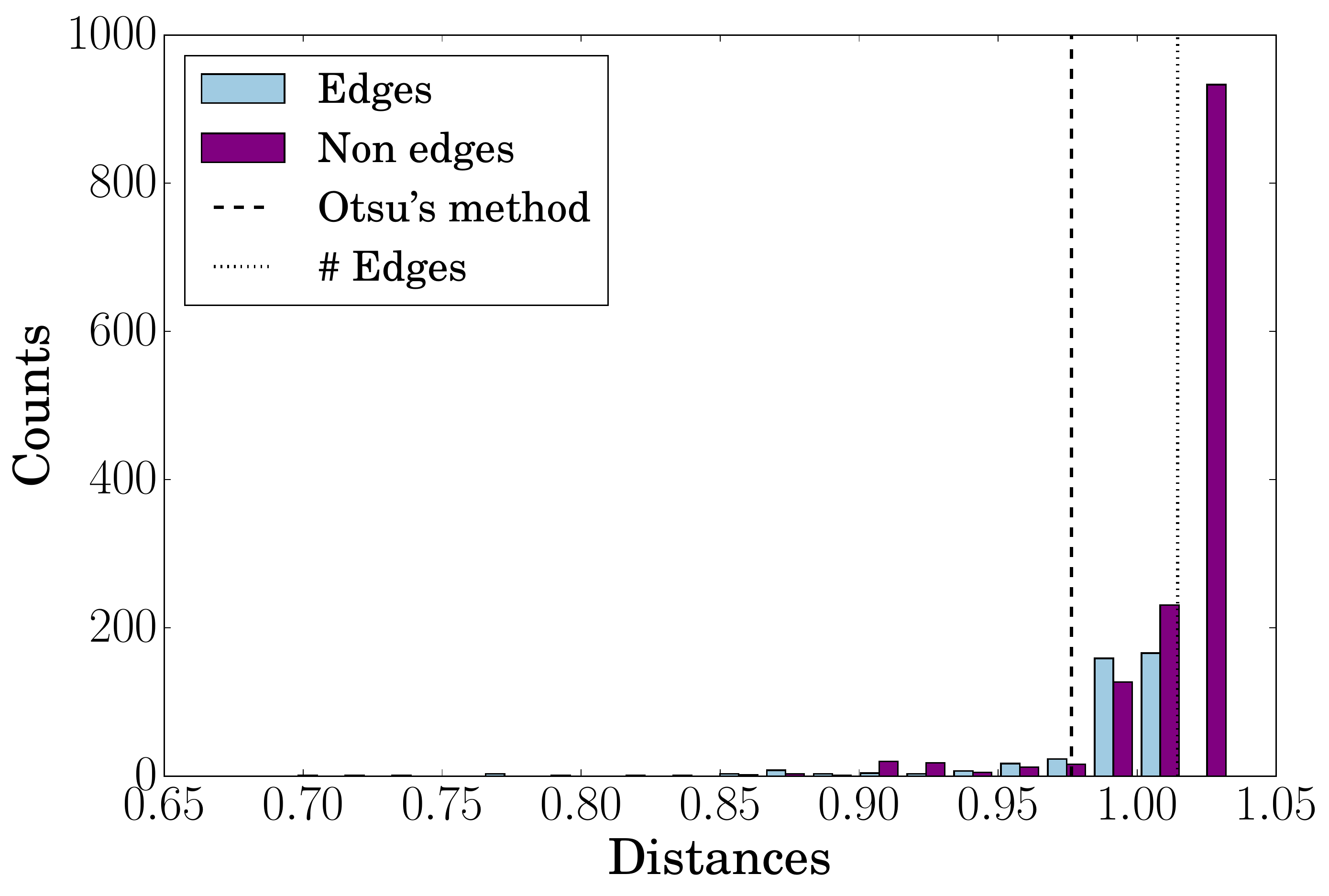}
}
  \subfloat[\label{subfig:it_illus_57_2}]{
  \includegraphics[width=0.24\textwidth]{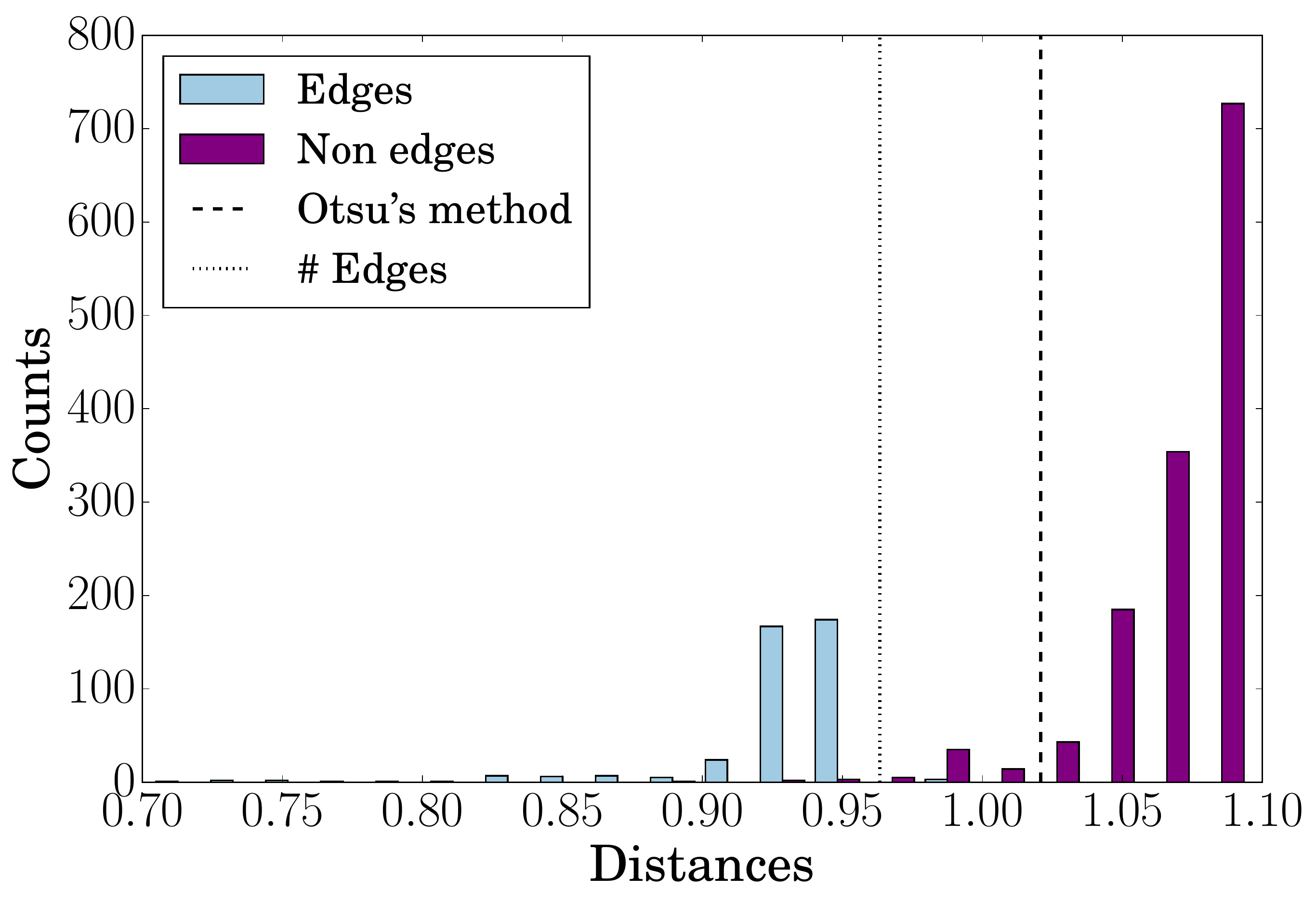}
}
  \subfloat[\label{subfig:it_illus_57_4}]{
  \includegraphics[width=0.24\textwidth]{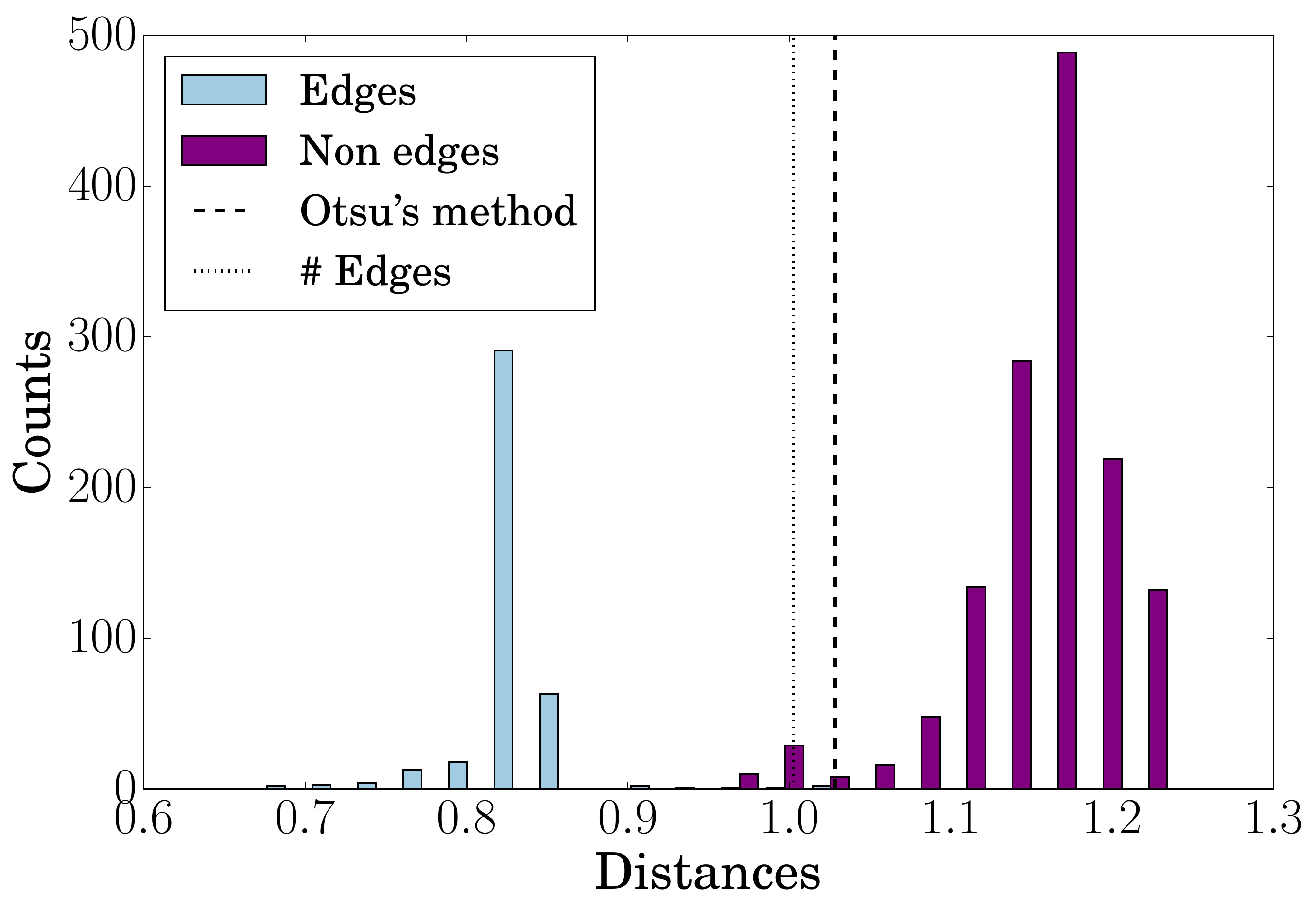}
}
\caption{\label{fig:IT_illustrations}Inverse transformation from a collection of
  signals obtained after transformation of an instance of the stochastic block
  model with four communities (\textbf{SBM60-4}). The figures show the
  histograms of distances between points
  $\{d(\bm{X})_{ij}\}_{i,j \in \{1, \hdots, n\}}$ in the Euclidean distances:
  light blue bars represent distances corresponding to edges while dark purple
  bars represent distances corresponding to non-edges. The dashed line
  represents the threshold obtained using Adapted Otsu's method (see
  Algorithm~\ref{algo:otsu_method}), while the dotted line represents the
  threshold obtained using the number of edges of the original graph. The
  signals are considered either non-degraded or degraded, obtained by removing
  the lowest-energy component. The computation of distances is performed without
  or with considering energies of components (a) Non-degraded case and
  unweighted distances: the distances are either equal to 1 or
  $w$. Discrimination between edges and non-edges is trivial. (b) Degraded case
  and unweighted distances: distribution of distances are mixed and no simple
  thresholding is feasible. (c) Degraded case and weighted distances with
  $\alpha=2$: the distributions are separated and thresholding is good (d)
  Degraded-case and weighted distances with $\alpha=4$: the distributions are
  better separated.}

\end{figure*}

We propose in the following section a robust inverse transformation, based on 
the thresholding of distances. Two enhancements are discussed to improve the 
separability of distances. Finally, we discuss two methods of thresholding.

\subsection{Robust inverse transformation}

The two contributions we propose in the following aim at addressing the drawback 
raised in the statement of the problem: the two distributions of distances 
representing edges and non-edges, can overlap when the collection of signals is 
degraded, preventing an efficient thresholding.

Let us consider a collection of signals $\bm{X}$ with $K$ components. The 
objective of the inverse transformation is to obtain an adjacency matrix 
$\tilde{\bm{A}}$, from the distances $\bm{D}(\bm{X})$.  This 
adjacency matrix describes the graph $\tilde{\G} = (\tilde{\V}, \tilde{\E})$.

\paragraph*{Energy-weighted distances}

\begin{figure*}
  \centering
  \subfloat[\label{subfig:it_illus2_A_0}Original adjacency matrix]{
  \includegraphics[width=0.25\textwidth]{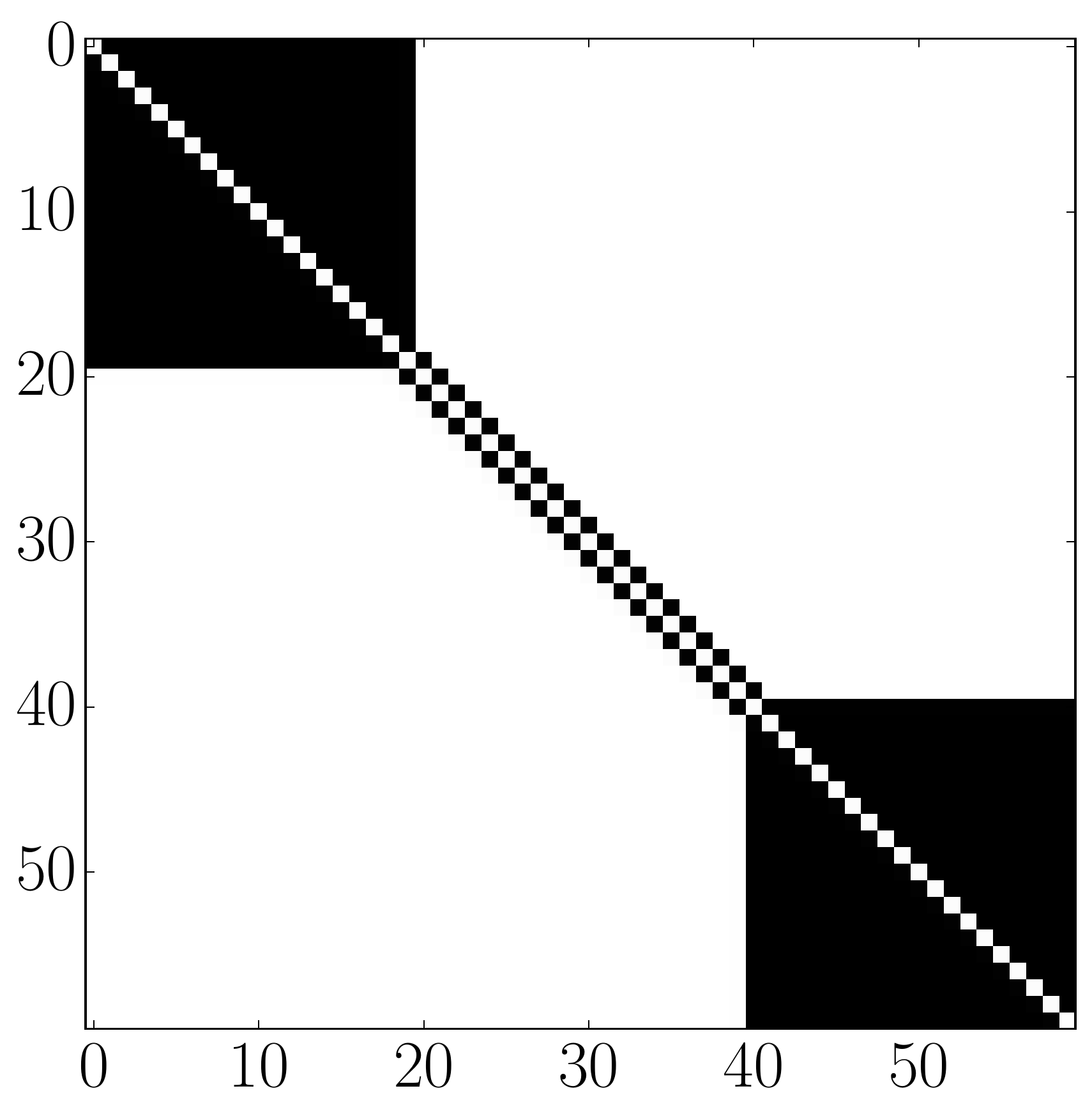}
}
  \subfloat[\label{subfig:it_illus2_A_F_10_0}Final adjacency matrix with 
  $K=10$]{
  \includegraphics[width=0.25\textwidth]{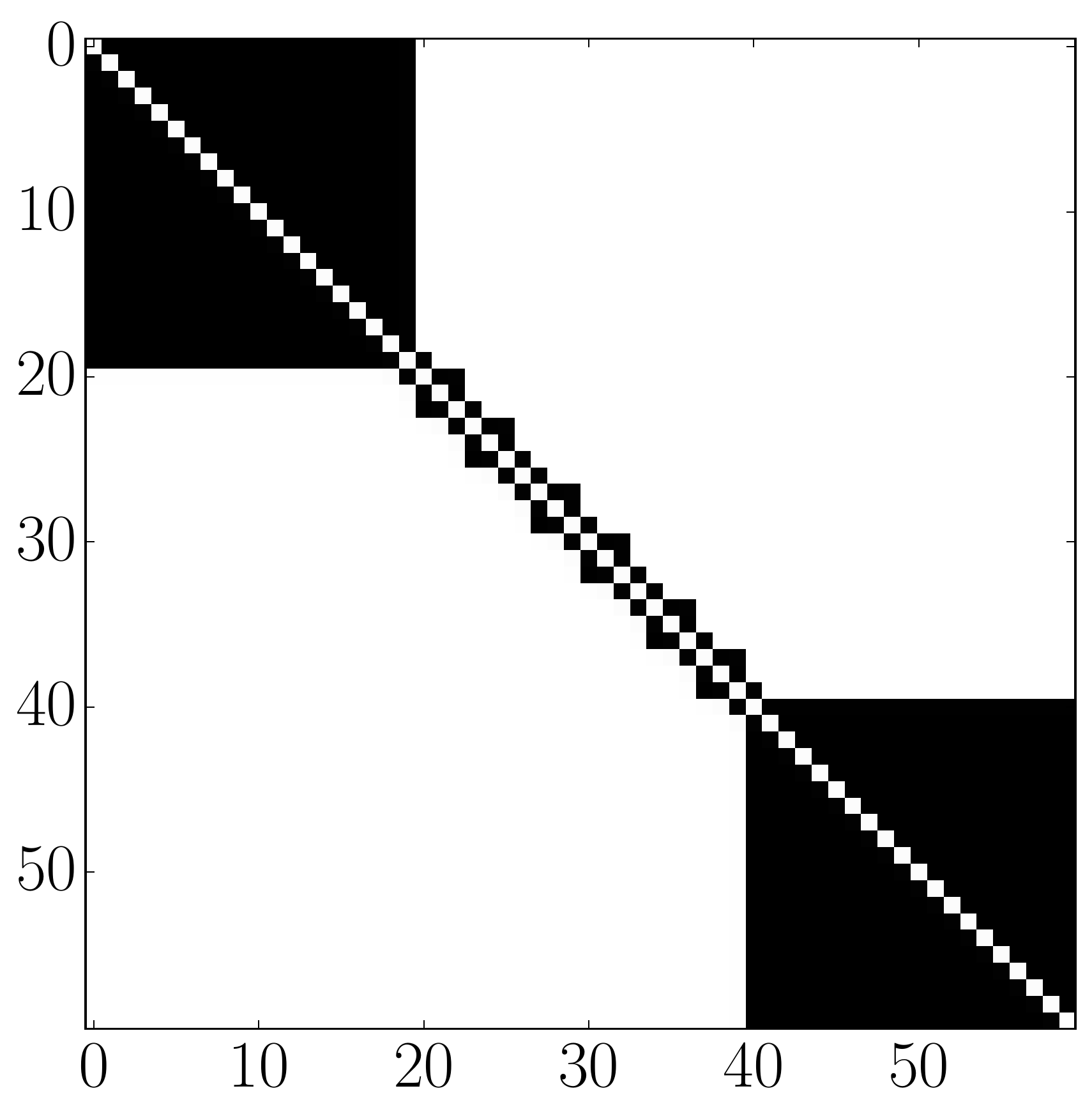}
}

  \subfloat[\label{subfig:it_illus2_A_10_0_2}Adjacency matrix after 
  considering the first three components: $\tilde{\bm{A}} = 
  \tilde{\bm{A}}_1 + \tilde{\bm{A}}_2 + \tilde{\bm{A}}_3$]{
  \includegraphics[width=0.25\textwidth]{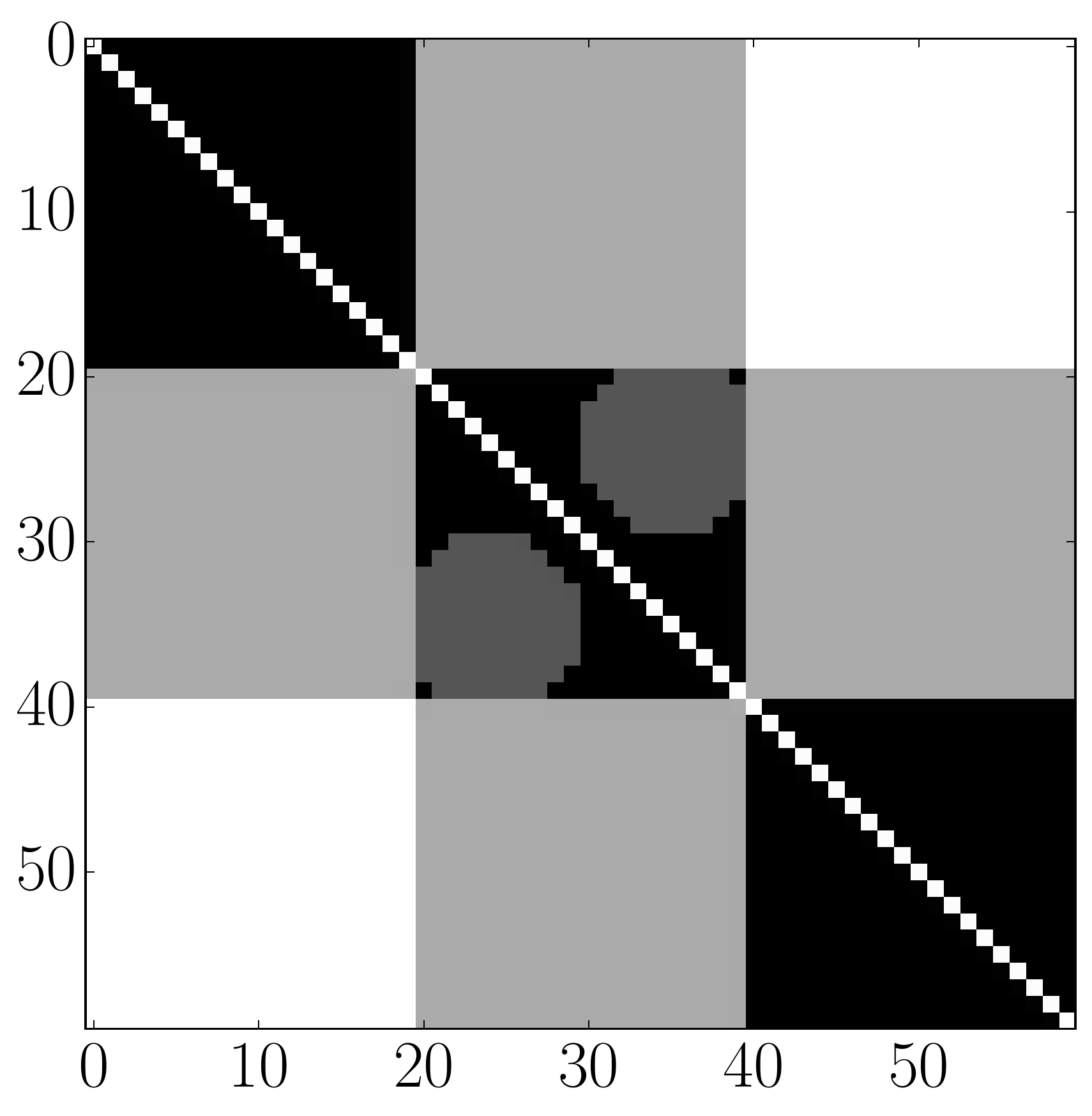}
}
  \subfloat[\label{subfig:it_illus2_A_10_0_9}Adjacency matrix 
  before thresholding at Step~6 of Algorithm~\ref{algo:sequential_update}]{
  \includegraphics[width=0.25\textwidth]{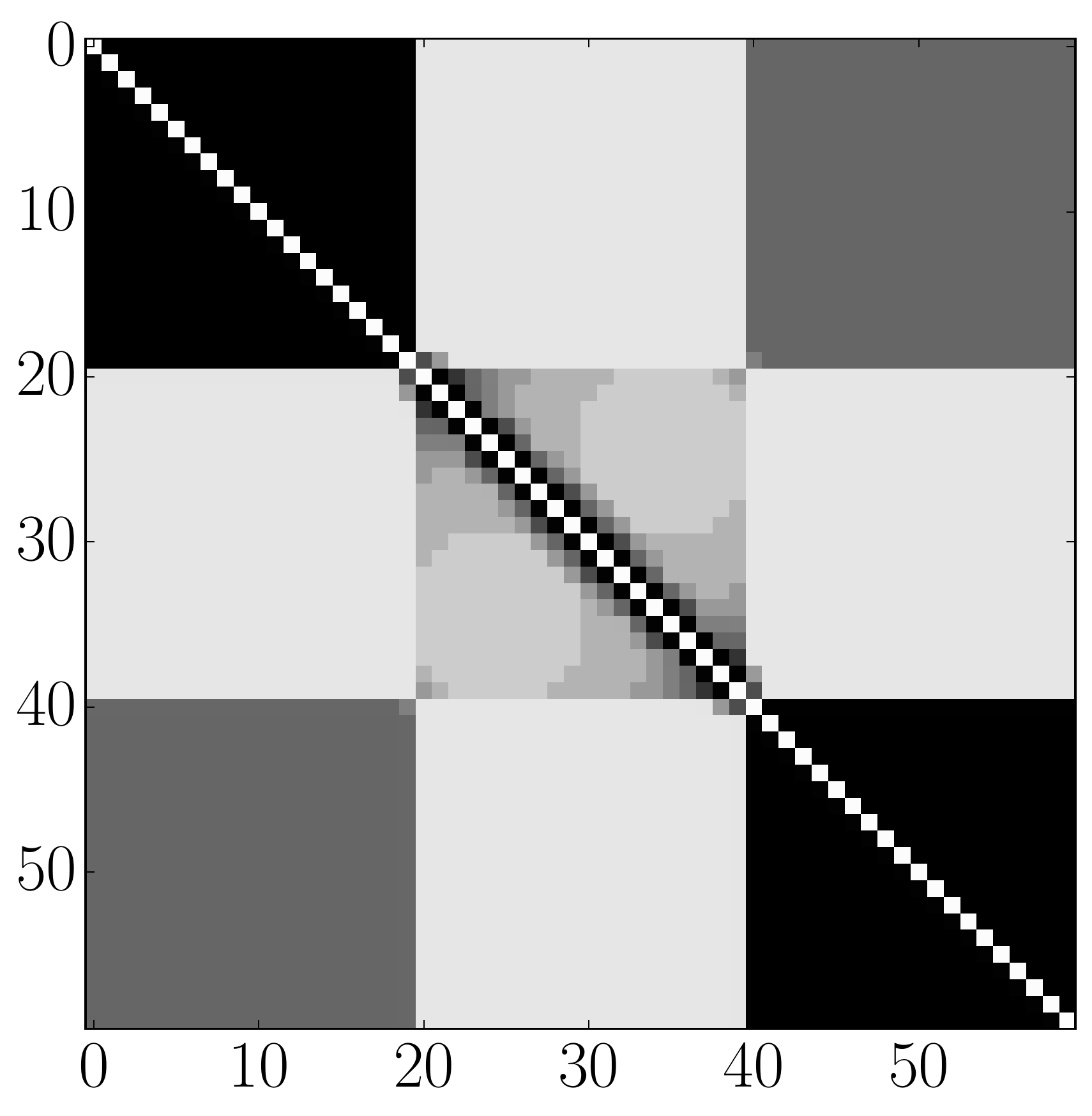}
}
  \caption{\label{fig:IT_illustrations_2}Illustration of the sequential update 
  of the adjacency matrix to retrieve the underlying graph from a collection of 
  signals. The original graph is an instance of the Barbell model, composed of two 
  cliques of $20$ vertices linked between them by a path of $20$ vertices 
  (\textbf{BAR60}). The collection of signals $\bm{X}$ is degraded by keeping only 
  the first $K=10$ components. The color codes the intensities, from white to 
  black.} 

\end{figure*}

An initial observation to improve the distinction between distances 
representing edges and those representing non-edges is that the energies of 
components are not taken into account in the processing of the distance matrix 
$\bm{D}(\bm{X})$. The high-energy components have though a strong influence in 
the description of the global topology of the graph: If the distance between 
two vertices $i$ and $j$ in a high-energy component is high, it means that the 
two vertices are likely to be distant in the graph. Conversely, if the distance 
in a high-energy component is low, then the two vertices are likely to be 
connected in the graph. Neglecting the importance of energies of components 
comes to forget the hierarchy of components in their description of the 
global structure of the graph. One way to add this information is to compute a 
distance weighted by the energies of the components: $d(\bm{X})_{ij} = 
\sqrt{\sum_{k=1}^K {e_k^\alpha(x_{ik} - x_{jk})^2}}$ with $\alpha \geq 0$,  
where $e_k$ is the energy of component $k$, computed as $e_k = \sum_{i=1}^n 
{x_{ik}^2}$, and normalized such that $\sum_{k=1}^K e_k^\alpha = 1$. The 
parameter $\alpha$ controls the importance of the weighting: if $\alpha$ is 
high, the high-energy components have a higher importance in the computation of 
distances compared to the low-energy components. 
Figures~\ref{subfig:it_illus_57_2} and \ref{subfig:it_illus_57_4} display the 
distribution of distances representing edges and those representing non-edges, 
using respectively $\alpha=2$ and $\alpha=4$. Compared with the distributions 
in Figure~\ref{subfig:it_illus_57_0}, the two distributions are better 
discriminated when $\alpha$ increases, enabling a better retrieval of the 
underlying graph. The right choice of $\alpha$ is nevertheless empirical and 
depends on the collection of signals.

\paragraph*{Sequential update of the adjacency matrix}

Weighted computation of distances enables the inverse transformation to take 
into account the energies of components, and produces, as shown in 
Figure~\ref{fig:IT_illustrations}, a better separation of distances. Energies of 
components could be taken into account in another way by considering 
intermediary adjacency matrices $\tilde{\bm{A}_k}$, where $\tilde{\bm{A}_k}$ is 
the adjacency matrix obtained by retaining only the first $k$ components using 
the method described in the introduction. Thus, these different structures, at 
different scales, are successively considered: the first components describe the 
structure at large scale, and adding one-by-one the components increase weights 
for the less dominating edges. Summing over all these intermediary states is 
based on the assumption that if an edge exists in the graph, it will be present 
in many intermediary states. The final adjacency matrix $\tilde{\bm{A}}$ will 
then have values between $0$ and $K$, and has to be thresholded to obtain a 
binary matrix. The choice of the threshold is arbitrarily set to $\frac{2K}{3}$. 
This choice has been guided by empirical studies, by looking at the distribution 
of distances. Intuitively, it means that a link has to be present in two thirds 
of the adjacency matrix to be considered as reliable. The algorithm for the 
sequential update of the adjacency matrix is then described in 
Algorithm~\ref{algo:sequential_update}.

\begin{algorithm}
  \caption{Sequential update of the adjacency matrix
  \label{algo:sequential_update}}
  \begin{algorithmic}[1]
    \ENSURE $\bm{X}$ the collection of signals
    \STATE Define $\tilde{\bm{A}}$ as an empty adjacency matrix
    \FOR{k from $1$ to $K$}
    \STATE Compute the adjacency matrix $\tilde{\bm{A}_k}$ from collection
    of signals reduced to the first $k$ components
    \STATE Add $\tilde{\bm{A}_k}$ to $\tilde{\bm{A}}$
    \ENDFOR
    \STATE Set values of $\tilde{\bm{A}}$ lower than $\frac{2K}{3}$ to 0 
    and greater than $\frac{2K}{3}$ to 1
    \RETURN $\tilde{\bm{A}}$
  \end{algorithmic}
\end{algorithm}

Figure~\ref{fig:IT_illustrations_2} gives an example of sequential construction 
of the adjacency matrix $\tilde{\bm{A}}$ on an instance of the Barbell model 
described in Section~\ref{subsec:gs_illustrations}, composed of two cliques of 
$20$ vertices linked by a path of $20$ vertices (\textbf{BAR60}). 
Figure~\ref{subfig:it_illus2_A_0} shows the adjacency matrix of the graph. The 
transformation is performed on a degraded collection of signals, obtained by 
keeping the first $10$ components of the original collection of signals, over 
$59$ components. Figure~\ref{subfig:it_illus2_A_10_0_2} gives the adjacency 
matrix $\tilde{\bm{A}}$ after the summing of adjacency matrices obtained by 
retaining successively the first, the first two and the first three components: 
$\tilde{\bm{A}} = \tilde{\bm{A}}_1 + \tilde{\bm{A}}_2 + \tilde{\bm{A}}_3$. The 
three communities appear visually, as well as the particularity of the middle 
one, which is actually a path. The retrieved topology is much more accurate when 
the sum of adjacency matrices goes on: Figure~\ref{subfig:it_illus2_A_10_0_9} 
plots the adjacency matrix at the end of the loop, before applying the 
thresholding in Step~6 of Algorithm~\ref{algo:sequential_update}: 
$\tilde{\bm{A}} = \sum_{k=1}^{10} {\tilde{\bm{A}}_k}$. After the thresholding, 
the final adjacency matrix $\tilde{\bm{A}}$ is binary, as plotted in 
Figure~\ref{subfig:it_illus2_A_F_10_0}: The topology of the obtained graph is 
very close to the original one, even if only a small portion of components is 
retained.

\begin{algorithm}
  \caption{Adapted Otsu's method\label{algo:otsu_method}}
  \begin{algorithmic}[1]
    \ENSURE $\bm{d}$ the distribution of distances
    \FOR{Each value of $\bm{d}$ set as threshold $\tau$}
    \STATE Compute the sets $\bm{d}_B$ and $\bm{d}_F$ as the values of 
    $\bm{d}$ respectively lower and greater than the threshold 
    \STATE Compute $w_B$ and $w_F$ as the proportion of elements in 
    respectively $\bm{d}_B$ and $\bm{d}_F$ 
    \STATE Compute the variances $v_B$ and $v_F$  of respectively 
    $\bm{d}_B$ and 	$\bm{d}_F$
    \STATE Store for the threshold $\tau$ the total variance within classes 
    $t_\tau = w_Bv_B + w_Fv_F$.
    \ENDFOR
    \STATE Select $\bar{\tau}$ such that $t_\tau$ is minimum.
    \RETURN $\bar{\tau}$
  \end{algorithmic}
\end{algorithm}

\paragraph*{Selection of the threshold}

The selection of the threshold is a crucial step in the differentiation of 
distances representing edges from those representing non-edges. A first 
approach, mentioned in \cite{Shimada2012}, is to preserve the number of edges 
in the reconstructed graph: the selection of the threshold is then computed by 
considering the smallest distances as edges until the number of edges in the 
original graph is reached. This approach could be restrictive if this piece of 
information is not available. More specifically, if the degraded collection of 
signals has a high impact on the topology of the underlying graph, then the 
number of edges in the reconstructed graph could be significantly different 
from the number of edges in the original graph. We propose in the following 
another approach to threshold the distances, analogous to binarization in image 
processing, where a gray-scale image is compressed in a black-and-white image: 
From a range of gray levels, a threshold is computed such that the details of 
the picture are preserved at best when the number of levels is reduced to $2$ 
(black and white). A well-known algorithm to perform such binarization has been 
proposed by Otsu \cite{Otsu1975} to segment gray-scale pictures: Considering 
two classes, the algorithm finds the threshold among all possible thresholds 
that minimizes the variance within classes. In our context, the distribution is 
composed of all distances between pairs of points, which differs from gray 
levels, since the number of possible distances might be equal to the number of 
distances, while in images, the number of levels is fixed (for instance 256 
levels for an 8-bit color images, whatever the size of the image). We propose 
an adaptation of the Otsu's method in Algorithm~\ref{algo:otsu_method}.

The choice of the number of edges in the constructed graph is then guided by 
the obtained distances, and is not dependent of the original structure of the 
graph. Figure~\ref{fig:IT_illustrations} shows for the four histograms of 
distances the obtained  thresholding using the number of edges (dotted line) and 
the adapted Otsu's method (dashed line).

\subsection{Performance}

\begin{figure*}

  \subfloat[\label{subfig:it_comp_rl-60-2}\textbf{RL60-2}]{
  \includegraphics[width=0.24\textwidth]{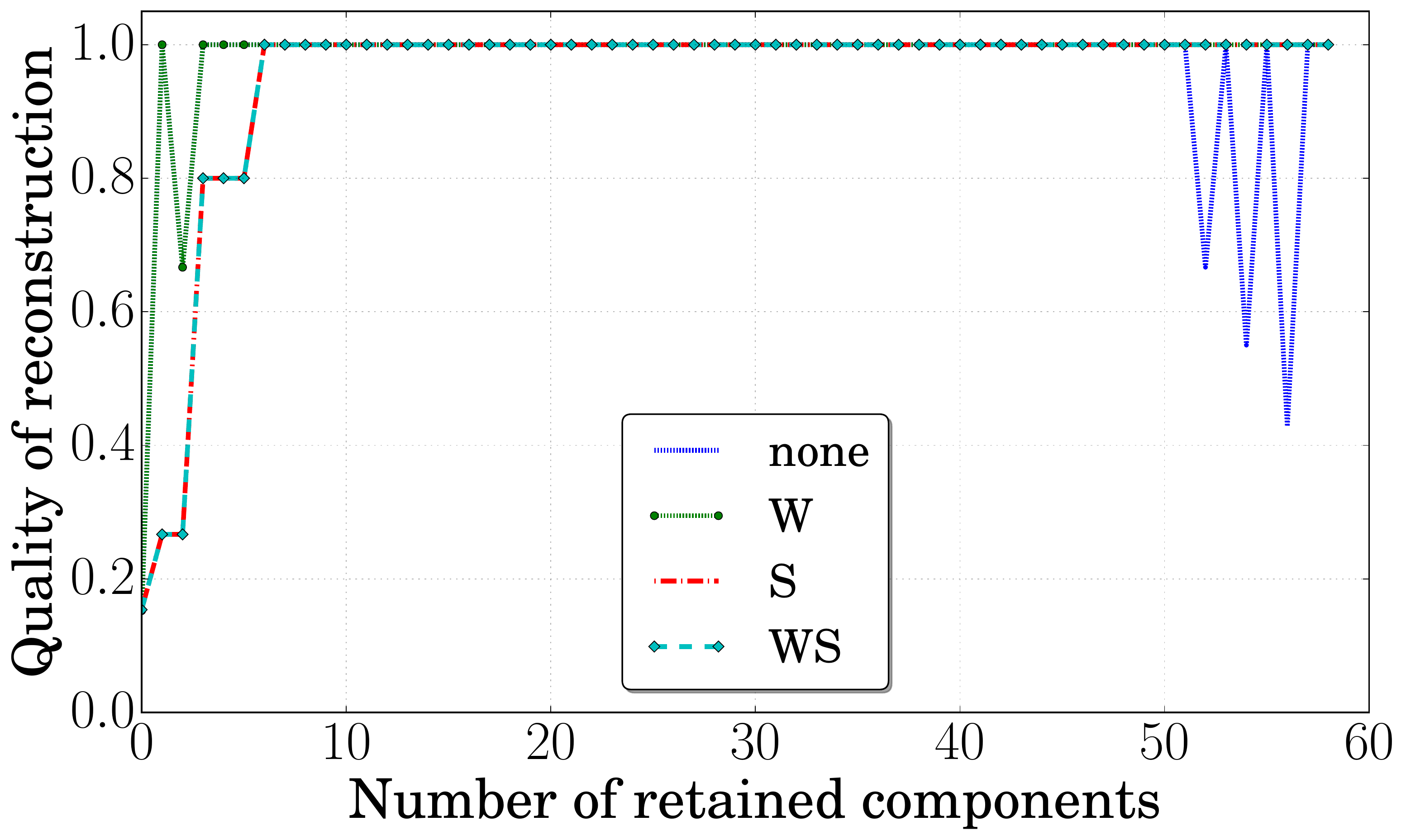}
  \includegraphics[width=0.24\textwidth]{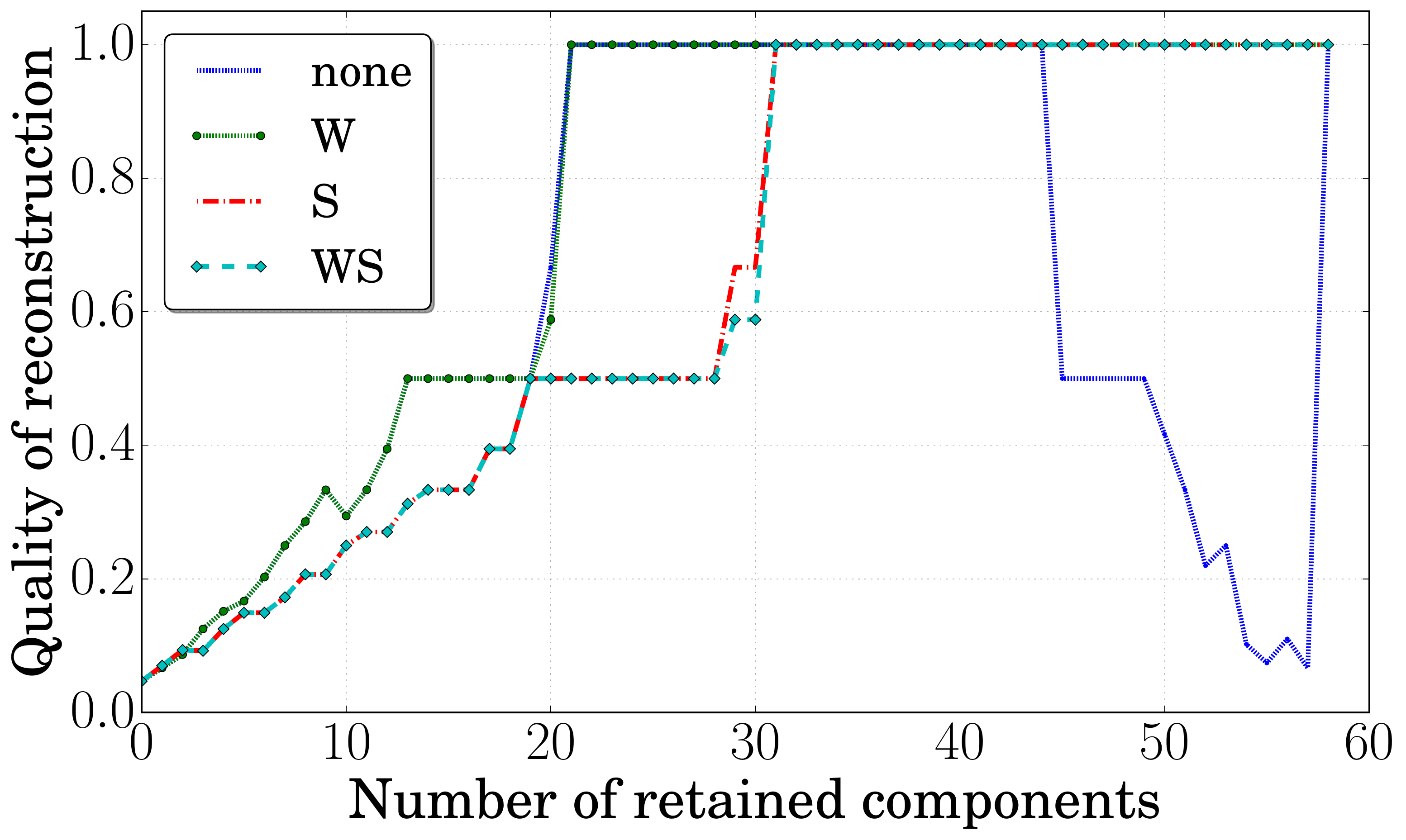}

}
  \subfloat[\label{subfig:it_comp_rl-60-10}\textbf{RL60-10}]{
  \includegraphics[width=0.24\textwidth]{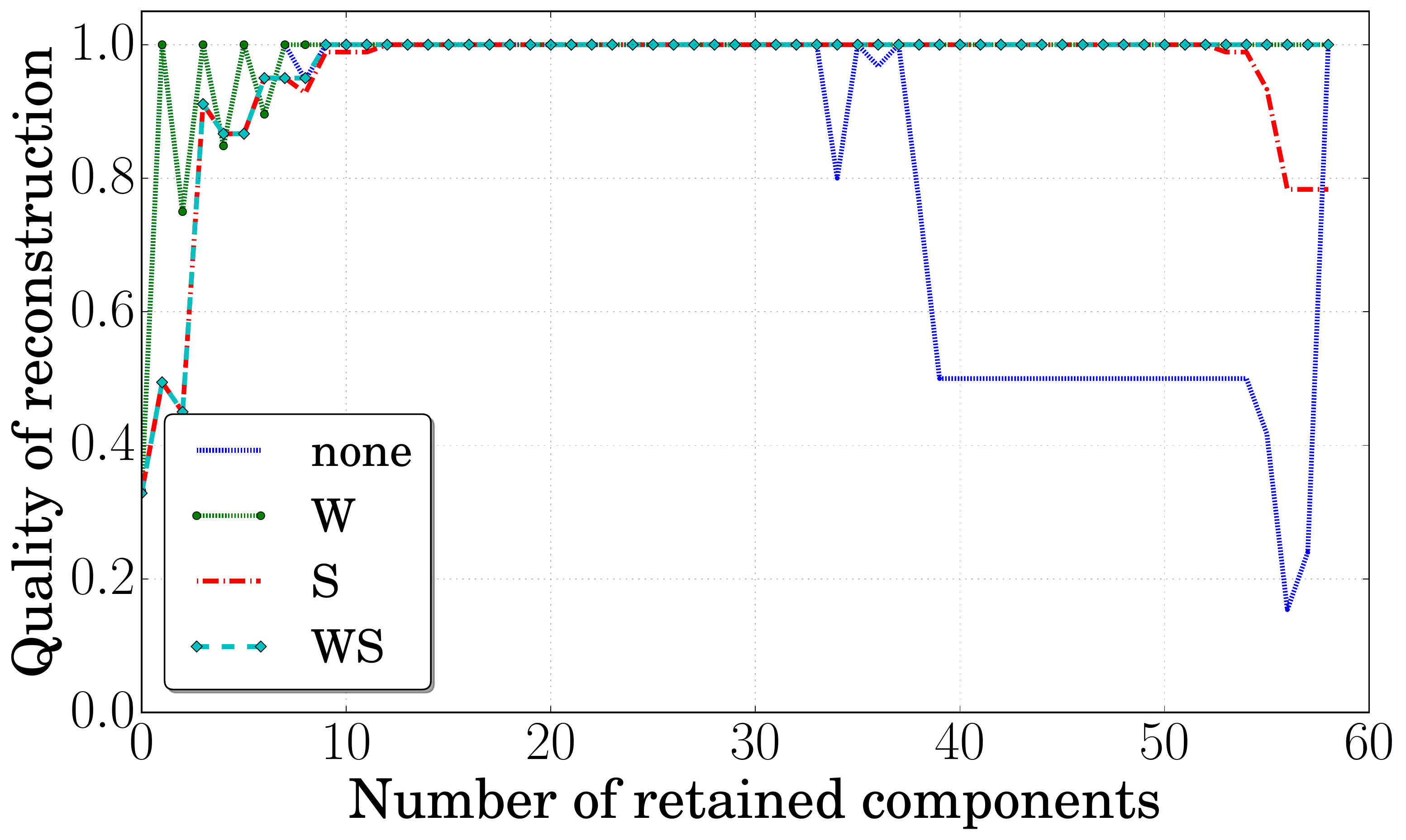}
  \includegraphics[width=0.24\textwidth]{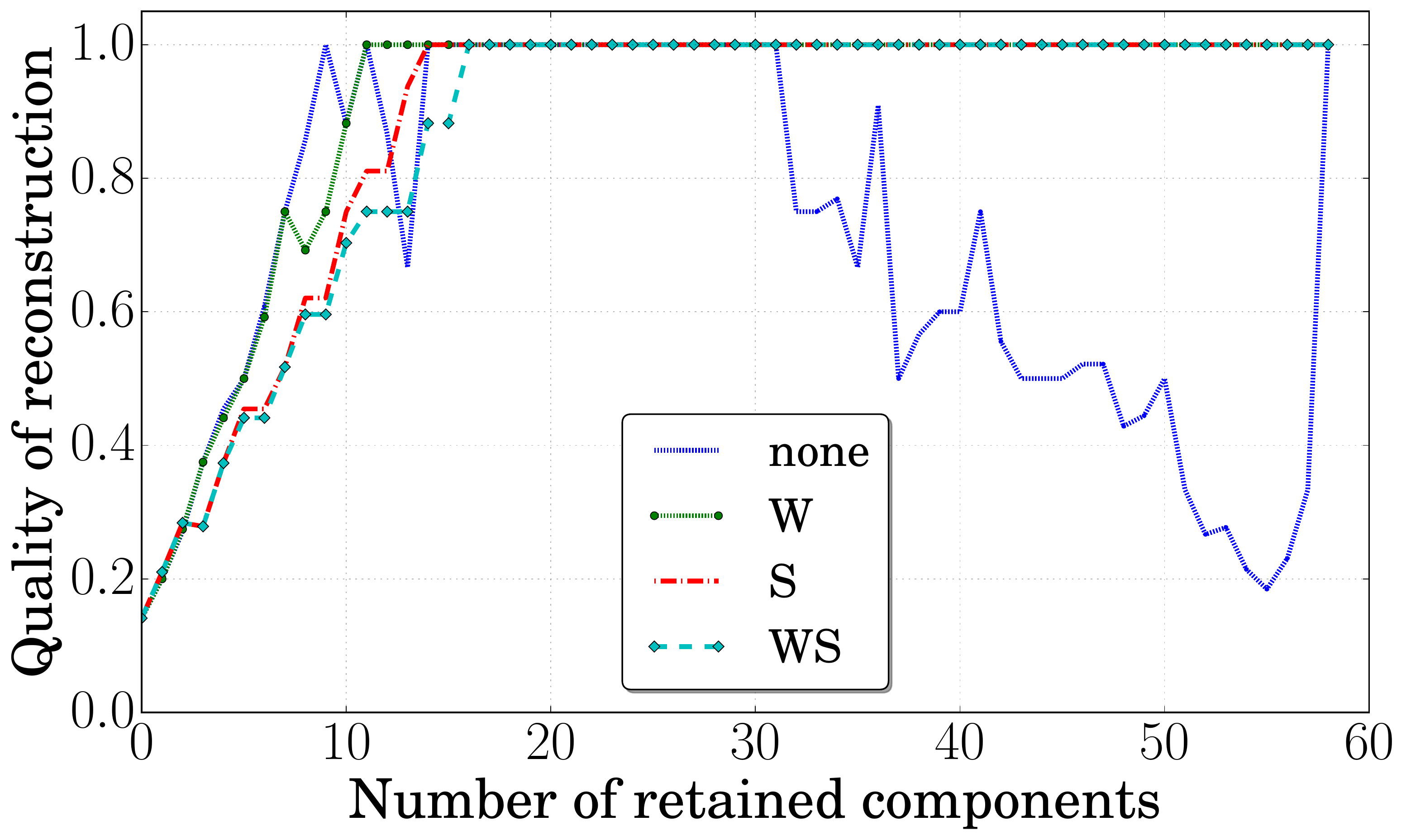}

}

  \subfloat[\label{subfig:it_comp_ws-60-2-1}\textbf{WS60-2-.1}]{
  \includegraphics[width=0.24\textwidth]{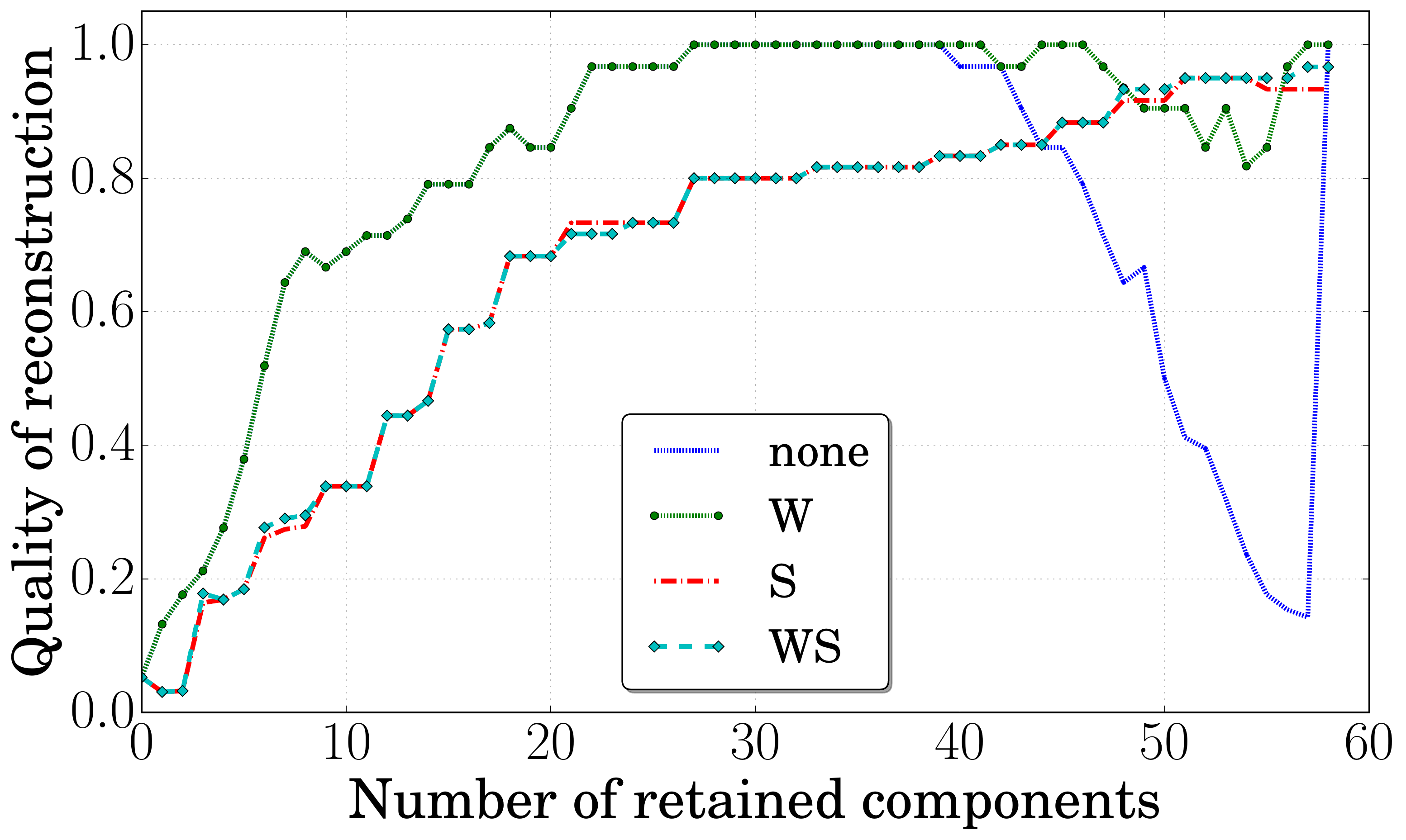}
  \includegraphics[width=0.24\textwidth]{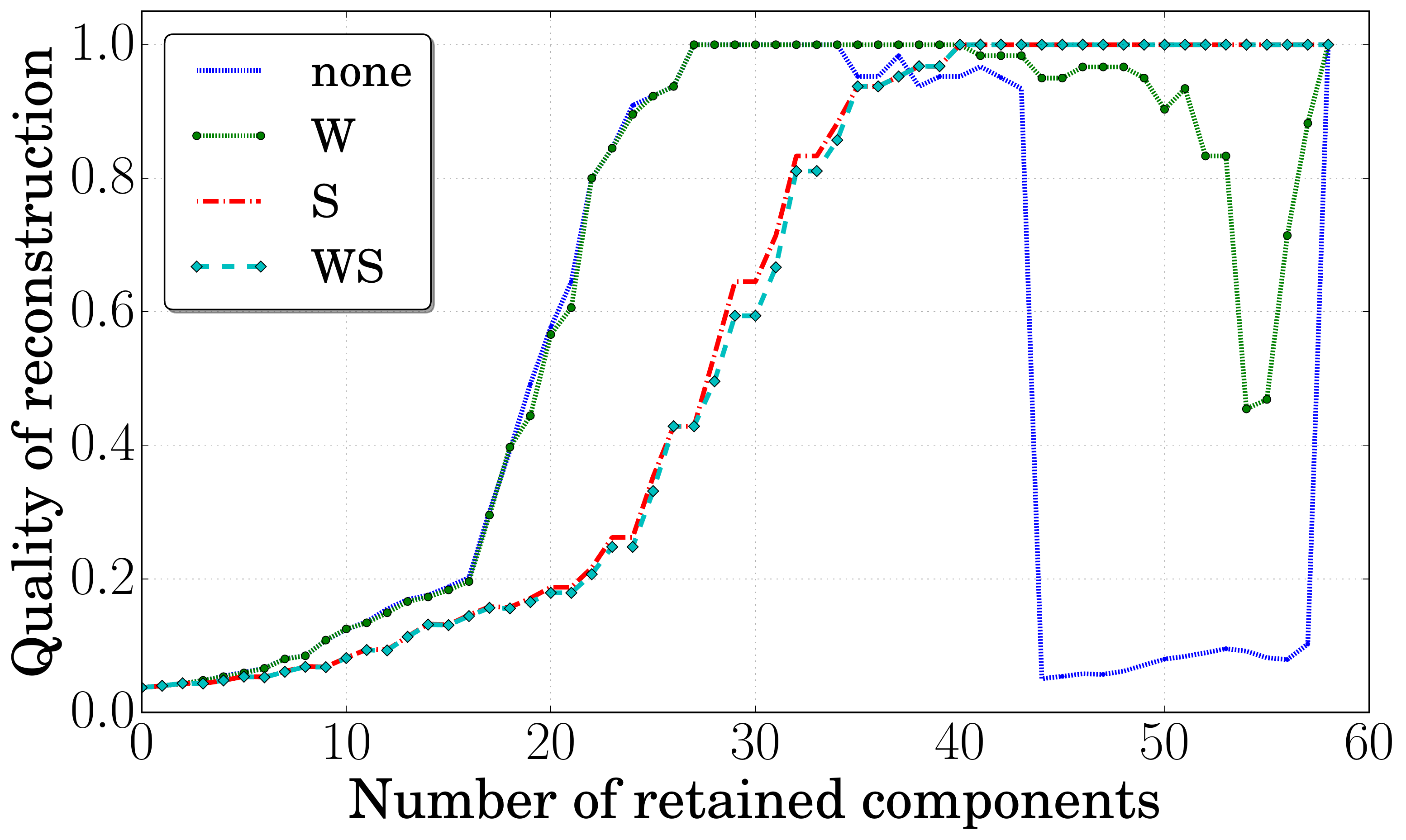}
}
  \subfloat[\label{subfig:it_comp_ws-60-10-1}\textbf{WS60-2-.1}]{
  \includegraphics[width=0.24\textwidth]{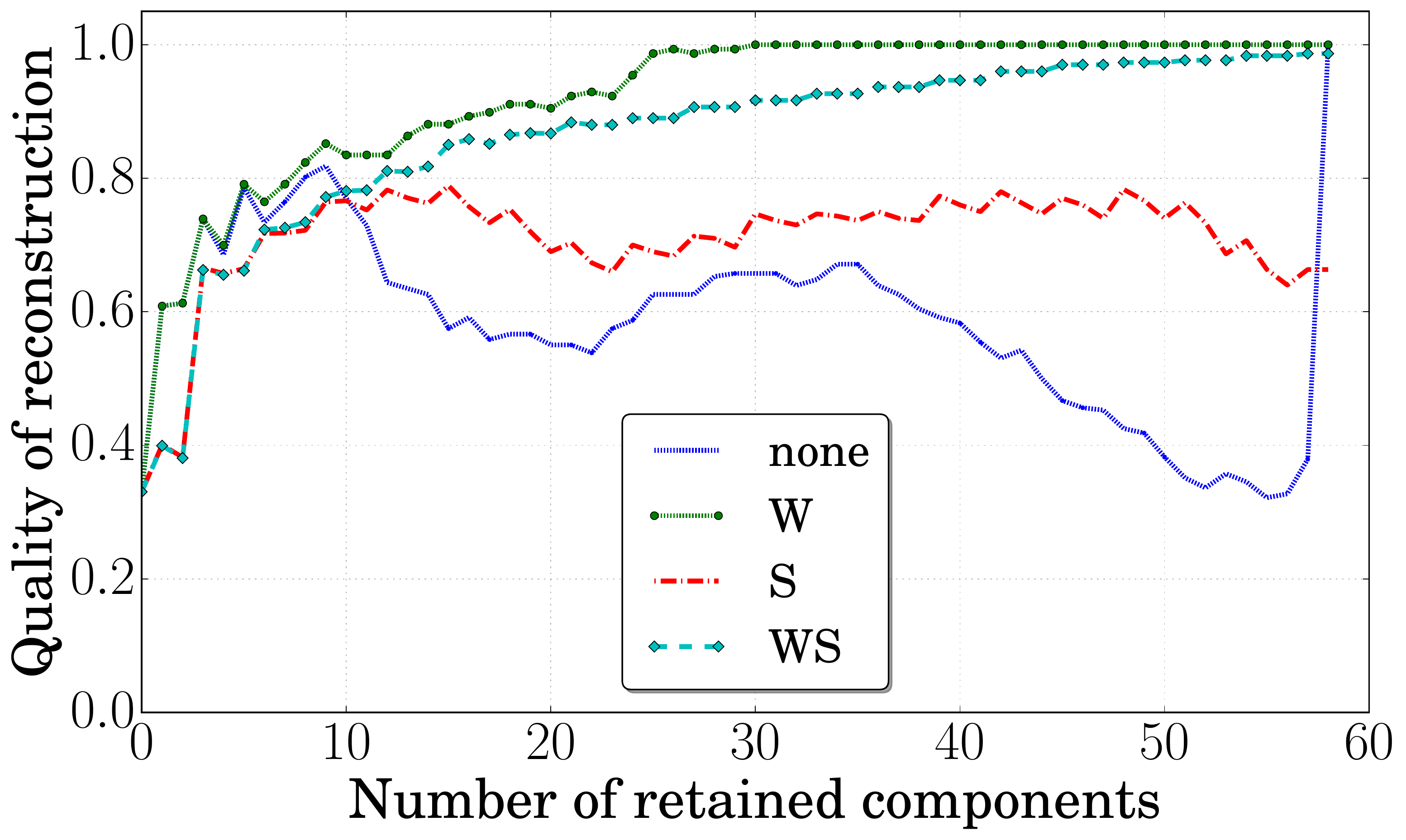}
  \includegraphics[width=0.24\textwidth]{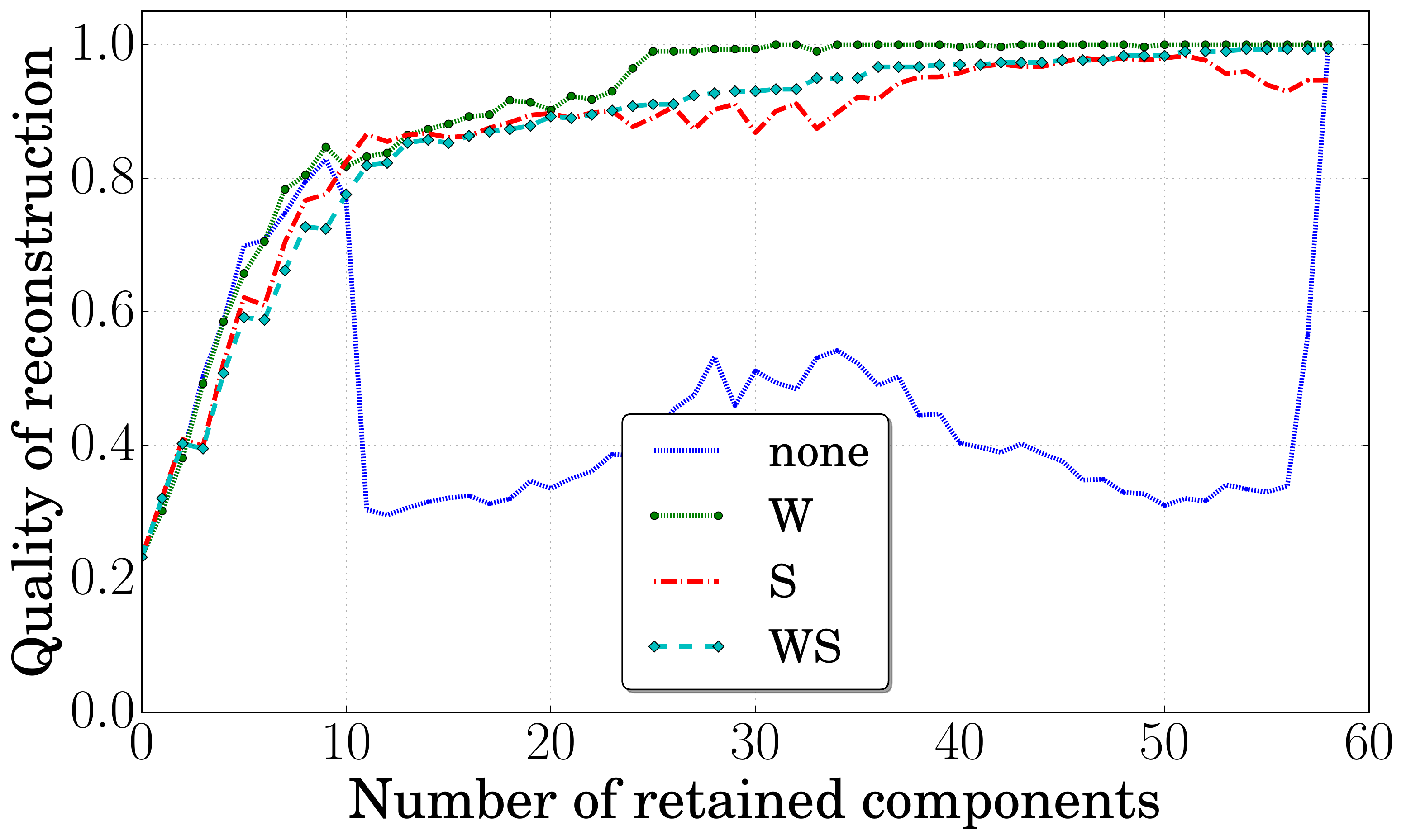}
}

  \subfloat[\label{subfig:it_comp_sbm-60-2-7-10}\textbf{SBM60-2}]{
  \includegraphics[width=0.24\textwidth]{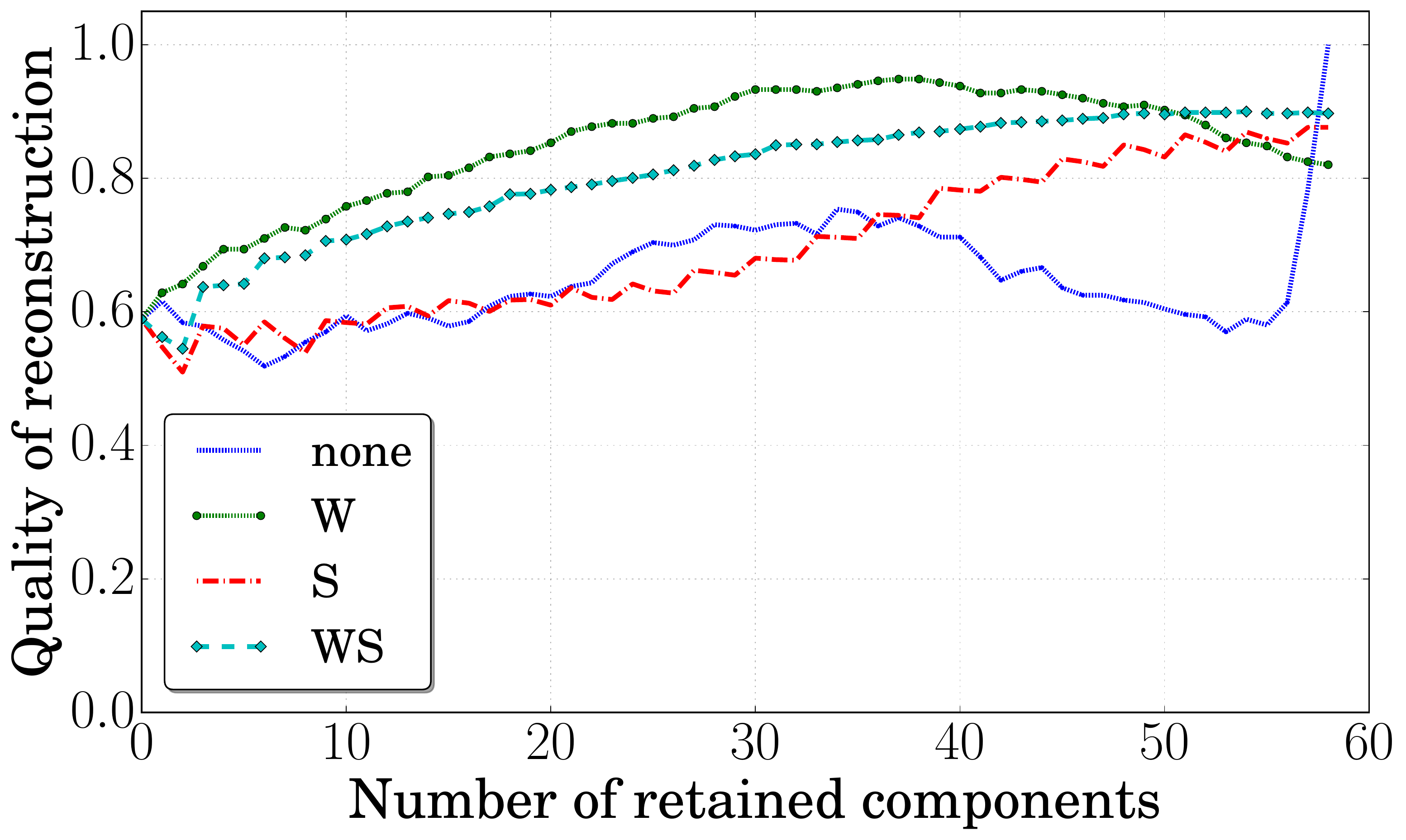}
  \includegraphics[width=0.24\textwidth]{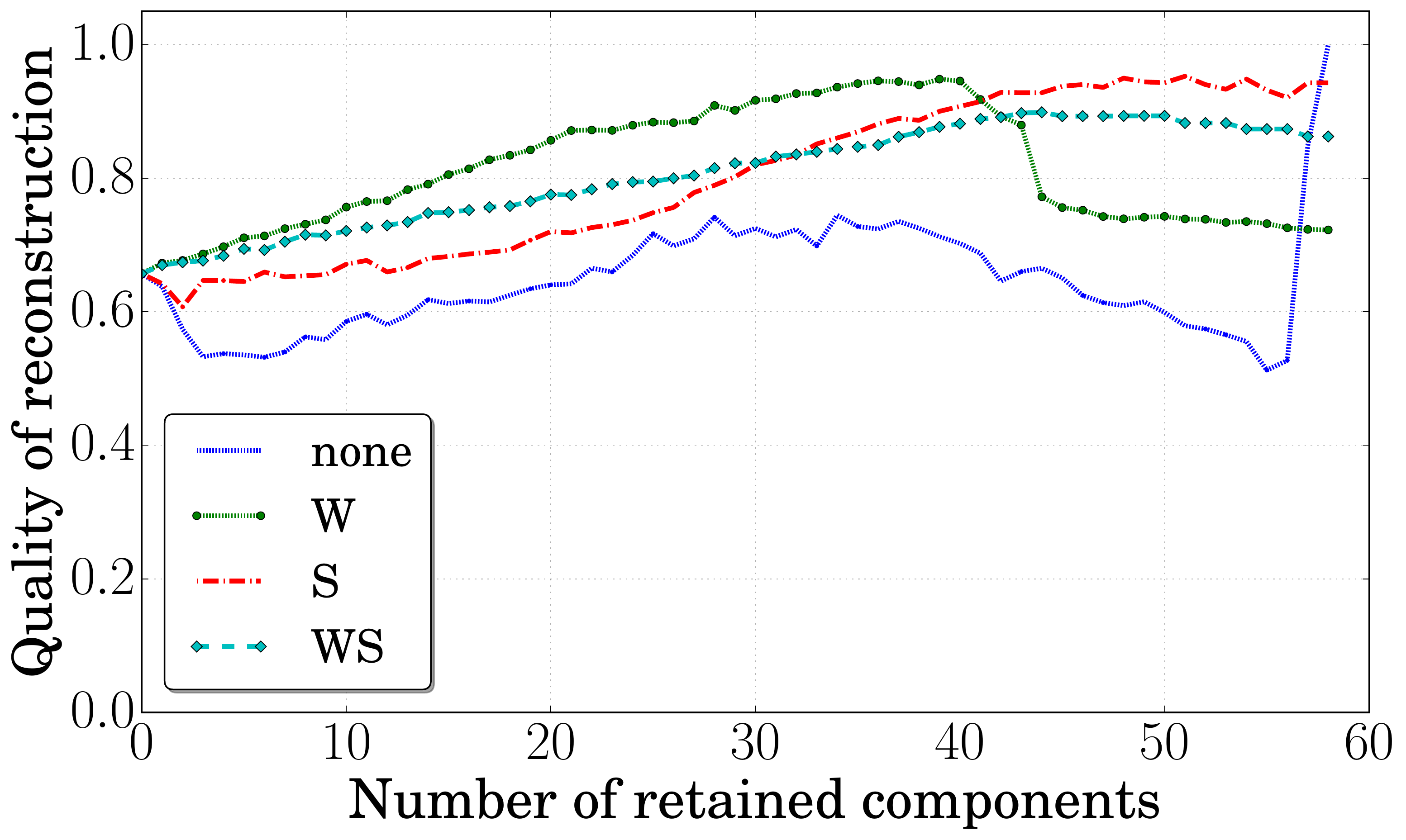}
}
  \subfloat[\label{subfig:it_comp_sbm-60-4-9-1}\textbf{SBM60-4}]{
  \includegraphics[width=0.24\textwidth]{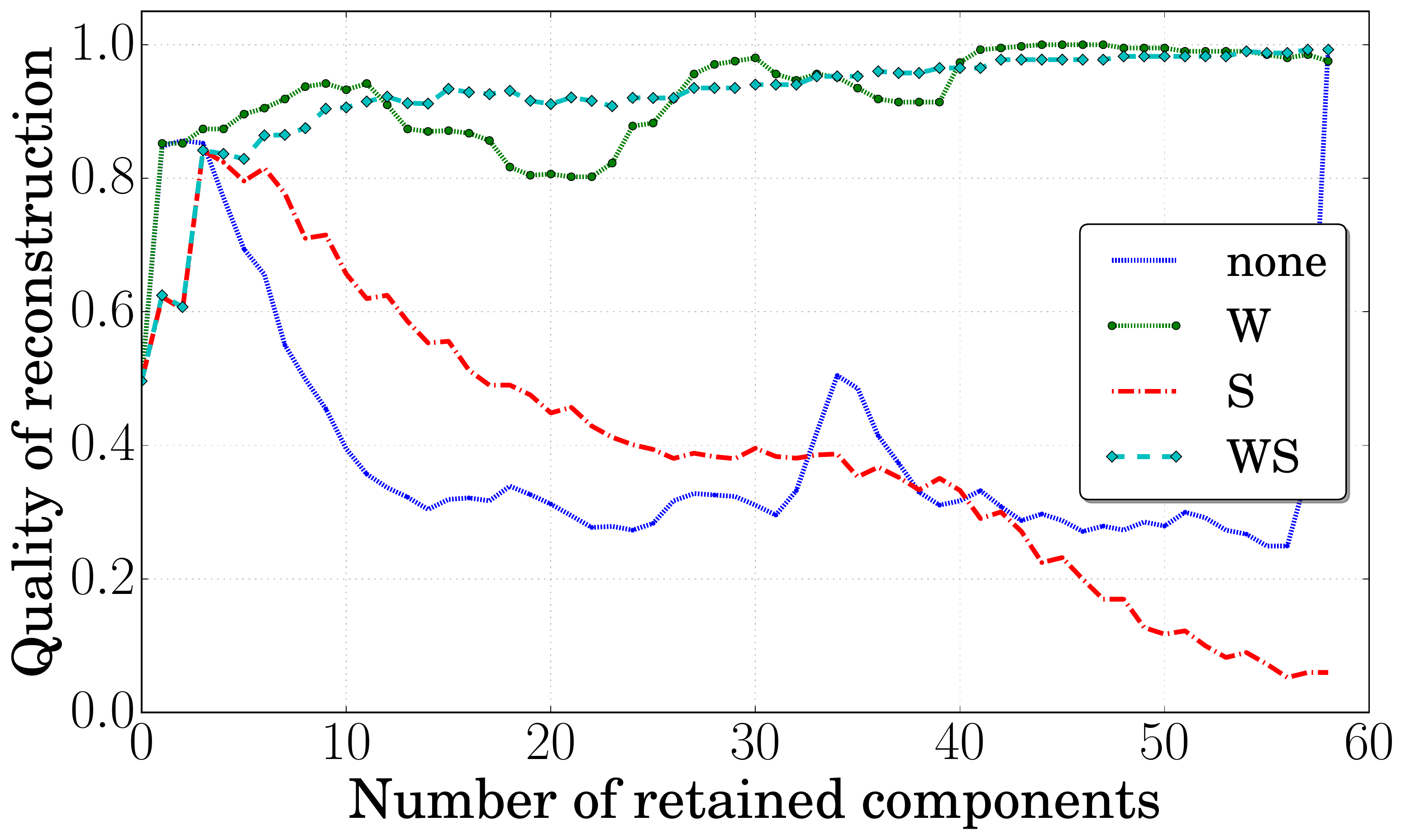}
  \includegraphics[width=0.24\textwidth]{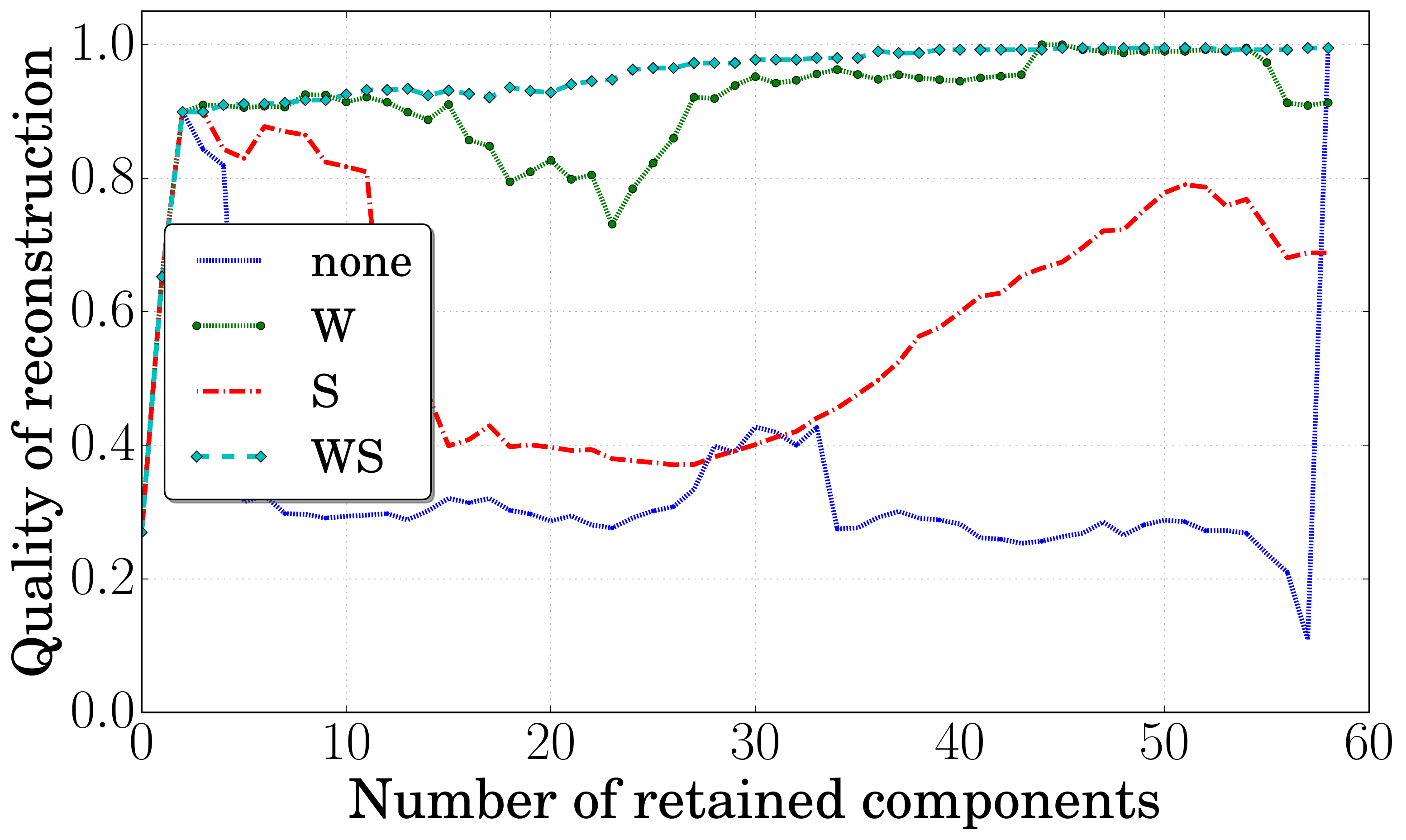}
}

  \subfloat[\label{subfig:it_comp_er-60-4}\textbf{ER60-.4}]{
  \includegraphics[width=0.24\textwidth]{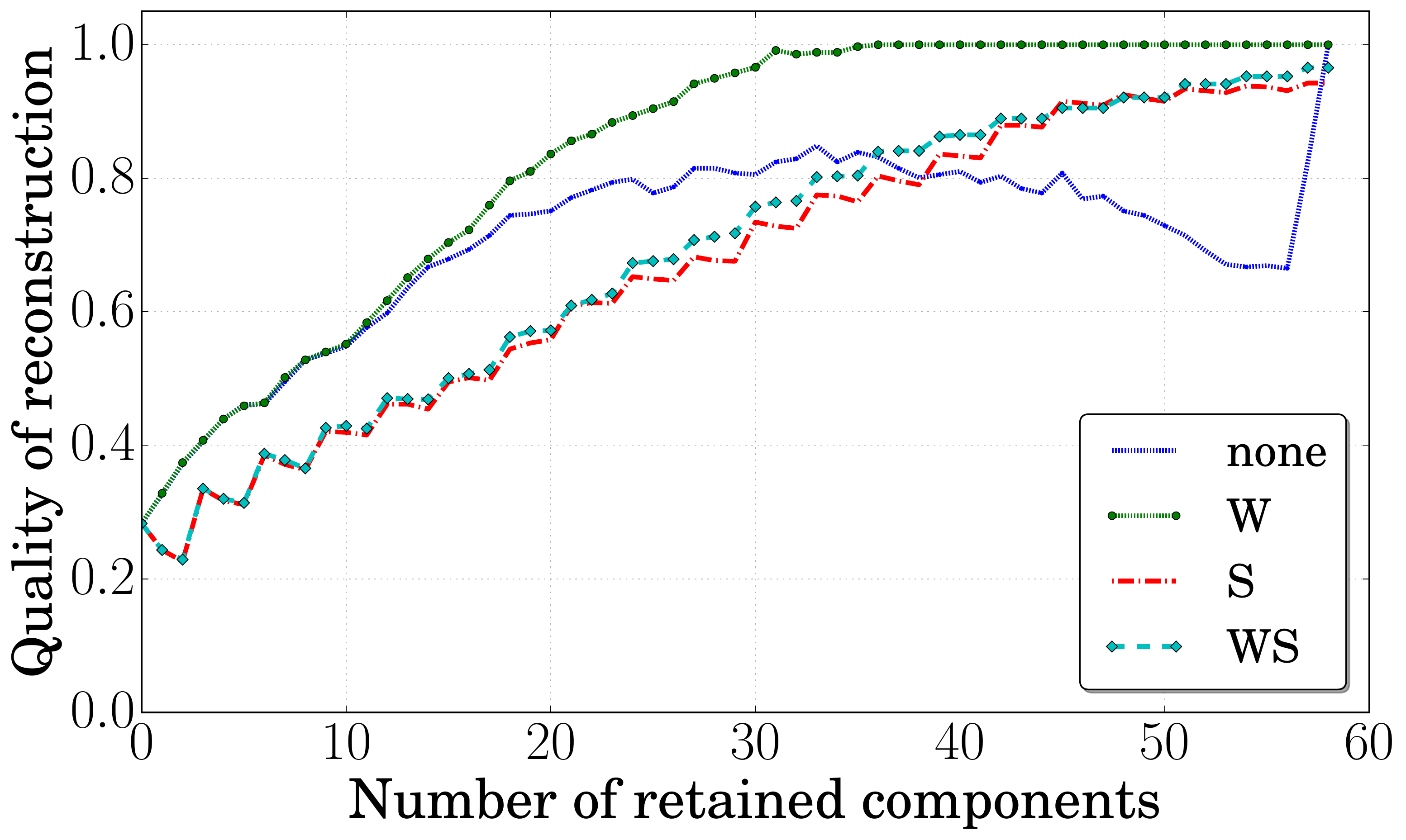}
  \includegraphics[width=0.24\textwidth]{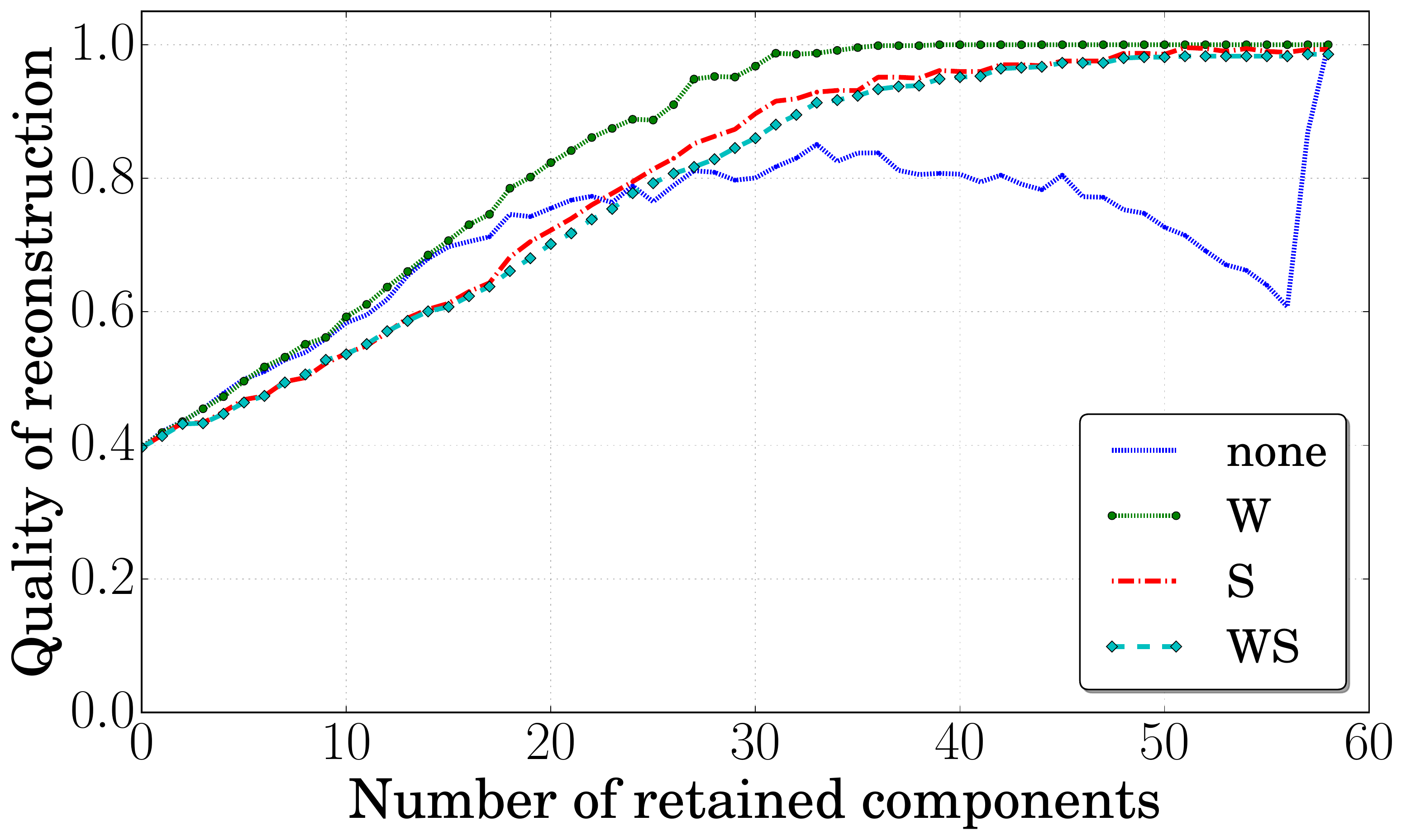}
}
  \subfloat[\label{subfig:it_comp_bar-60}\textbf{BAR60}]{
  \includegraphics[width=0.24\textwidth]{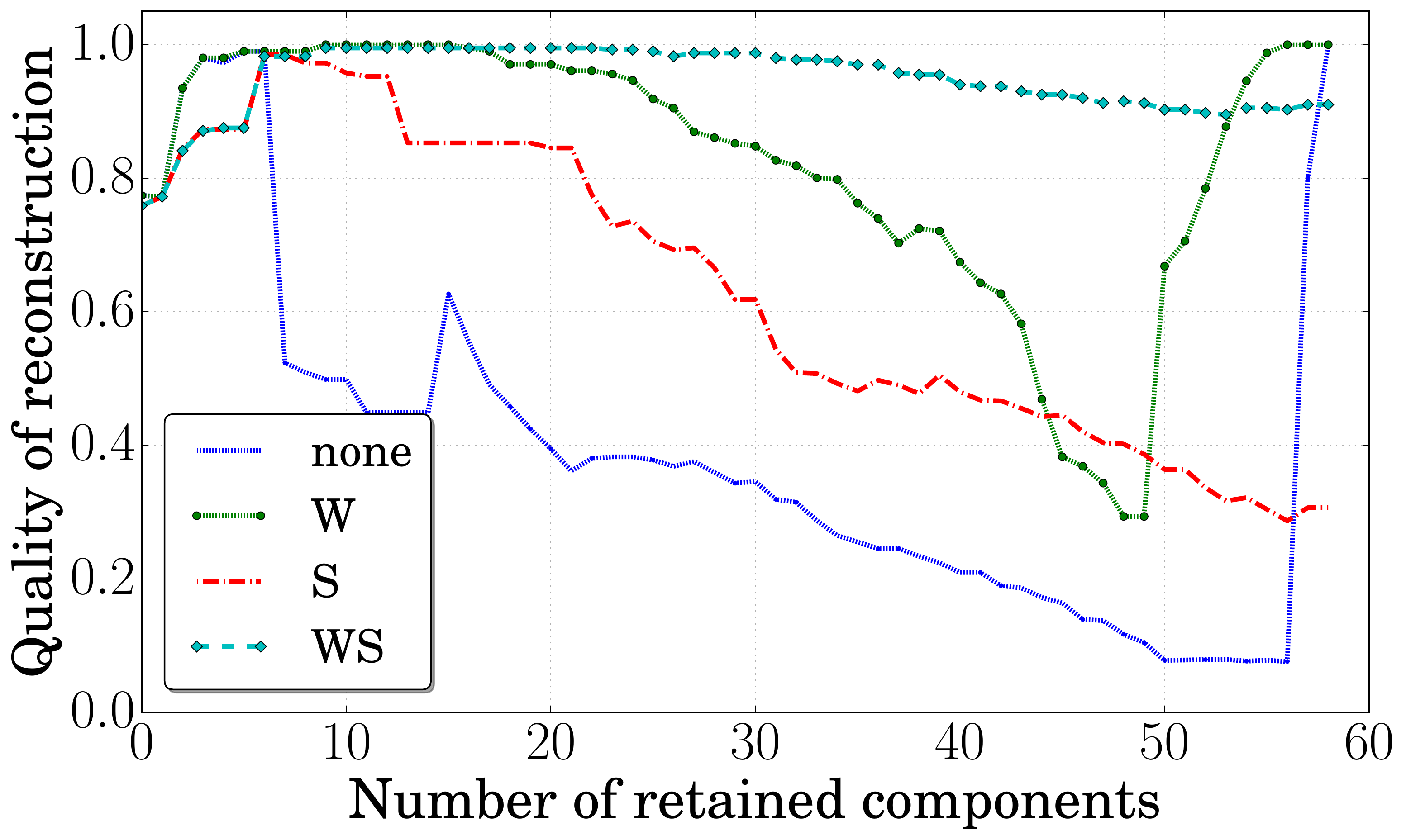}
  \includegraphics[width=0.24\textwidth]{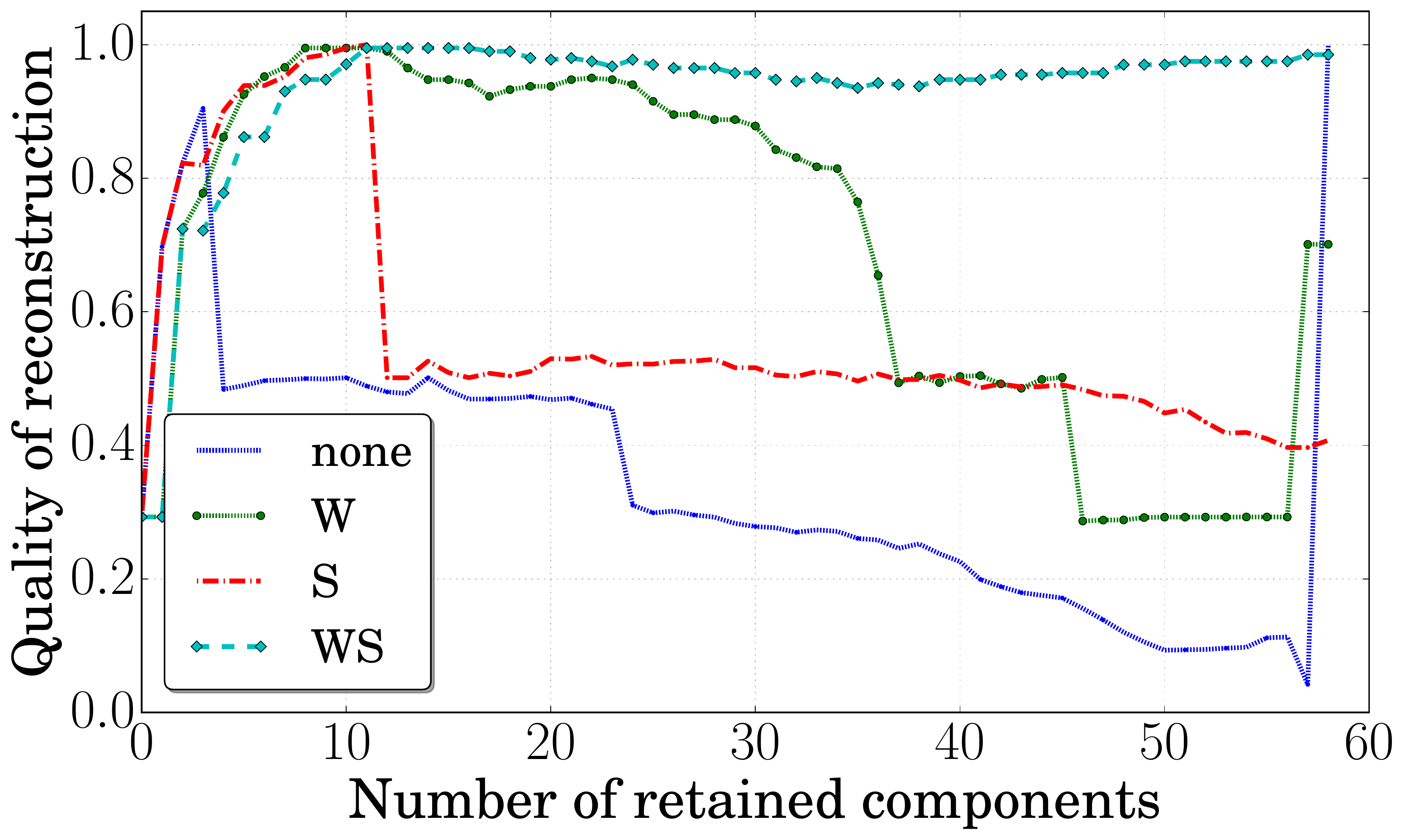}
}

  \caption{\label{fig:IT_results_comp}Results of the inverse transformation 
  of a degraded collection of signals, obtained by retaining only a portion of 
  the components. For each number of retained components, a quality of 
  reconstruction is measured by comparing the obtained graph with the original 
  one.  Each sub-figure shows the results for one instance, whose name refers to 
  those given in Figure~\ref{fig:GS_illustrations}. For each sub-figure, the left 
  plot shows the results using thresholding based on the number of edges, while 
  the right plot shows the results using the Adapted Otsu's method. Four 
  configurations of the inverse transformation are studied. \textbf{none}: no 
  enhancement is performed; \textbf{W}: weighted computation of distances only 
  with $\alpha=4$; \textbf{S} sequential update of the adjacency matrix only; 
  \textbf{WS} both enhancements.}
\end{figure*}

\begin{figure*}

  \subfloat[\label{subfig:it_noise_rl-60-2}\textbf{RL60-2}]{
  \includegraphics[width=0.24\textwidth]{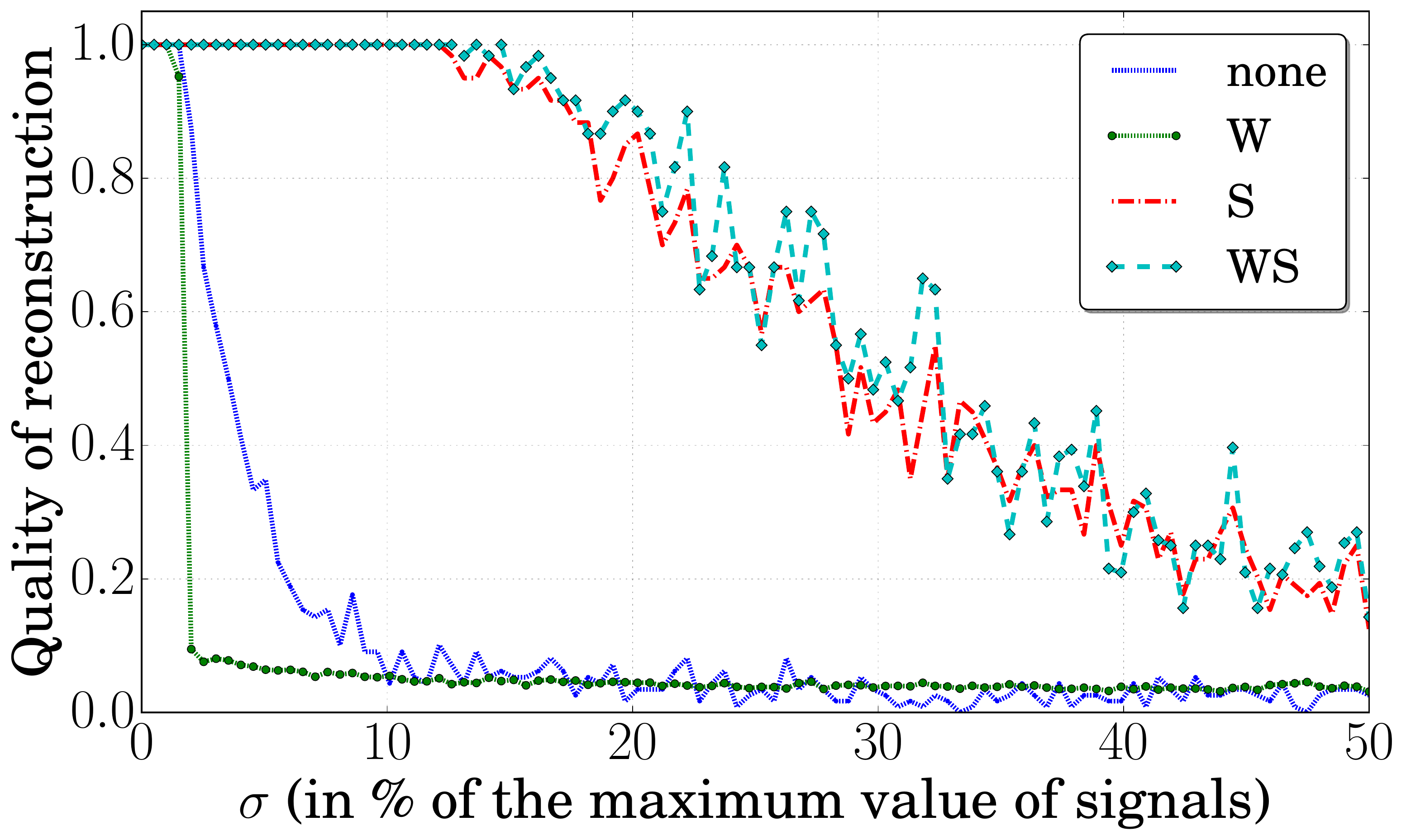}
  \includegraphics[width=0.24\textwidth]{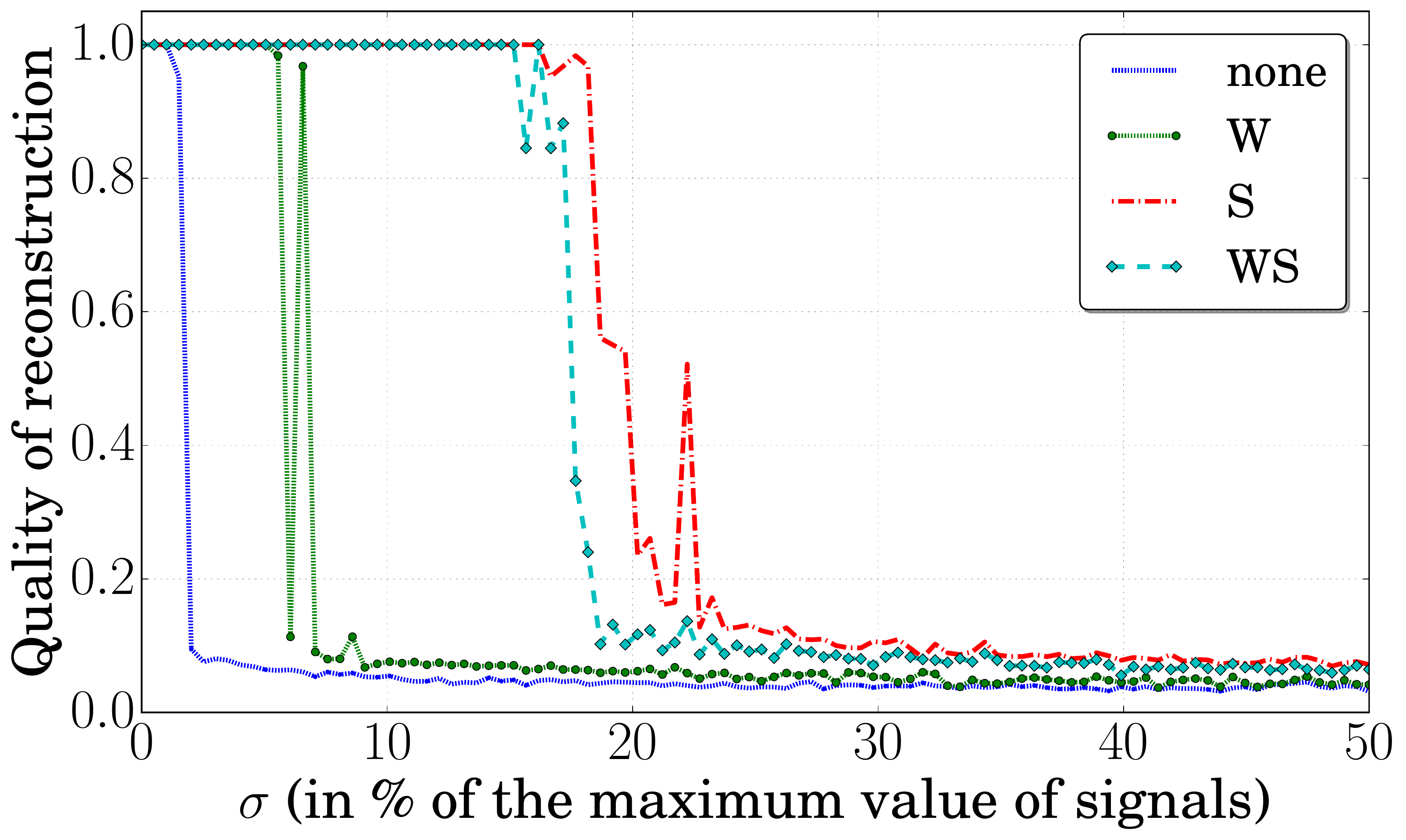}

}
  \subfloat[\label{subfig:it_noise_rl-60-10}\textbf{RL60-10}]{
  \includegraphics[width=0.24\textwidth]{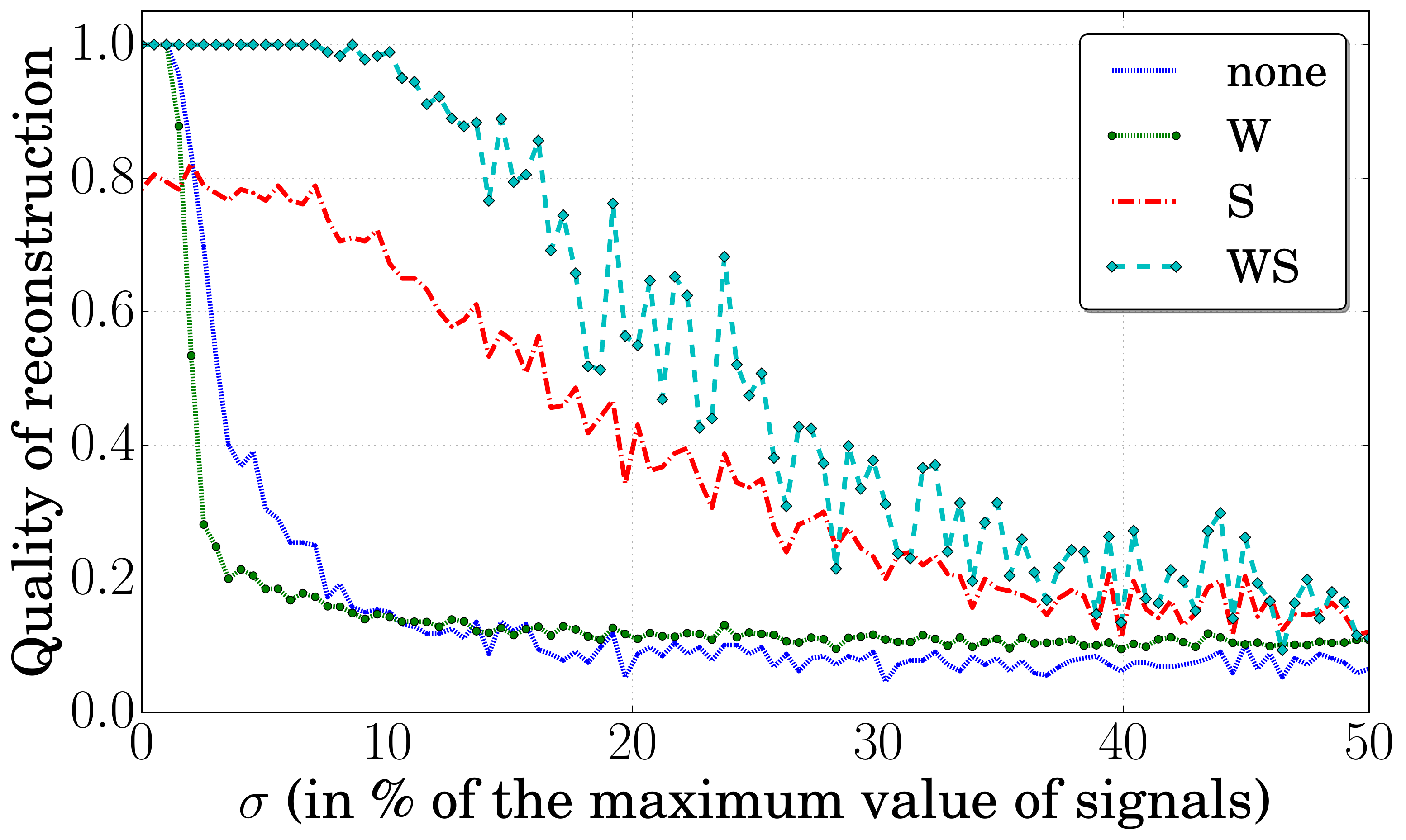}
  \includegraphics[width=0.24\textwidth]{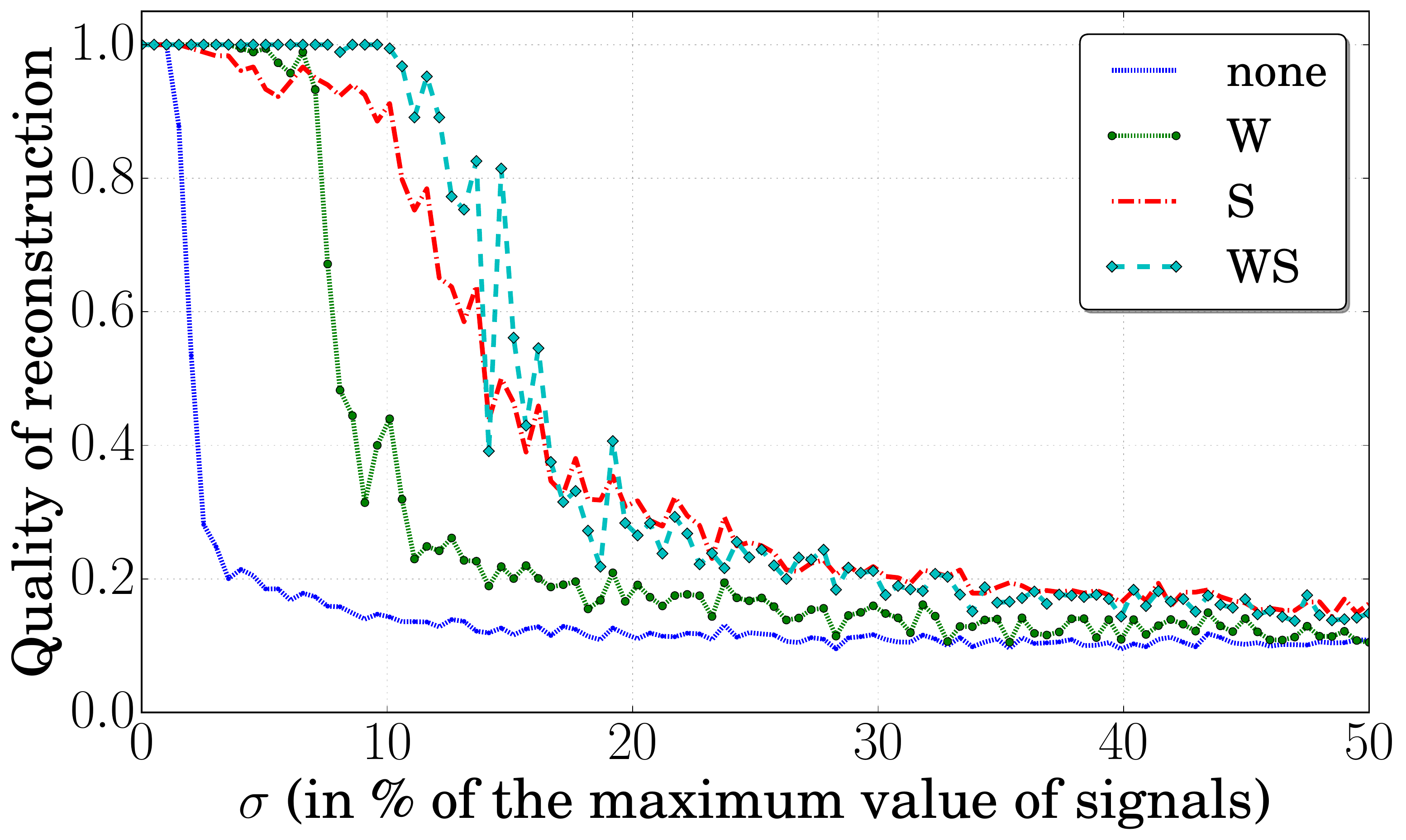}

}

  \subfloat[\label{subfig:it_noise_ws-60-10-1}\textbf{WS60-2-.1}]{
  \includegraphics[width=0.24\textwidth]{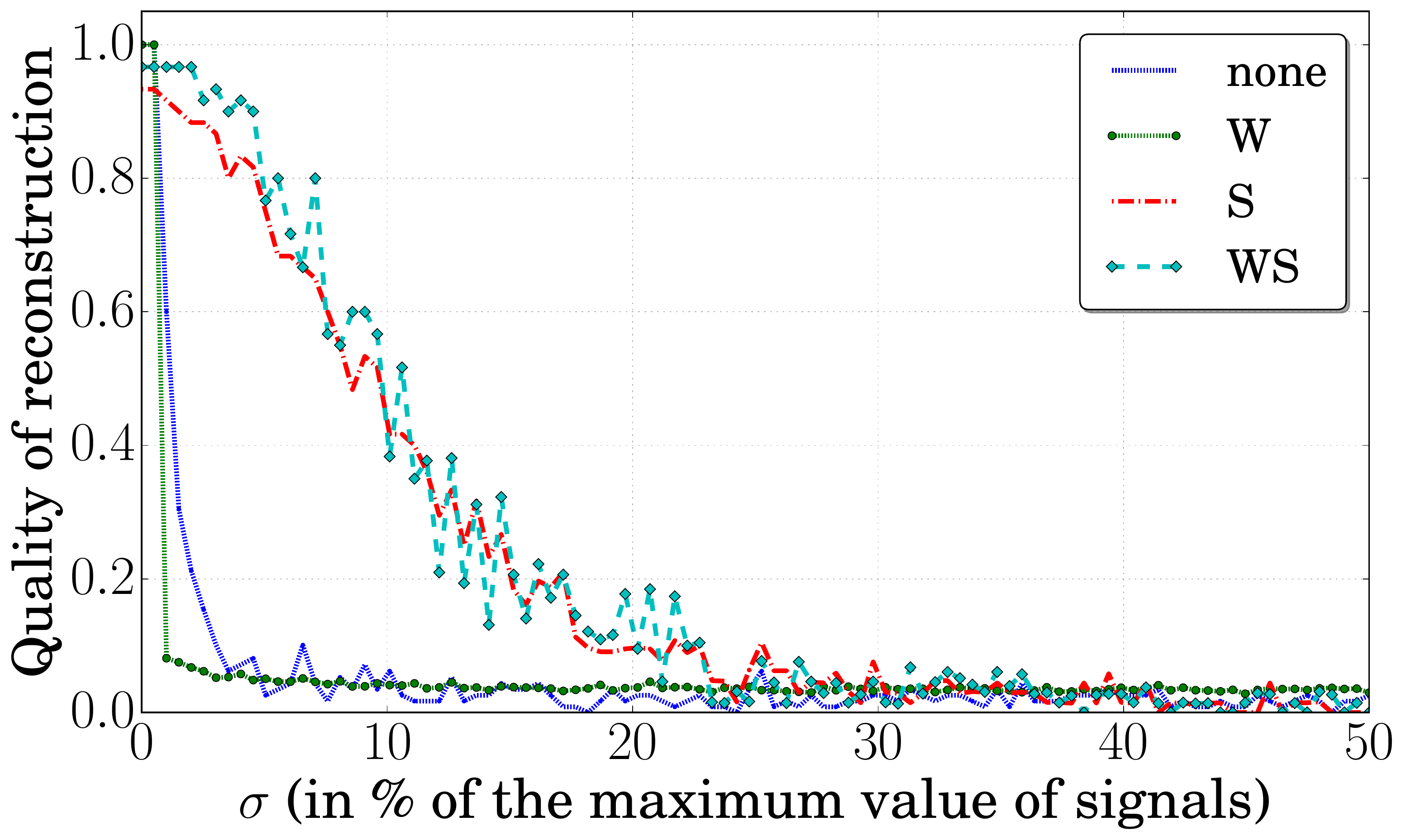}
  \includegraphics[width=0.24\textwidth]{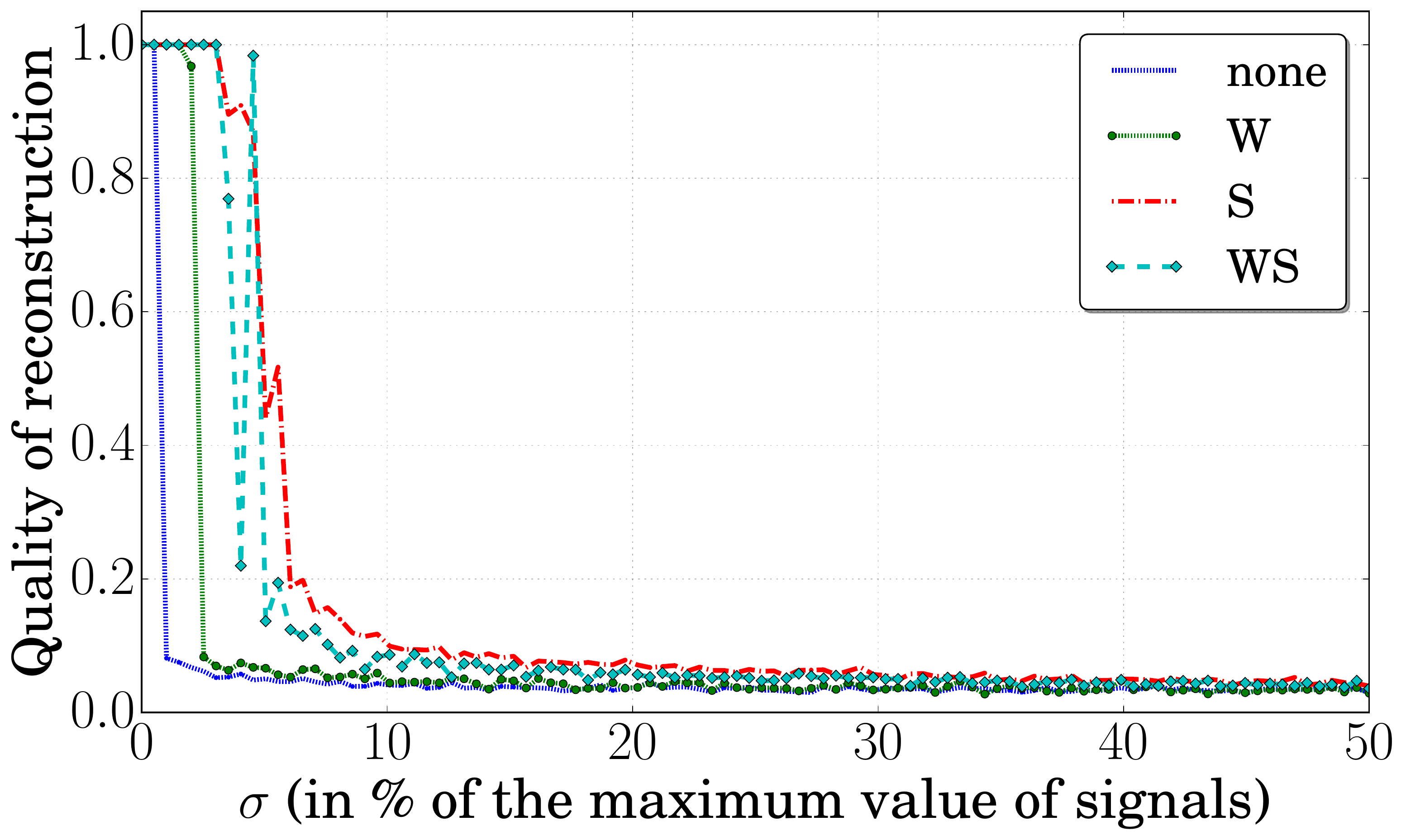}
}
  \subfloat[\label{subfig:it_noise_ws-60-2-1}\textbf{WS60-2-.1}]{
  \includegraphics[width=0.24\textwidth]{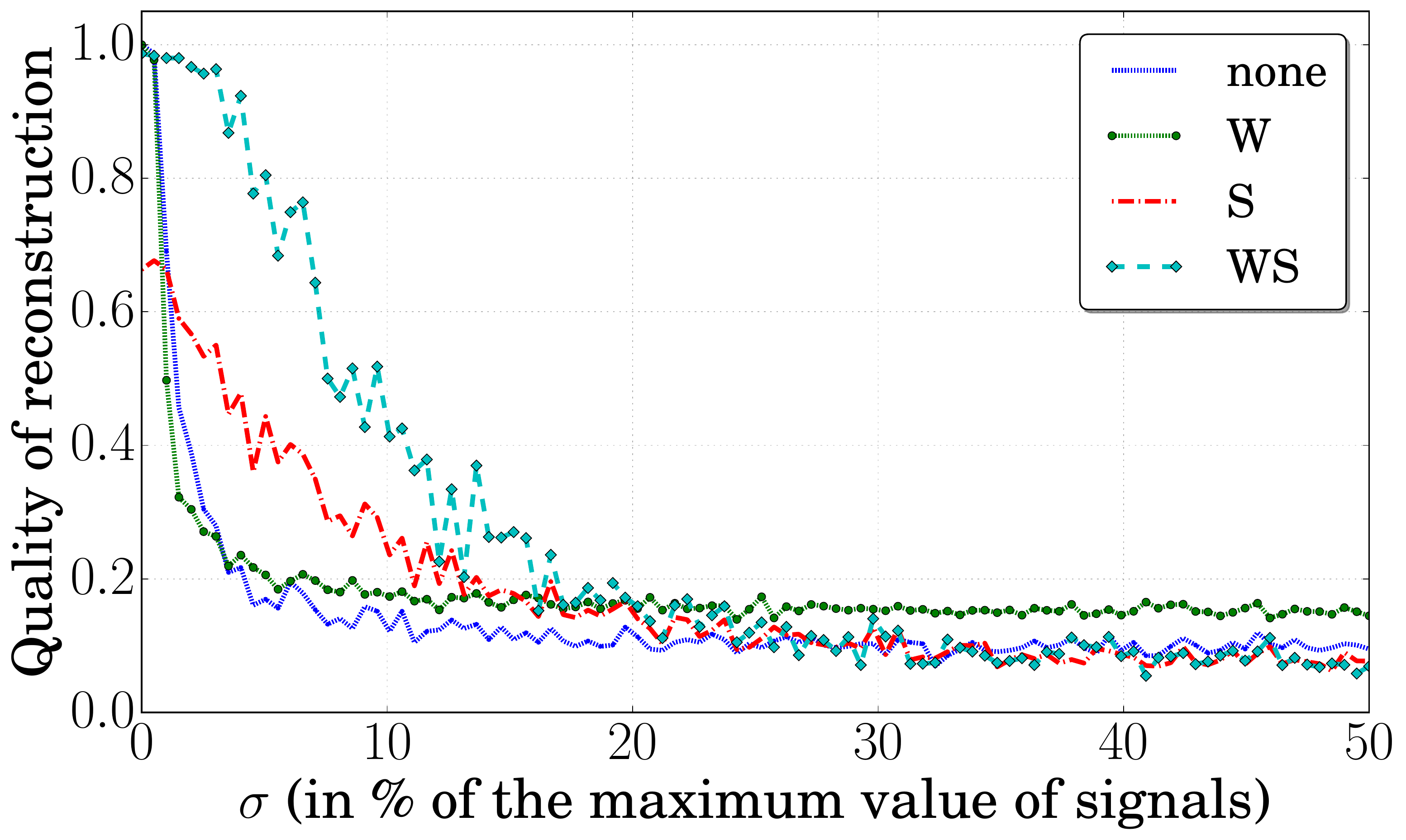}
  \includegraphics[width=0.24\textwidth]{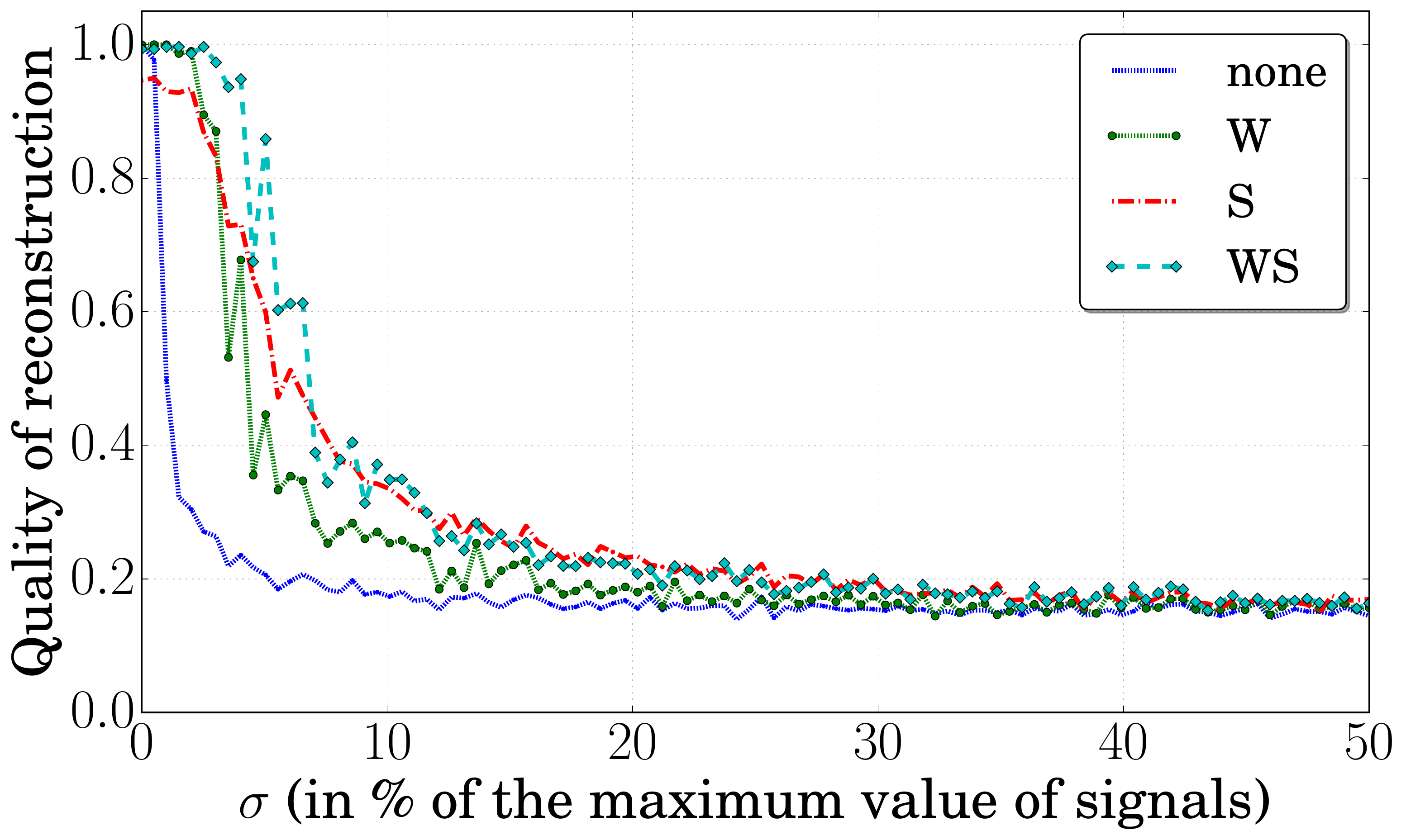}
}

  \subfloat[\label{subfig:it_noise_sbm-60-2-7-10}\textbf{SBM60-2}]{
  \includegraphics[width=0.24\textwidth]{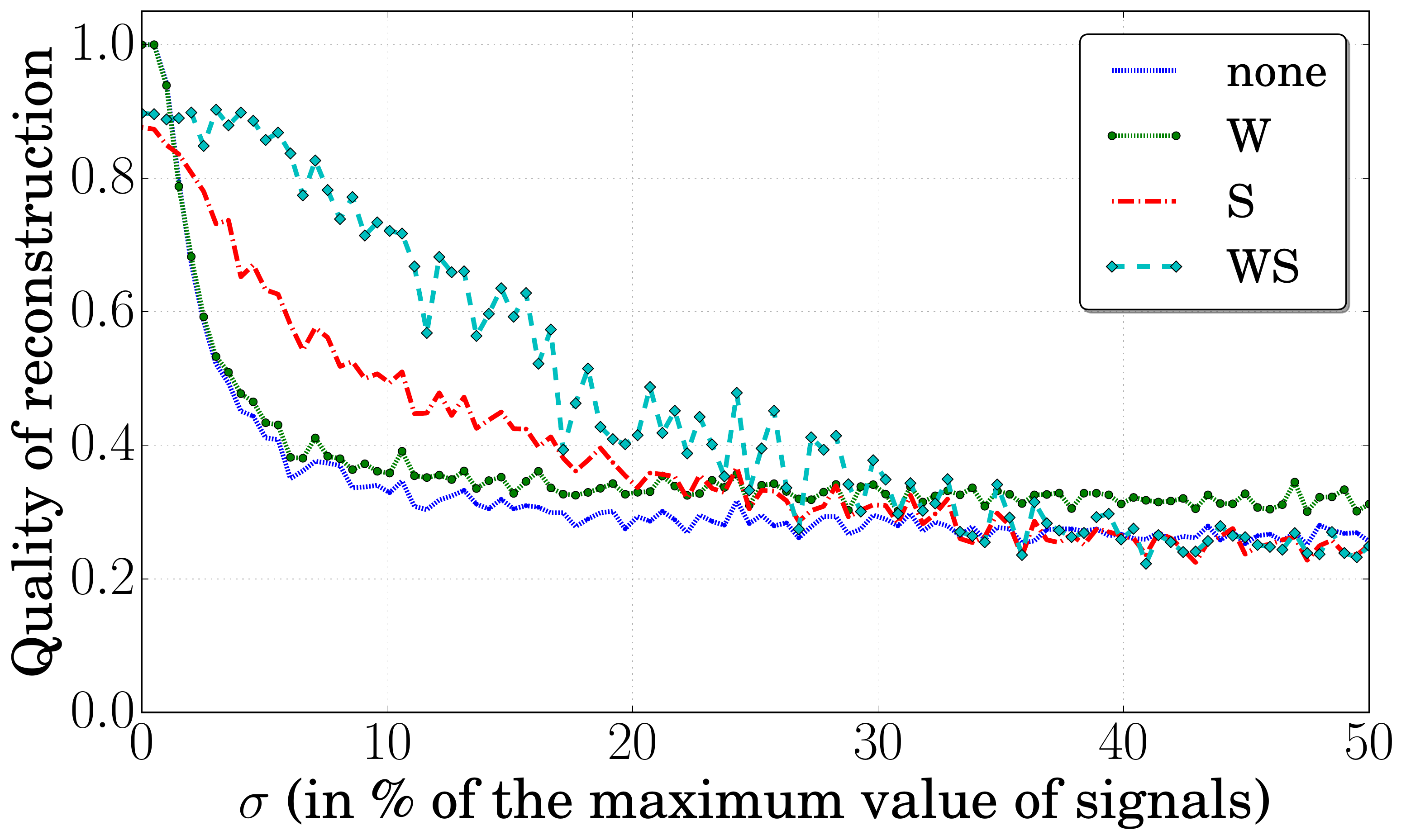}
  \includegraphics[width=0.24\textwidth]{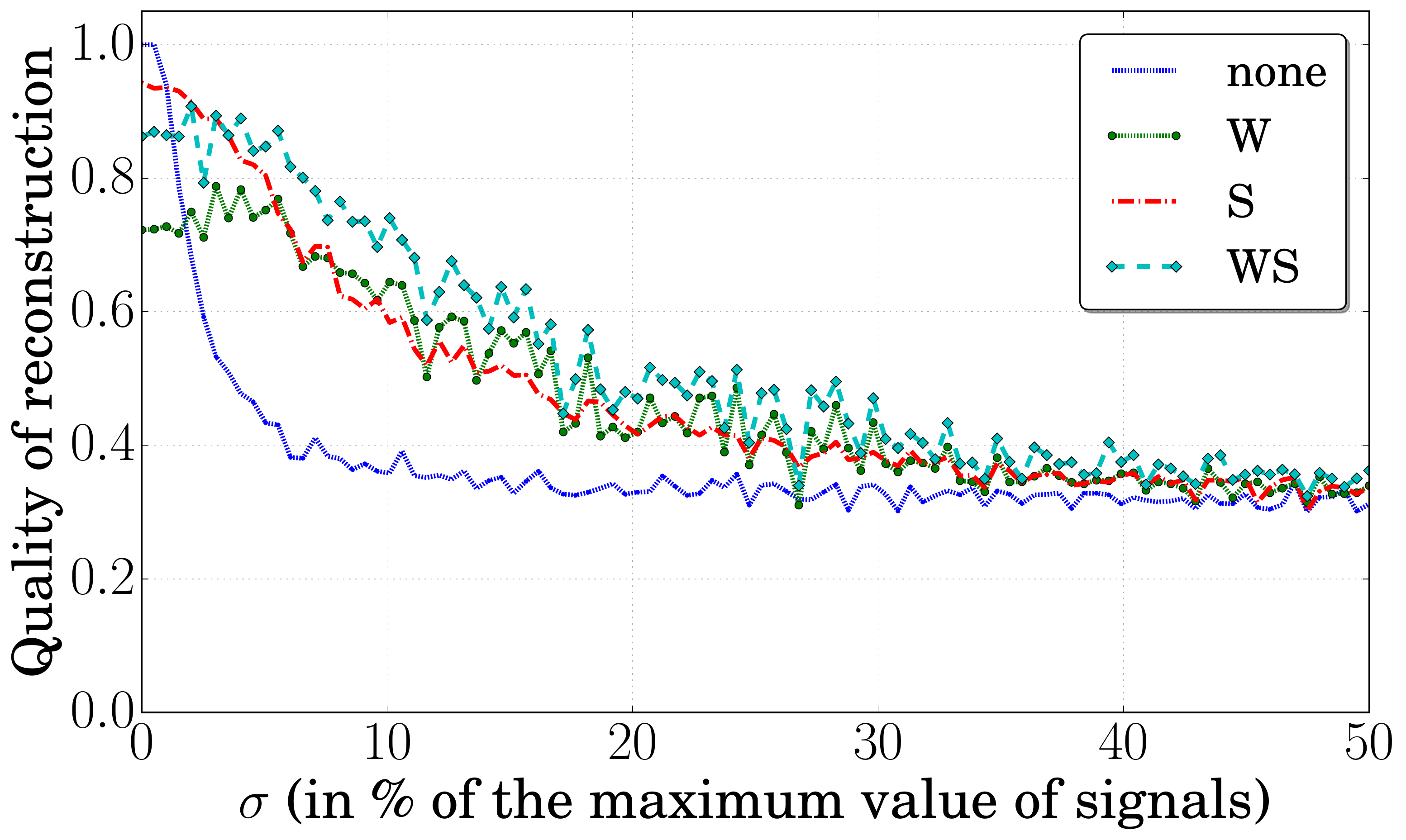}
}
  \subfloat[\label{subfig:it_noise_sbm-60-4-9-1}\textbf{SBM60-4}]{
  \includegraphics[width=0.24\textwidth]{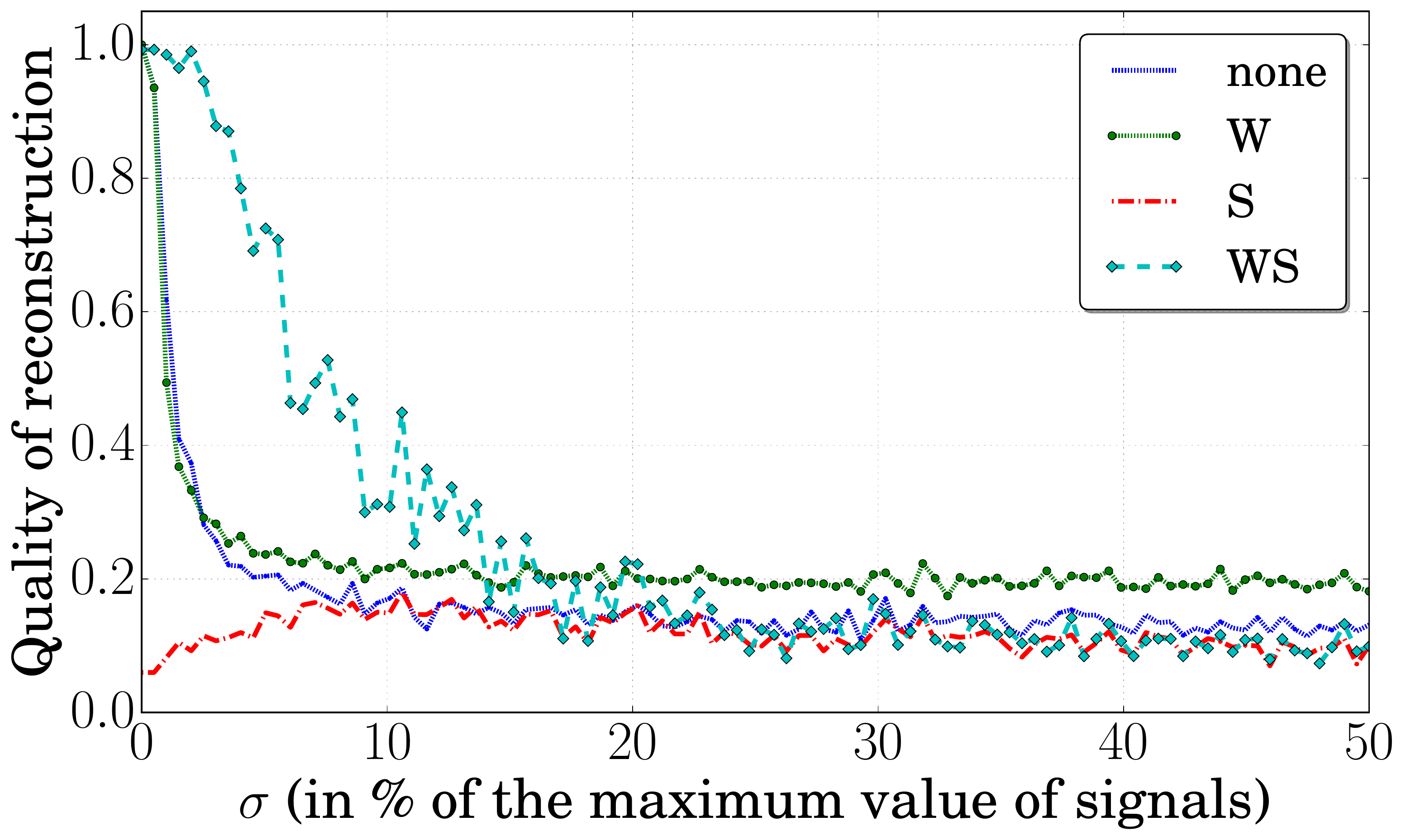}
  \includegraphics[width=0.24\textwidth]{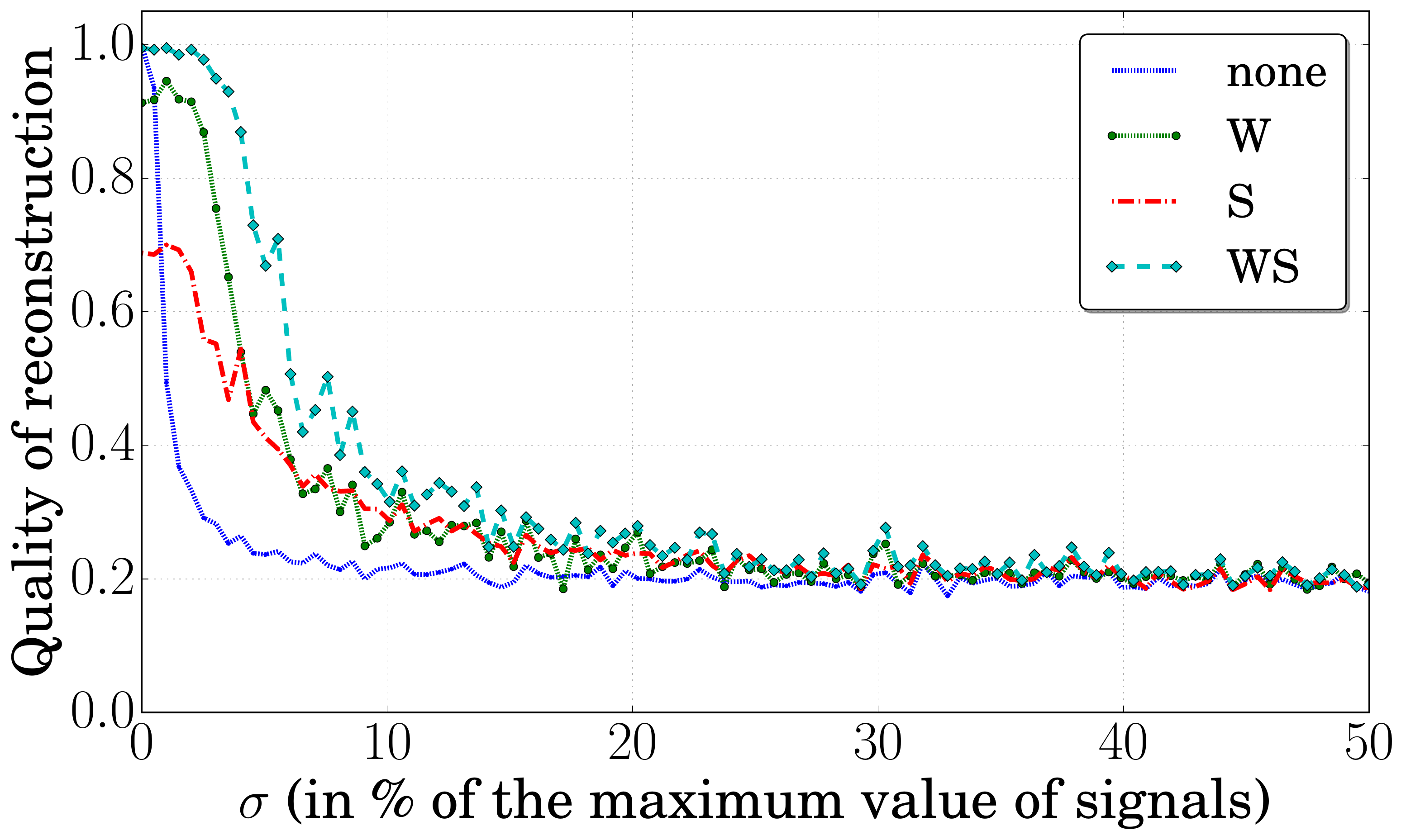}
}

  \subfloat[\label{subfig:it_noise_er-60-4}\textbf{ER60-.4}]{
  \includegraphics[width=0.24\textwidth]{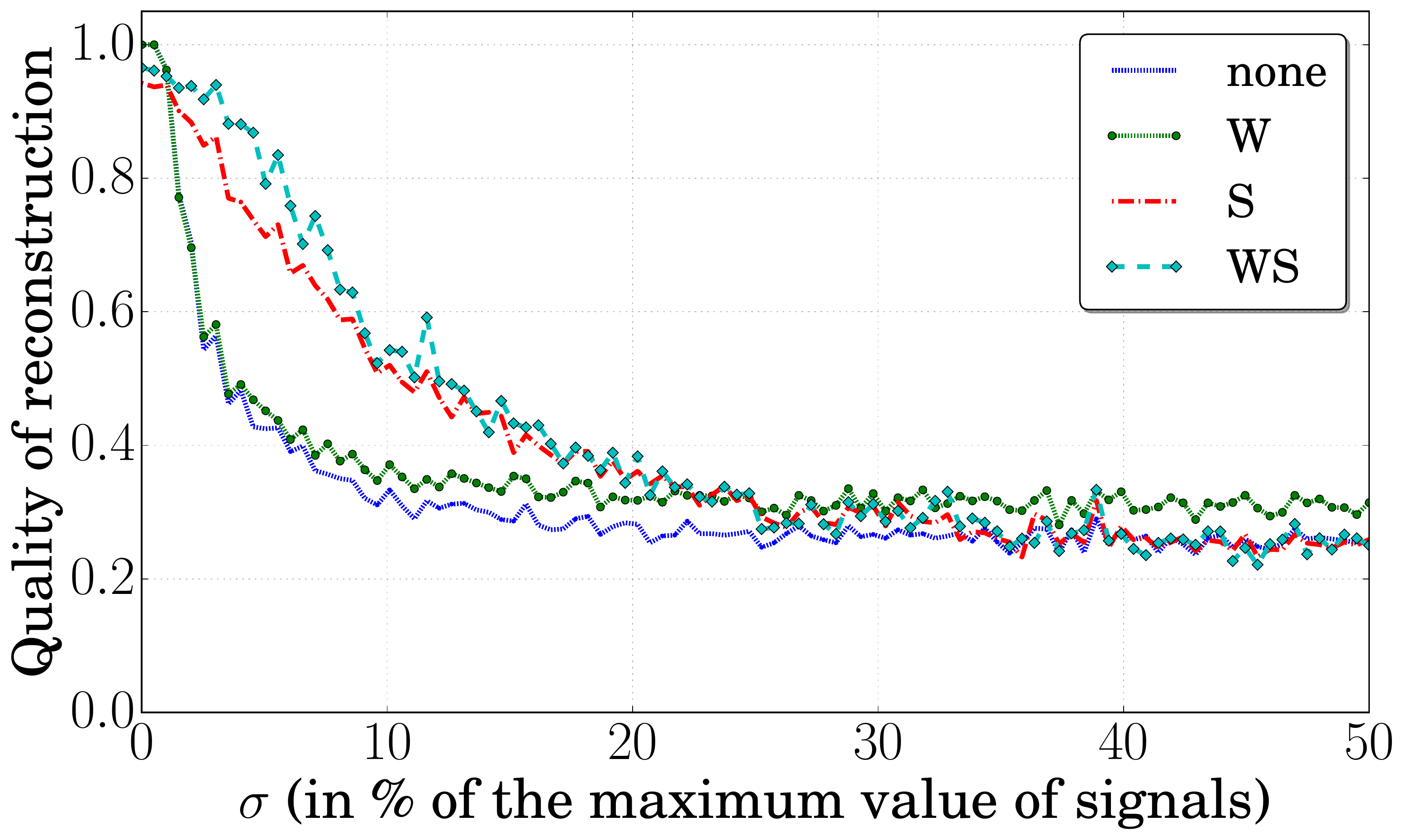}
  \includegraphics[width=0.24\textwidth]{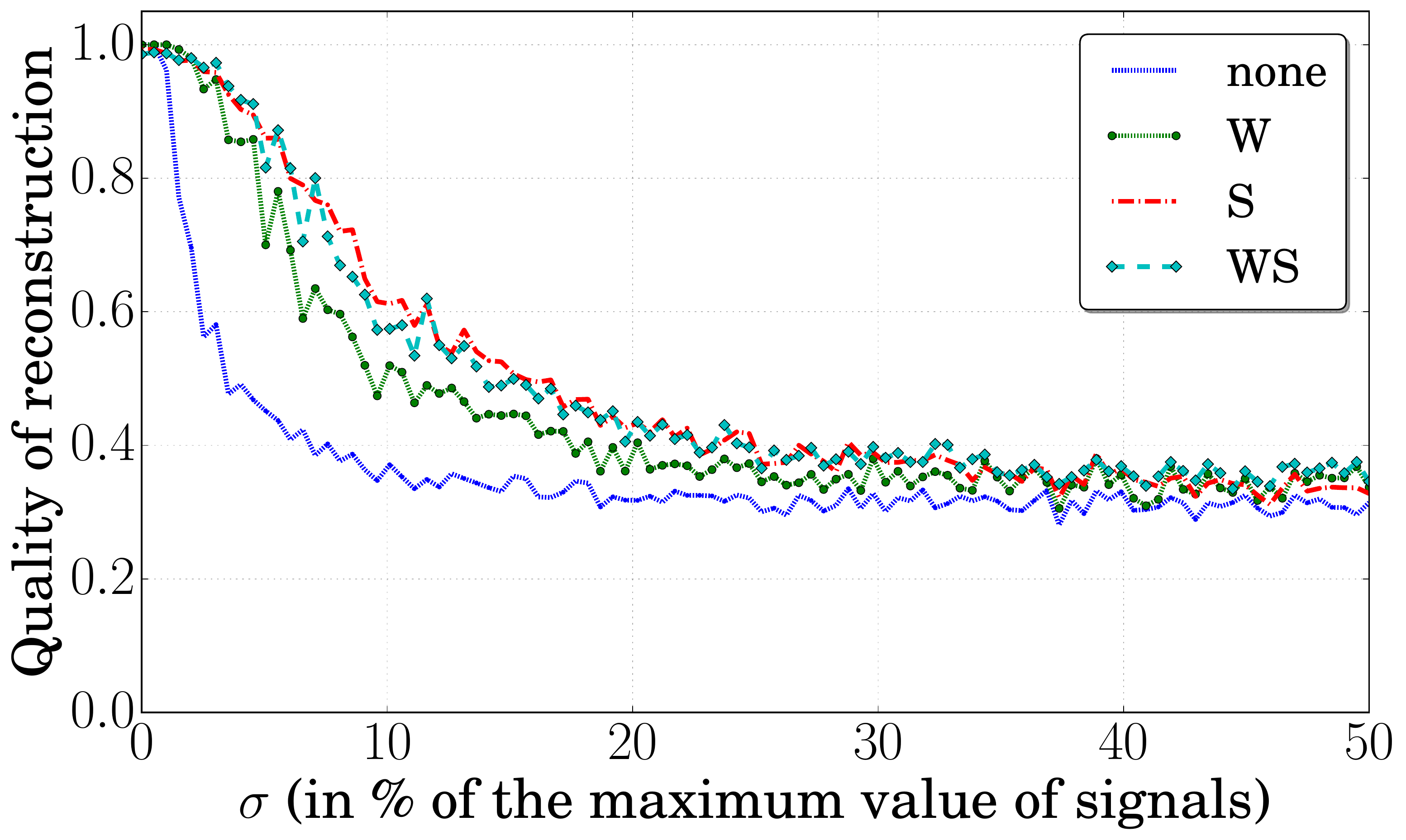}
}
  \subfloat[\label{subfig:it_noise_bar-60}\textbf{BAR60}]{
  \includegraphics[width=0.24\textwidth]{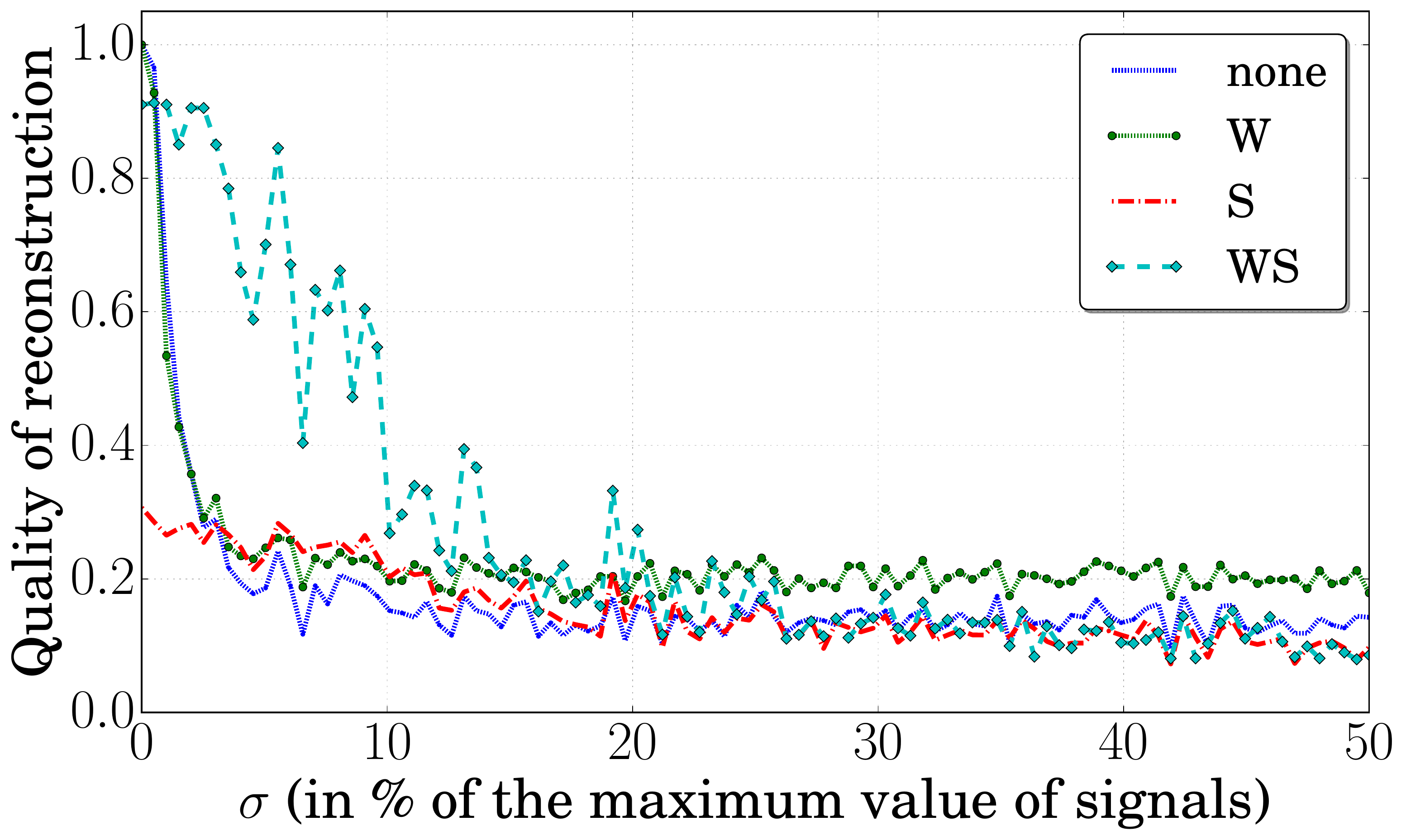}
  \includegraphics[width=0.24\textwidth]{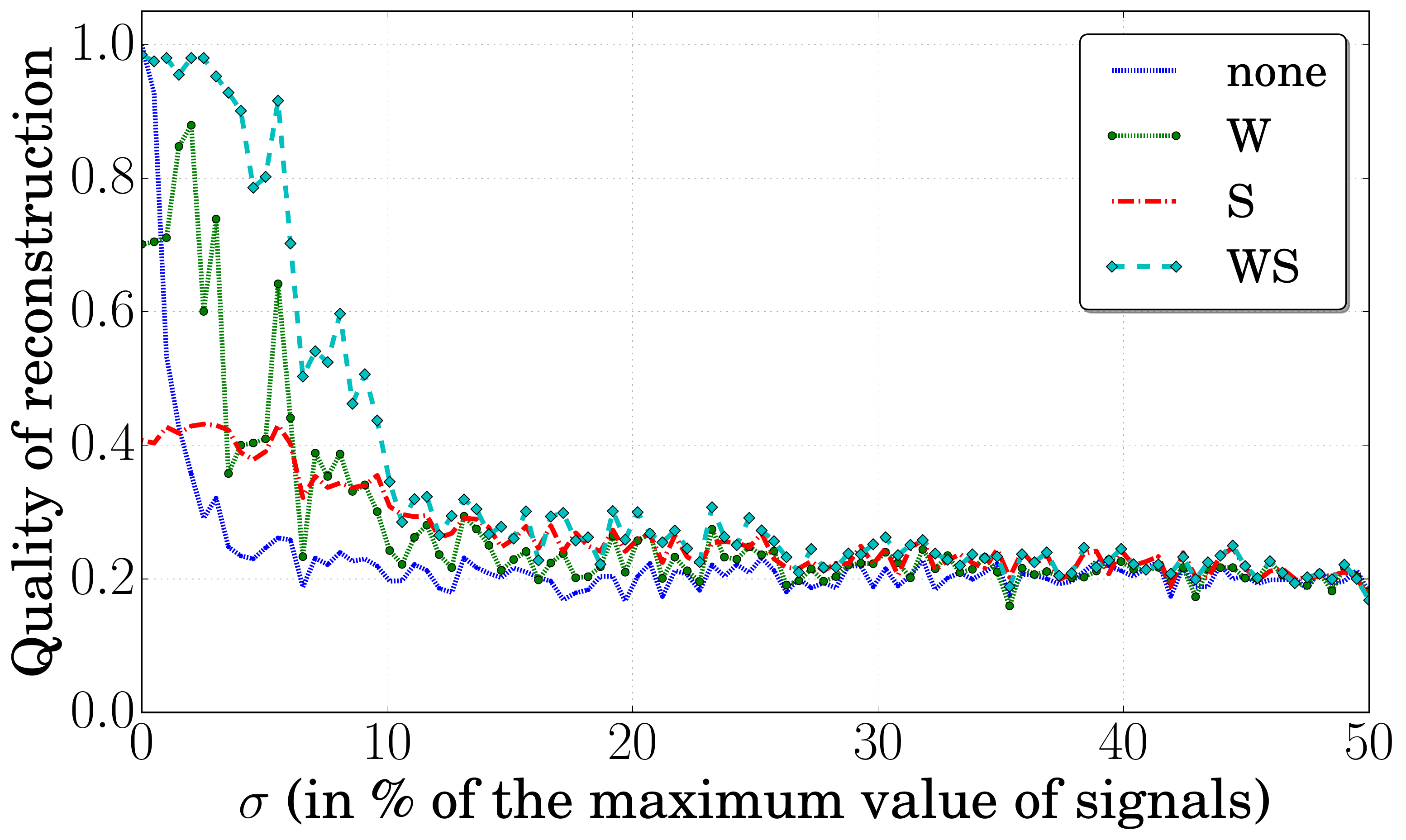}
}

  \caption{\label{fig:IT_results_noise}Results of the inverse transformation of a 
  degraded collection of signals, obtained by adding Gaussian noise with mean $0$ 
  and variance ranged from $0\%$ to $50\%$ of the maximal value of the signals. 
  For each value of $\sigma$, a quality of reconstruction is measured by 
  comparing the obtained graph with the original one.  Each sub-figure shows the 
  results for one instance, whose name refers to those given in 
  Figure~\ref{fig:GS_illustrations}. For each sub-figure, the left plot shows the 
  results using thresholding based on the number of edges, while the right plot 
  shows the results using the Adapted Otsu's method. Four configurations of the 
  inverse transformation are studied. \textbf{none}: no enhancement is performed; 
  \textbf{W}: weighted computation of distances only with $\alpha=4$; \textbf{S} 
  sequential update of the adjacency matrix only; \textbf{WS} both enhancements.}

\end{figure*}

\paragraph*{Experimental setup}

The performance of the robust inverse transformation using the proposed 
enhancements is evaluated in this section. A thorough evaluation is almost 
impossible: if the collection of signals is directly obtained using a 
transformation from graphs to signals as introduced in 
Section~\ref{sec:transformation}, then the inverse transformation is immediate and exact, 
and it does not require any sophisticated method. Conversely, if the collection of 
signals is not the representation of a graph, or if the signals are modified, 
there is no indication of the correct graph which is described at best by this 
collection. We propose nonetheless to assess the method by comparison of a graph 
with reconstructed versions after applying perturbations on the collection of 
signals. 

Let us consider a graph $\G = (\V, \E)$, described by a collection of signals 
$\bm{X}$. The signals are perturbed to obtain a degraded collection of signals 
$\tilde{\bm{X}}$, whose inverse transformation gives a graph $\tilde{\G} = 
(\tilde{\V}, \tilde{\E})$. $\G$ and $\tilde{\G}$ are compared by using an index 
of similarity, noted $Q(\G, \tilde{\G})$, based on the Jaccard index 
\cite{Jaccard1901} defined as $Q = \frac{|\E \cap \tilde{\E}|}{|\E \cup 
\tilde{\E} | }$, which compares the ratio of common edges between $\G$ and 
$\tilde{\G}$ over the total number of edges. $Q(\G, \tilde{\G})$ is comprised 
between $0$, if all the edges are different, and $1$, if both graphs have the 
same edge set. 

Two perturbations of the matrix $\bm{X}$ are studied. A first perturbation 
consists of removing low-energy components of a collection of signals: Starting 
from the first component, the components are successively added one by one in 
decreasing order of energy, until the collection is complete. The second 
perturbation consists in adding noise to the components: a Gaussian noise is 
added with mean $0$ and variance $\sigma$, ranged from $0\%$ to $50\%$ of the 
maximal value of the signals.

The inverse transformation is performed using the different enhancements of the 
method, in order to assess the contributions of each one. Four cases are then 
distinguished, according to the presence or absence of the weights in the 
computation of distances (with $\alpha=4$), and of the sequential update of 
the adjacency matrix: \textbf{none} (no enhancement), \textbf{W} (weighted 
computation of distances only), \textbf{S} (sequential update of the adjacency 
matrix only) and \textbf{WS} (both enhancements). Besides, the two methods of 
thresholding are tested separately in similar conditions. The experiments are 
performed on instances of graph models described in 
Section~\ref{sec:transformation}, in order to browse a wide variety of graph 
structures. 

\paragraph*{Results}

Figures~\ref{fig:IT_results_comp} and \ref{fig:IT_results_noise} show the
results of the evaluation of the inverse transformation. Each sub-figure shows 
the results for one instance, whose name refers to those given in 
Figure~\ref{fig:GS_illustrations}. Since the Adapted Otsu's method is not 
itself an enhancement but an alternative method to find a threshold of 
distances, the results have been divided for readability in two plots: For each 
sub-figure, the left plot shows the results without using thresholding based on 
the number of edges, while the right plot shows the results using the Adapted 
Otsu's method.

Let us consider the case where the perturbation of the collection of signals 
$\bm{X}$ is obtained by retaining only a reduced number of components 
(Figure~\ref{fig:IT_results_comp}). We first focus on the left plots of each 
sub-figure, where the threshold is chosen by using the number of edges. Two 
configurations emerge from these results: \textbf{W} and \textbf{WS}. The gain 
in quality of reconstruction when using weighted distances is obvious, and 
reveals that this feature is the best contribution in the retrieval of the 
original graph. Adding \textbf{S} in combination with \textbf{W} does not affect 
significantly the score obtained with \textbf{W} alone, but without any 
improvement of the reconstruction, except for the instance (\textbf{BAR60}) in 
Figure~\ref{subfig:it_comp_bar-60}: In this case, only the combination of both 
enhancements gives good results. The comparison of the results in left and right 
plots shows that the choice of method of thresholding has a slight impact on the 
results when using the inverse transformation in configuration \textbf{WS}: the 
quality of reconstruction is quite similar in both cases. The Adapted Otsu's 
method seems however less efficient when the number of retained components is 
low. An immediate explanation is that considering the number of edges of the 
original graph in the reconstruction gives undoubtedly a higher similarity 
between the reconstructed and the original graph than retaining only distances 
based on a discrimination, as it is performed in the Adapted Otsu's method. 
Considering nonetheless that the latter method gives bad results when the number 
of retained components is low might be a cursory glance: with only few retained 
components, the actual graph described by $\tilde{\bm{X}}$ is likely to be quite 
far from the one described by $\bm{X}$. The Adapted Otsu's method gives then a 
graph $\tilde{\G}$ which is not similar to $\G$, but which reflects the actual 
number of vertices.

Figure~\ref{fig:IT_results_noise} shows the results for the second experiment, 
where the degraded collection of signals $\tilde{\bm{X}}$ is obtained by adding 
Gaussian noise to $\bm{X}$. Unexpectedly, the balance between the importance of 
\textbf{W} and \textbf{S} is the reverse: using enhancements \textbf{S} alone 
leads to better results than using \textbf{W} alone. The combination of both 
enhancements is nevertheless still the best method to reconstruct the graph 
from degraded collection of signals. The comparison between the two methods of 
thresholding is the same as in the previous case: Generally, results are a bit 
better when using the number of edges to select the threshold, especially when 
the perturbation is high.

To conclude this section about the inverse transformation, two contributions 
have been done to the method of inverse transformation to highly improve the 
quality of restoration from a degraded collection of signals. These results 
also show that not having the knowledge about the original graph does not 
penalize the inverse transformation, as it is possible to retrieve the correct 
number of edges from the signals. Besides, results give reasons to believe that 
the Adapted Otsu's method leads to a graph which is closer to the actual graph 
represented by the perturbed collection of signals.


\section{Analysis of signals representing graphs}
\label{sec:spectral_analysis}

The last section of this article is dedicated to the analysis of the collection
of signals obtained after transformation from a graph. The aim of this part is
to show how spectral analysis can be used to identify specific graph structures.
As seen in Figure~\ref{fig:GS_illustrations}, signals present specific shapes
which can be linked with the graph structure. Characterizing these shapes using
spectral analysis enables us to associate frequency patterns with graph
topology. Before that, we introduce the method of analysis by addressing an
issue related to the indexation of signals.

\subsection{Indexation of signals}

So far, the importance of indexation of signals has been hidden since only 
relations, described by the Euclidean distances, have been of interest. Hence, 	
the order in which we considered the vertices in the inverse transformation 	
has no significance. However, this order is essential to study some spects of 
the signals, especially when using spectral analysis of the signals, as they are 
indexed by this vertex ordering. In Figure~\ref{fig:GS_illustrations}, the 
vertex ordering has been suitably defined to highlight specific shapes of 
signals: looking at the numbering of vertices in the graph representation shows 
that numbers closely follow the topology of the graph, in a ``natural'' order.  
Ordering randomly the vertices does not change the value assigned to each 
vertex, but would lead to abrupt variations in the representation of signals: 
Specific frequency properties, clearly observable in signals, will no longer be 
visible. Unfortunately, the suitable ordering is usually not available, 
especially when dealing with real-world graphs. To address this issue, we 
proposed in \cite{Hamon2015a} to find a vertex ordering that reflects the 
topology of the underlying graph, based on the following assumption: if two 
vertices are close in the graph (by considering for instance the length of the 
shortest path between them), they have to be also close in the ordering. The 
method consists of the study of a related labeling problem, called cyclic 
bandwidth sum problem \cite{Jianxiu2001}, defined in the more general framework 
of graph labeling problems \cite{Chung1988}. These problems seek a mapping from 
vertices to integers, in such a way that an objective function, defined as the 	
sum of the distances in the vertex ordering between all pairs of connected 
vertices, is minimized. The heuristic we proposed in \cite{Hamon2015a} consists 
of a two-step algorithm. The first step performs local searches in order to find 
a collection of independent paths with respect to the local structure of the 
graph, while the second step determines the best way to arrange the paths such 
that the objective function of the cyclic bandwidth sum problem is minimized. 
Details of the algorithm and results about the consistency between the obtained 
vertex ordering and the topology of the graph are covered in \cite{Hamon2015a}.

\begin{figure*}[htp]
  \centering

  \subfloat[\label{subfig:nas_cycle}2-ring lattice]{
  \includegraphics[width=0.4\textwidth]{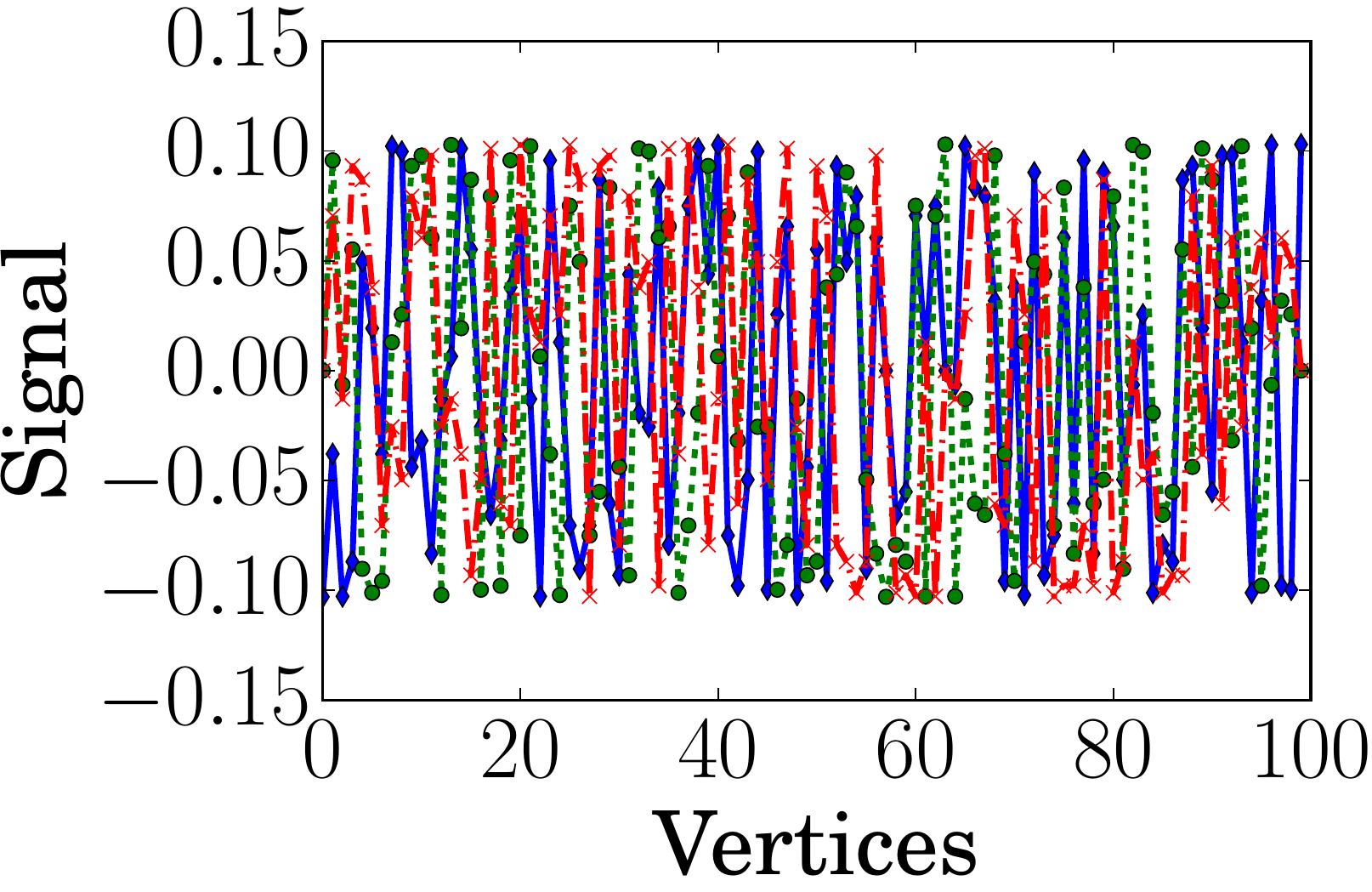}
  \includegraphics[width=0.4\textwidth]{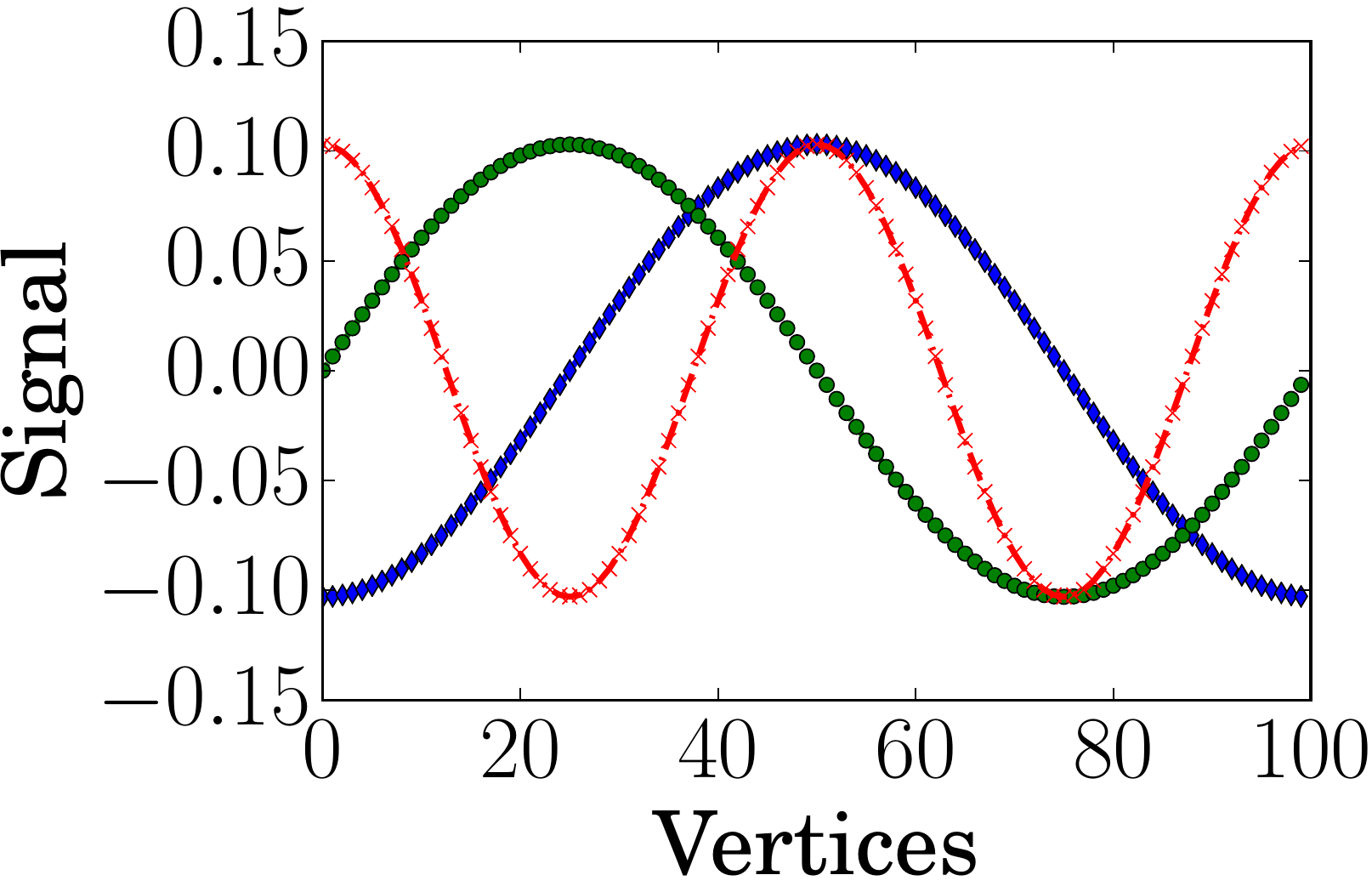}
}

  \subfloat[\label{subfig:nas_com}Stochastic block model with 4 communities]{
  \includegraphics[width=0.4\textwidth]{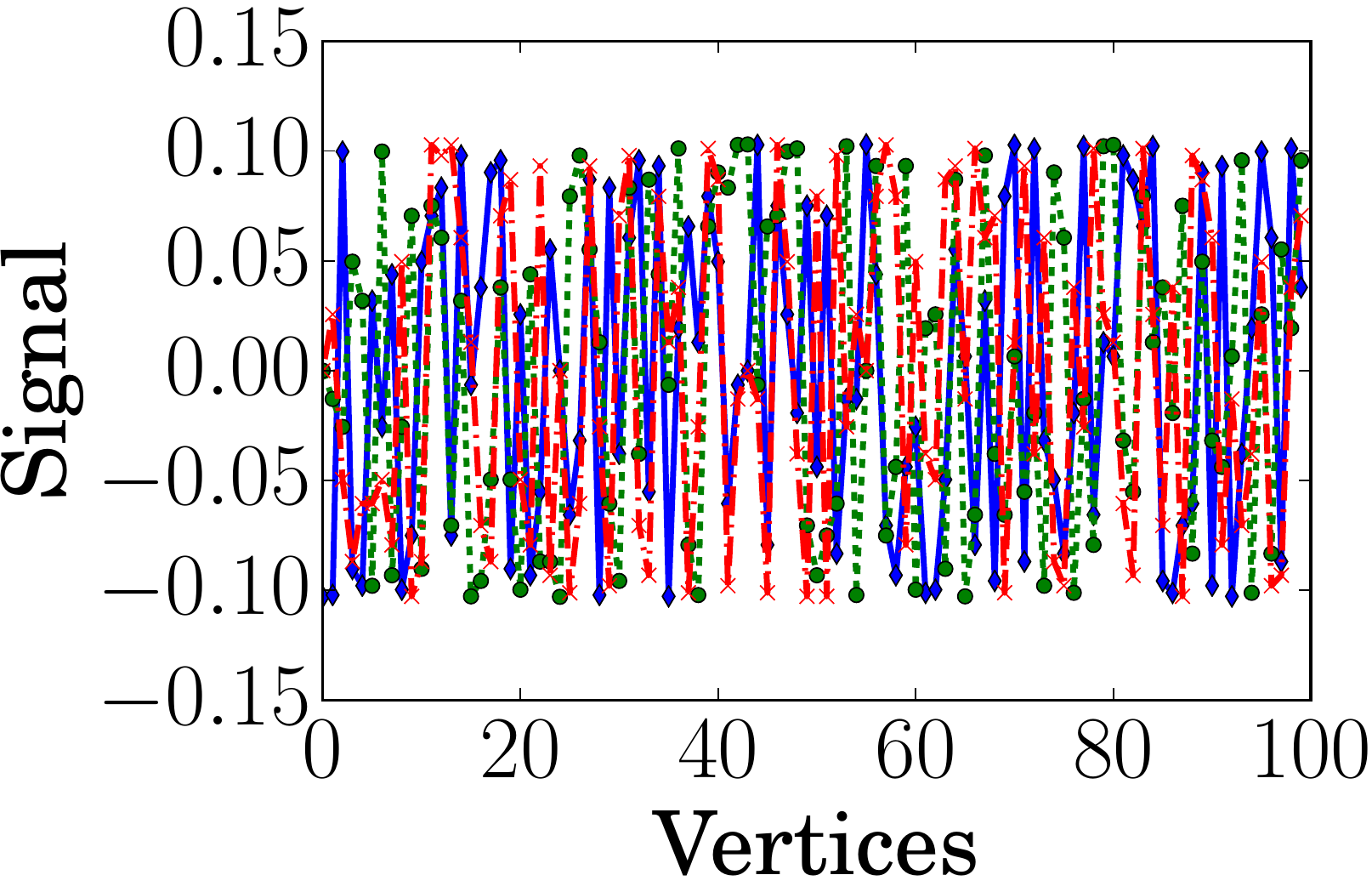}
  \includegraphics[width=0.4\textwidth]{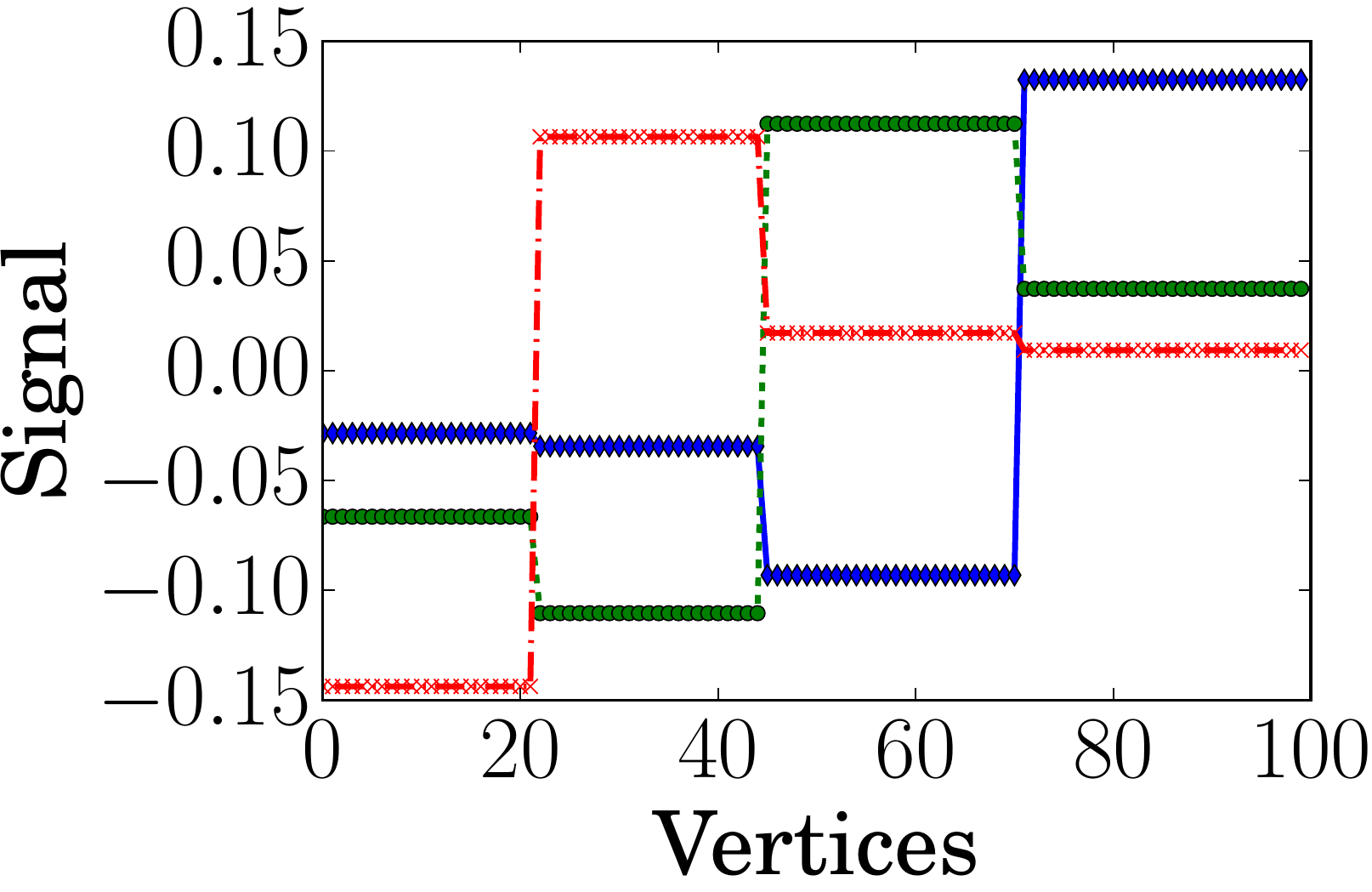}
}

  \caption{\label{fig:nas}Examples of transformation of two particular networks 
  into signals, indexed by the vertices. The resulting collection of signals is 
  indexed by the vertices. Three signals are displayed on each subplot as follows: 
  Random ordering of vertices (Left), suitable ordering of vertices (Right).}

\end{figure*}

Figure~\ref{fig:nas} shows two examples of transformation of two particular 
networks into signals. The resulting collection of signals is indexed by the 
vertices. Three signals are displayed on each subplot as follows: Random 
ordering of vertices (Left), suitable ordering of vertices (Right). It clearly 
shows that the ordering of the vertices has an influence on the spectral 
properties of the signals.

\subsection{Connection between frequency patterns of signals and graph 
structures}

\begin{figure*}

  \subfloat[\label{subfig:sa_illus_rl60-2}\textbf{RL60-2}]{
  \includegraphics[width=0.24\textwidth]{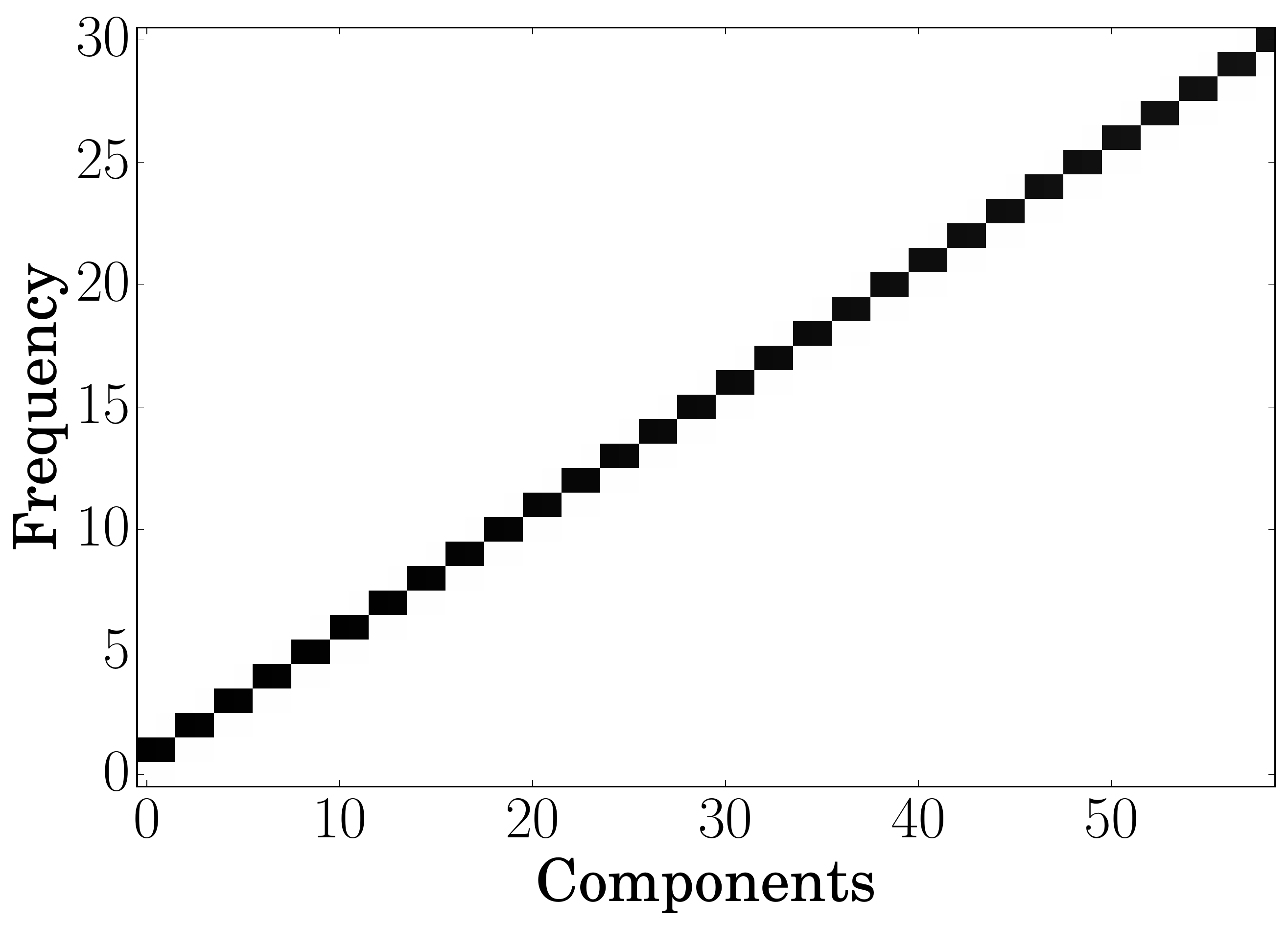}		
}
  \subfloat[\label{subfig:sa_illus_rl60-10}\textbf{RL60-10}]{
  \includegraphics[width=0.24\textwidth]{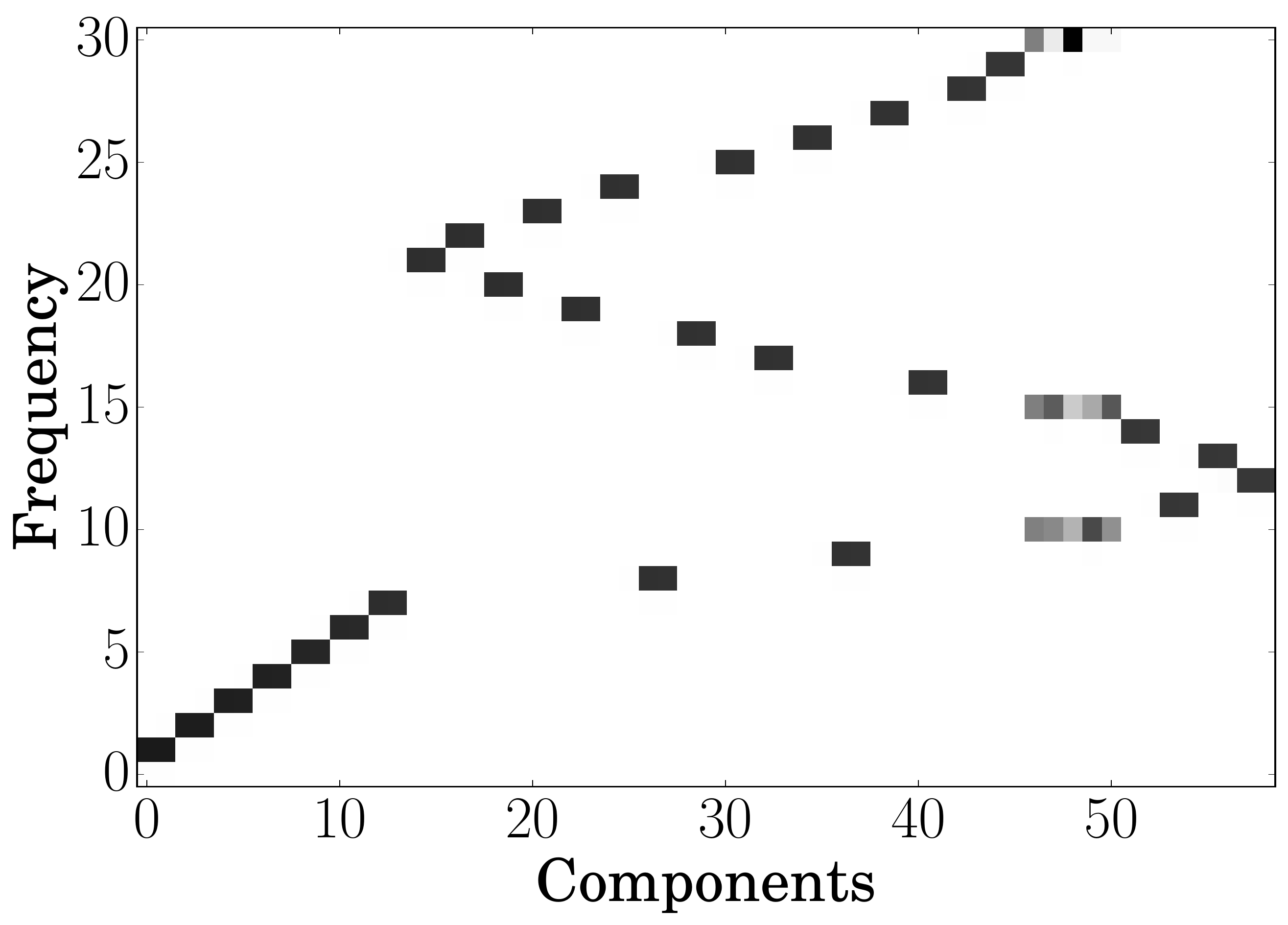}
}
  \subfloat[\label{subfig:sa_illus_ws60-2-1}\textbf{WS60-2-.1}]{
  \includegraphics[width=0.24\textwidth]{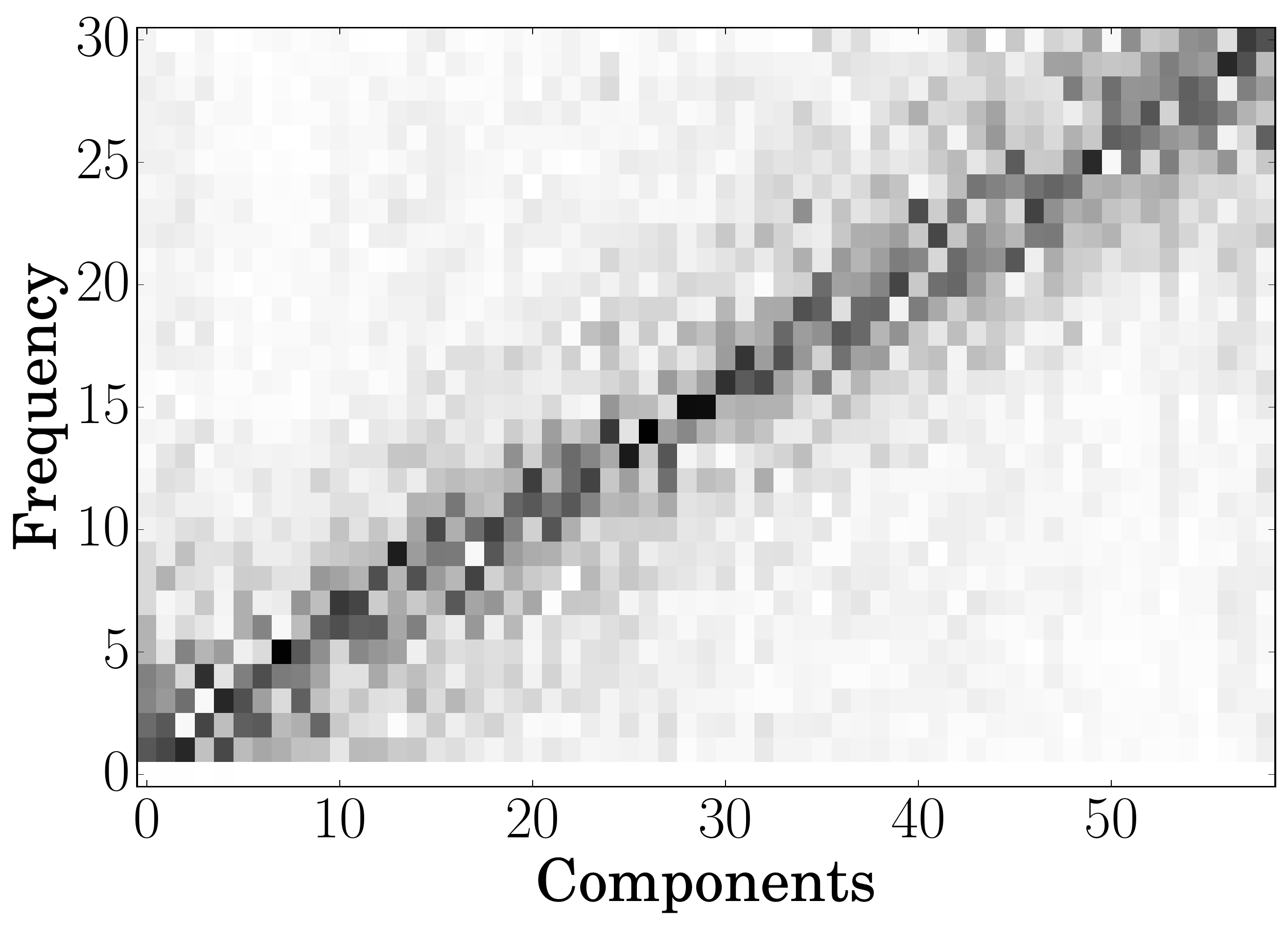}
}
  \subfloat[\label{subfig:sa_illus_ws60-10-1}\textbf{WS60-10-.1}]{
  \includegraphics[width=0.24\textwidth]{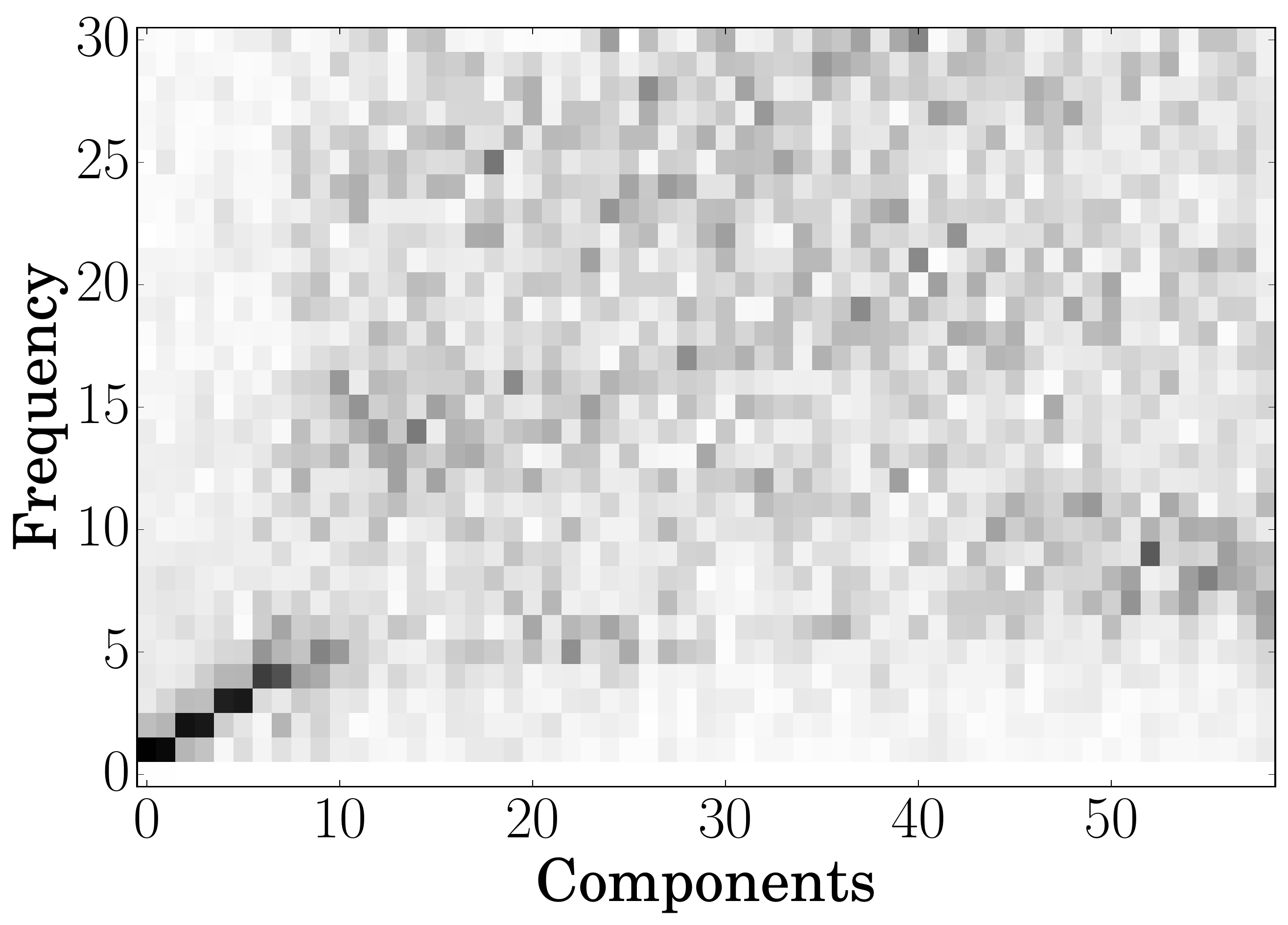}
}

  \subfloat[\label{subfig:sa_illus_sbm60-2-7-10}\textbf{SBM60-2}]{
  \includegraphics[width=0.24\textwidth]{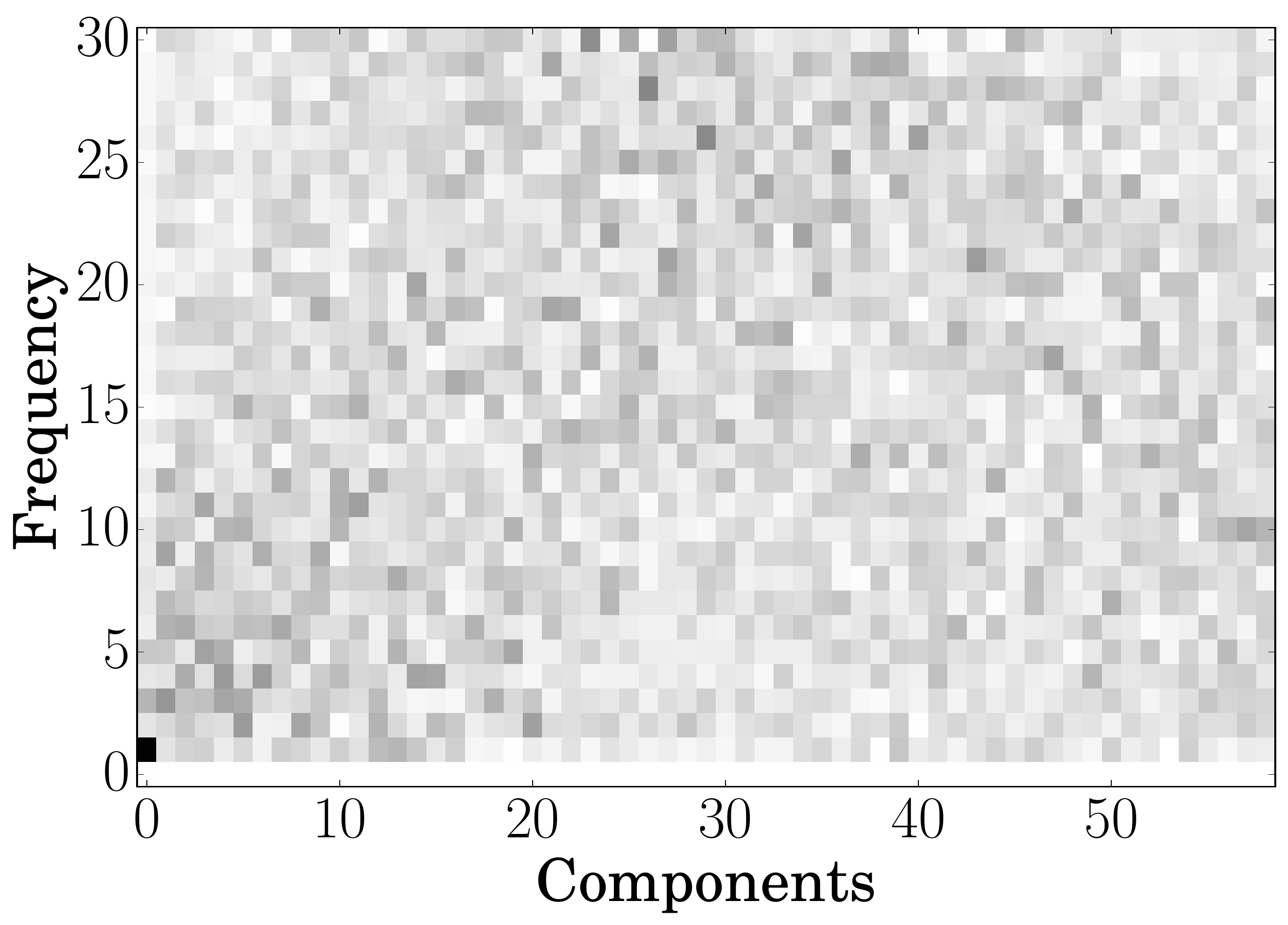}
}
  \subfloat[\label{subfig:sa_illus_sbm60-4-9-1}\textbf{SBM60-4}]{
  \includegraphics[width=0.24\textwidth]{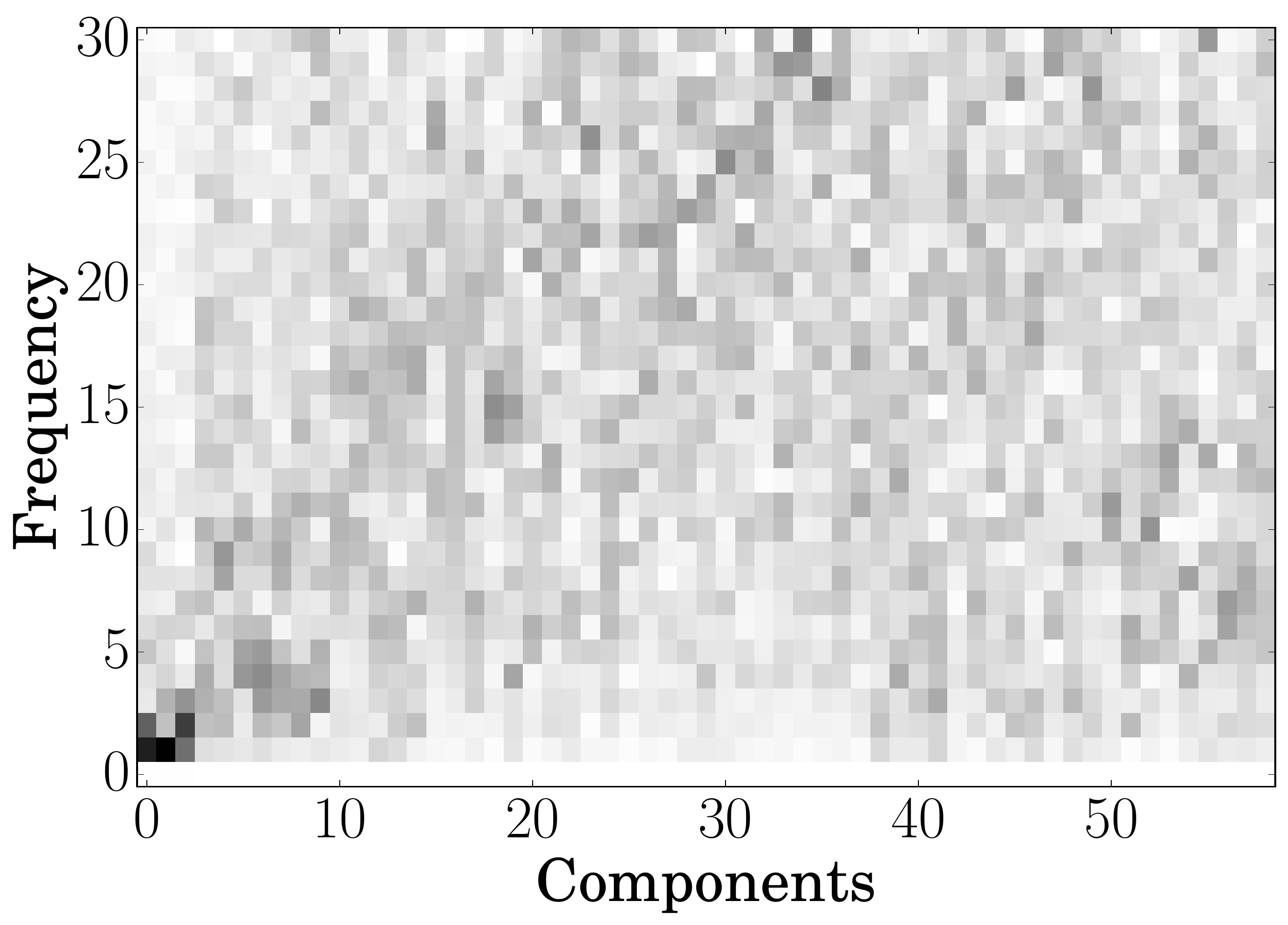}
}
  \subfloat[\label{subfig:sa_illus_er60-4}\textbf{ER60-.4}]{
  \includegraphics[width=0.24\textwidth]{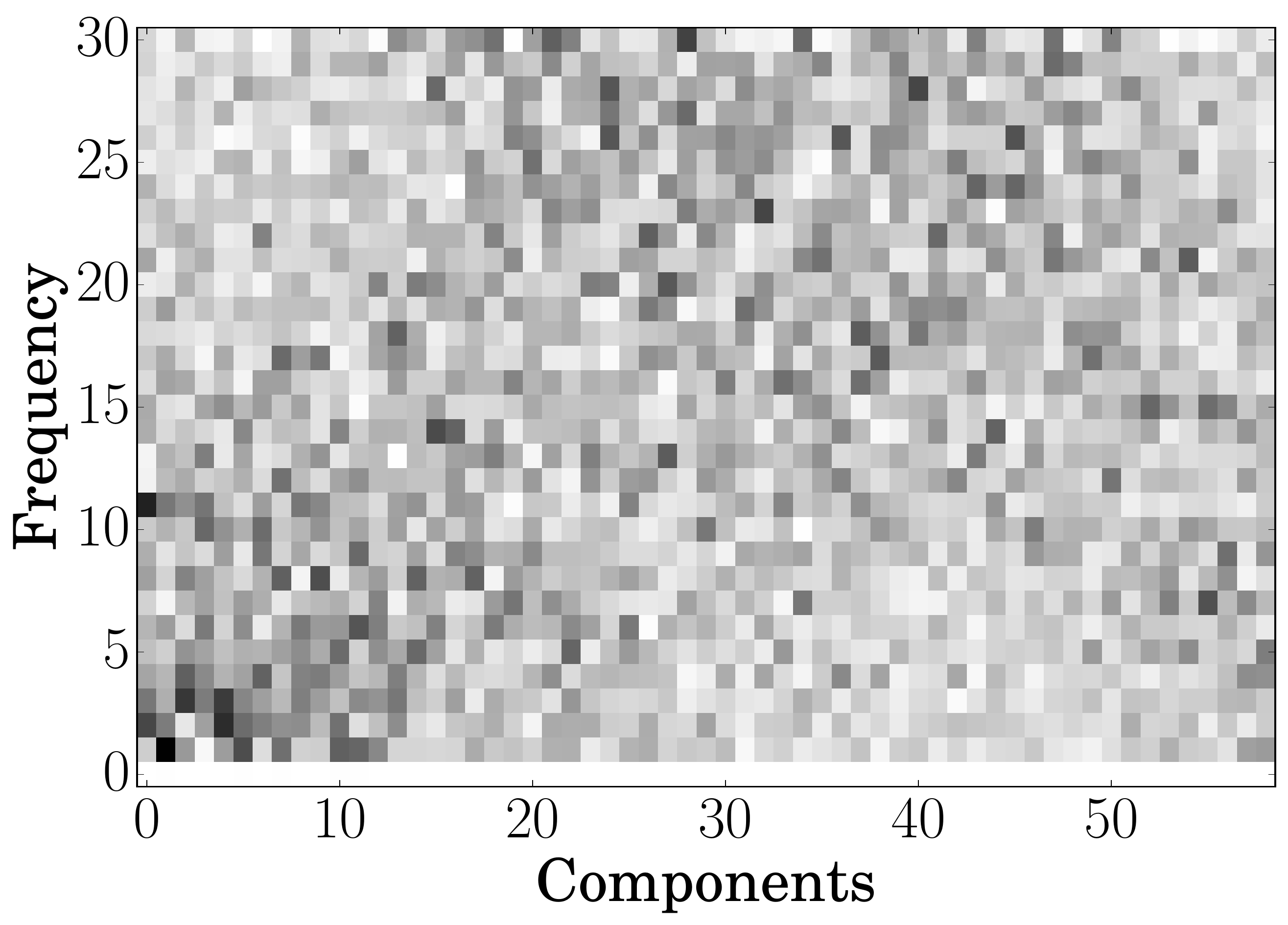}
}
  \subfloat[\label{subfig:sa_illus_bar60}\textbf{BAR60}]{
  \includegraphics[width=0.24\textwidth]{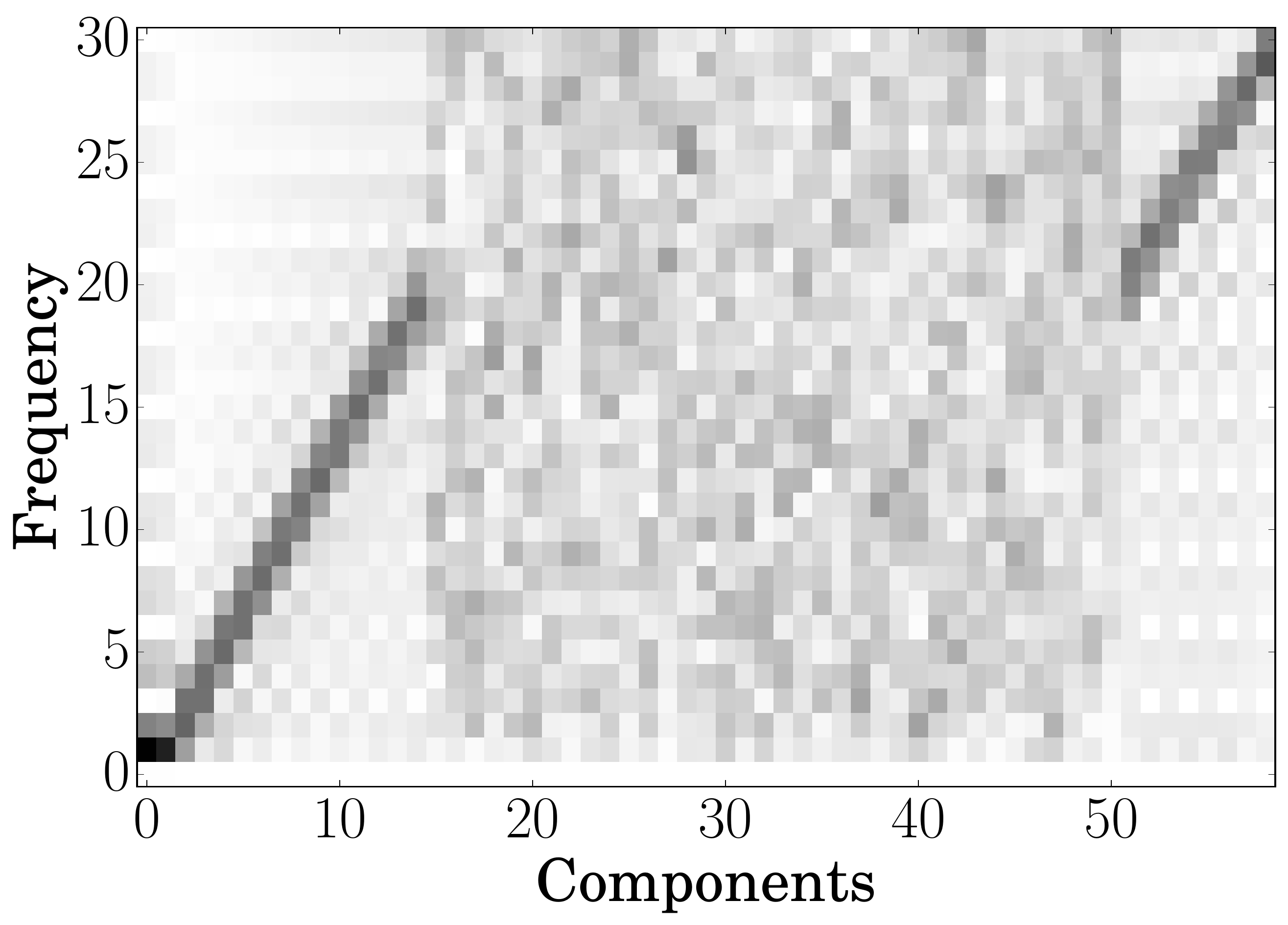}
}

  \caption{\label{fig:SA_illustrations}Frequency patterns obtained on 
  instances of graphs defined in Section~\ref{sec:transformation}. Each pattern 
  can be linked with the topology of the underlying graph. The color codes the 
  intensities, from white to black.}
\end{figure*}

\begin{figure*}[!ht]
  \centering
  \includegraphics[width=0.9\textwidth]{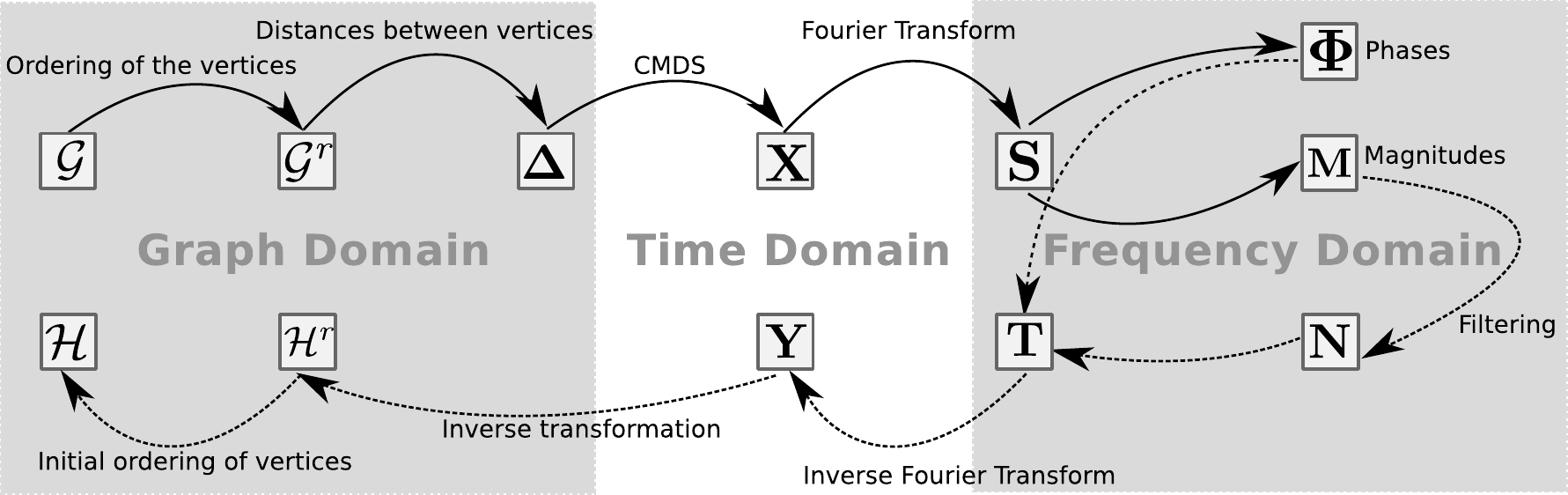}

  \caption{\label{fig:SA_filter_schema}This diagram describes the process 
  implemented to perform filtering of graphs using duality between graph and 
  signals. The boxes describe the objects, while arrows represent the operations. 
  The procedure is the following: From a graph $\G$, a collection of signals 
  $\bm{X}$ is obtained by performing a good indexation of vertices and applying 
  the CMDS method. A spectral analysis of the signals is then performed, and a 
  matrix of magnitudes $\bm{M}$ is obtained. In the frequency domain, frequencies 
  are truncated, leading to a matrix  $\bm{N}$, which gives a new collection of 
  signals $\bm{Y}$ by inverse Fourier transform. This process comes to apply a 
  crude low-pass filtering on $\bm{X}$. A graph $\mathcal{H}$ is then computed 
  from $\bm{Y}$, using the inverse transformation defined in 
  Section~\ref{sec:inverse_transformation}. Each object belongs to a specific 
  domain, either graph domain, time domain or frequency domain. Operations 
  performs some processing of objects, such as filtering, as well as the mapping 
  between the three domains.}

\end{figure*}

Spectral analysis is performed using standard signal processing methods: Let a 
collection $\bm{X}$ of $K$ signals indexed by $n$ vertices, the spectrum 
$\bm{S}$ gives the complex Fourier coefficients whose elements are obtained by 
applying the Fourier transform on each of the $K$ components of $\bm{X}$: 
$s_{kf} = \mathcal{F}\bm{X}^{(k)}(f)$ estimated, for positive frequencies, on $F 
= \frac{n}{2} + 1$ bins, $\mathcal{F}$ being the Fourier transform and $k \in 
\{1, \hdots, K \}$. From the spectrum $\bm{S}$, the magnitudes $\bm{M}$ of each 
frequency $f$ for each component read as: $M(k, f) = |s_{kf}|$. The matrix $\bm{M}$ 
is studied as a frequency-component map, exhibiting patterns in direct relation 
with the topology of the underlying graph. The phases of signals are stored in a 
matrix $\bm{\phi}$ to be used in the inverse Fourier transformation, when the 
collection of signals has to be retrieved from $\bm{M}$.

Fig. \ref{fig:SA_illustrations} shows the frequency patterns obtained for the 
graphs defined in Figure~\ref{fig:GS_illustrations}. Each graph highlights a 
specific frequency pattern, linked with its topology. Regular $k$-lattices 
display single-frequency components, whose order depends on the value of $k$: 
when $k$ is higher than $2$, the sorting of components is no longer consistent 
with the increasing order of frequencies, as described in 
Section~\ref{subsec:gs_theory}. (Figures~\ref{subfig:sa_illus_rl60-2} and 
\ref{subfig:sa_illus_rl60-10}). Adding noise to the graph also affects the 
patterns, as can be observed in Figures~\ref{subfig:sa_illus_ws60-2-1} and 
\ref{subfig:sa_illus_ws60-10-1}. Graphs with communities are associated with 
highly-localized features: high-energy components, formed by plateaus 
corresponding to the communities, appear with high magnitudes for low 
frequencies in the first components (Figures~\ref{subfig:sa_illus_sbm60-2-7-10} 
and \ref{subfig:sa_illus_sbm60-4-9-1}). These latter patterns are visible in the 
frequency-component map of the Barbell model in 
Figure~\ref{subfig:sa_illus_bar60}, denoting the combination of both structures 
in the graph. Finally, the magnitudes of random graphs, in 
Figure~\ref{subfig:sa_illus_er60-4}, do not look like white noise spectra as the 
signals are re-indexed, and this removes the i.i.d. property and adds higher 
magnitudes on low frequencies for the first components than expected. 
Nonetheless, there is no specific pattern, which is consistent with the random 
structure of this graph.

\subsection{Application to filtering of graphs}
\label{subsec:filtering}

Throughout this article, we described tools to transform a graph into a 
collection of signals, analyze these signals to describe the graph, and finally 
transform back these signals into a graph, even if the collection of signals is 
not exactly the same. We propose here an application of these tools to perform 
filtering of graphs: in the same way as classical signals, graphs are often 
built from measurements, inducing noise, i.e., presence of undesirable edges or 
absence of existing edges. Figure~\ref{fig:SA_filter_schema} proposes a 
framework to use duality between graph and signals as a way to easily perform 
filtering on graph by filtering signals representing the graph. 

The procedure is the following: From the matrix of magnitudes $\bm{M}$ obtained 
from a graph $\G$, a matrix $\bm{N}$ with the same shape is derived by retaining 
only the lowest frequencies. Using the matrix of phases $\bm{\phi}$, a degraded 
collection of signals $\bm{Y}$ is obtained by inverse Fourier transform. This 
process is tantamount to applying a crude low-pass filter on $\bm{X}$. A graph 
$\mathcal{H}$ is then computed from $\bm{Y}$, using the weighted distances with 
$\alpha=2$, the sequential update of the adjacency matrix and a thresholding 
using the Adapted Otsu's method, as described in 
Section~\ref{sec:inverse_transformation}.

\begin{figure*}
  \centering
  \subfloat[\label{subfig:sa_filter_ws60-10-1}\textbf{WS60-10-.1}]{
  \includegraphics[width=0.25\textwidth]{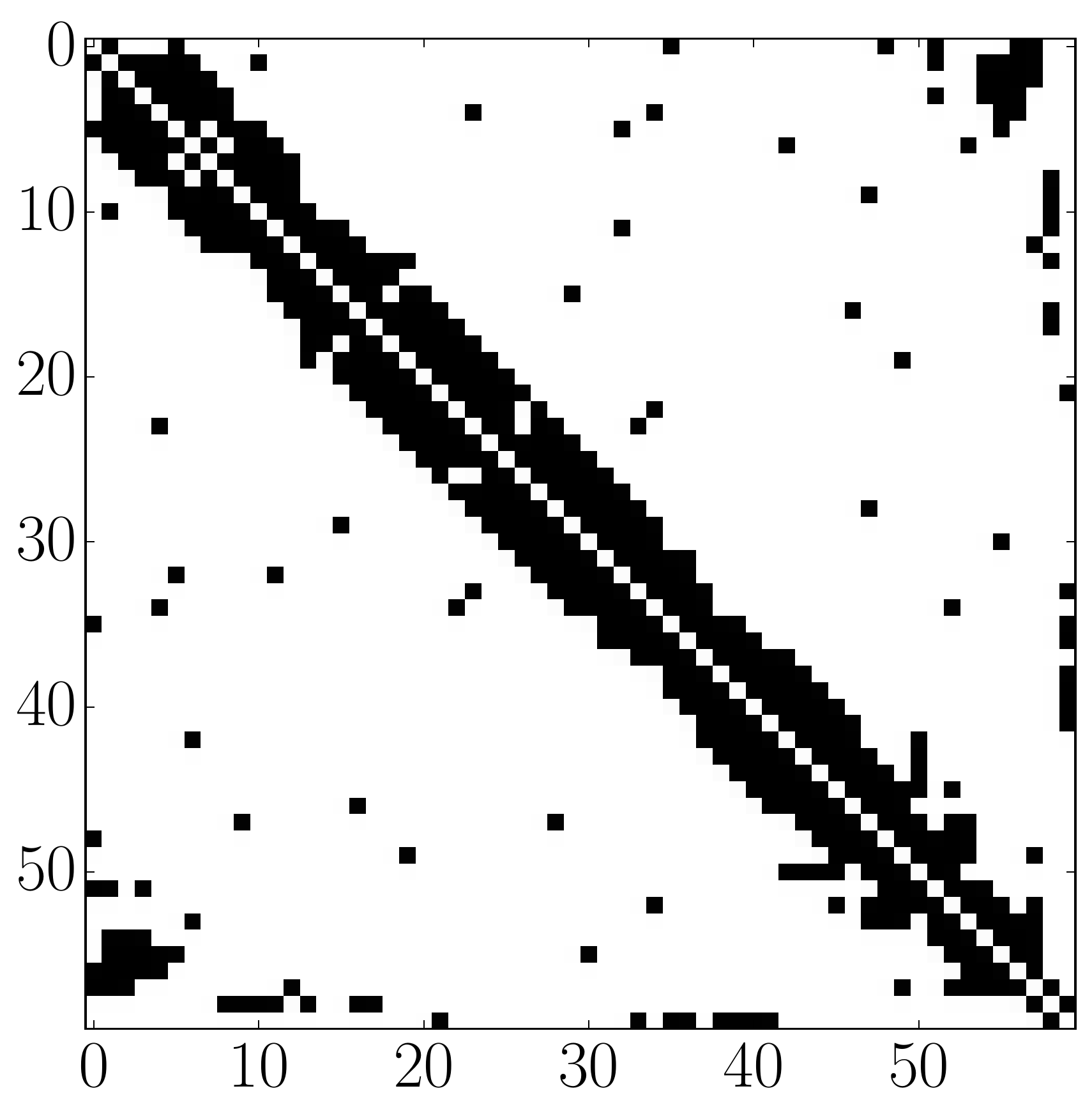}		
  \includegraphics[width=0.25\textwidth]{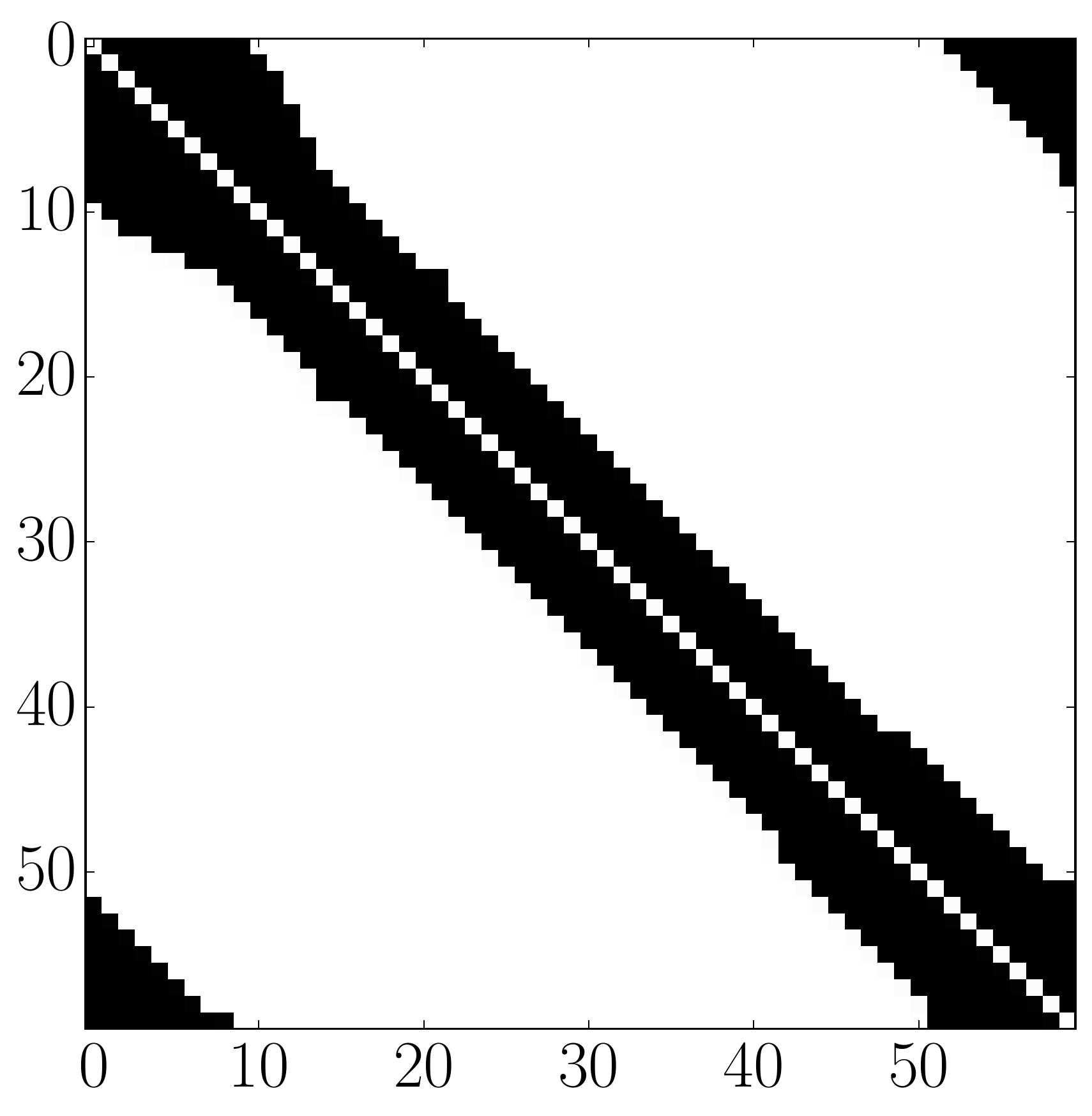}		
}
  \subfloat[\label{subfig:sa_filter_sbm60-2-7-10}\textbf{SBM60-2}]{
  \includegraphics[width=0.25\textwidth]{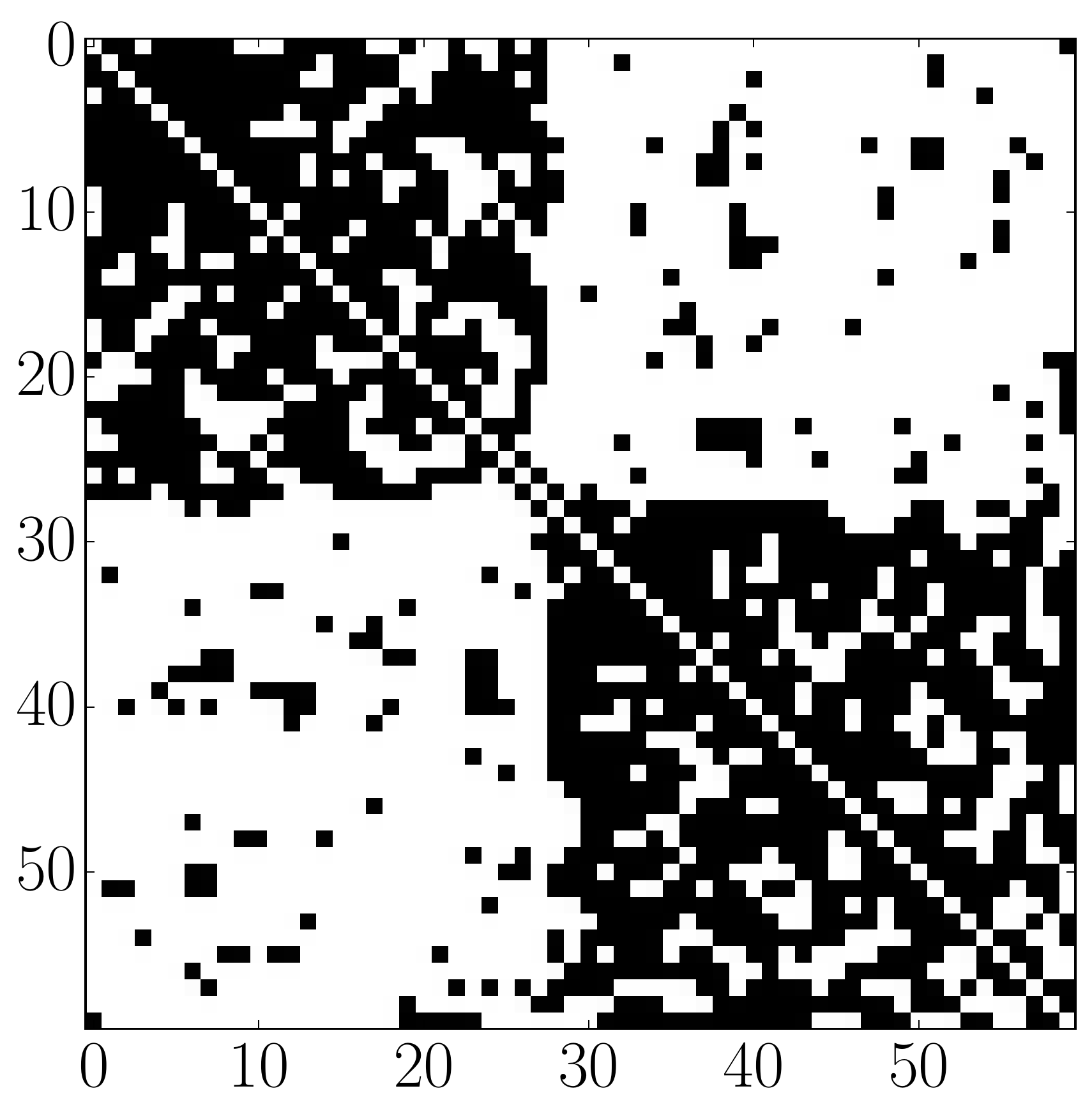}	
  \includegraphics[width=0.25\textwidth]{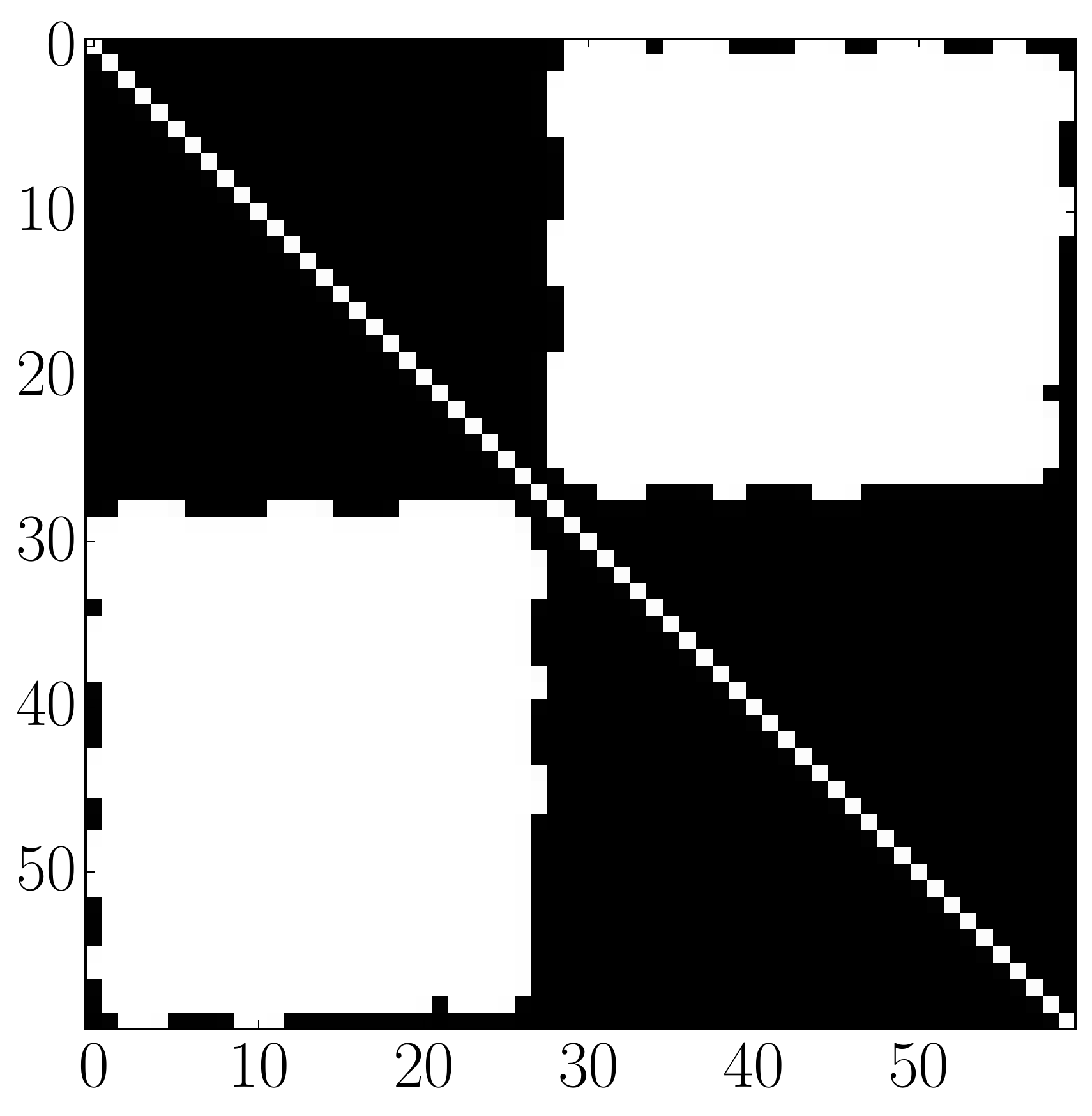}	
}

  \caption{\label{fig:SA_filtering_illus}Illustrations of denoising of 
  graphs, by applying the process described in Figure~\ref{fig:SA_filter_schema}. 
  The left plot gives the adjacency matrix of the graph $\G$, before filtering, 
  while the right plot gives the adjacency matrix of the graph $\Hg$, after 
  filtering. (a) The five lowest frequencies have been retained (b) The ten 
  lowest frequencies have been retained.}
\end{figure*}

Figure~\ref{fig:SA_filtering_illus} shows two examples of graph denoising, with 
on the left the adjacency matrix of the graph $\G$, and on the right the 
adjacency matrix of the graph $\Hg$, obtained after applying the process 
described in Figure~\ref{fig:SA_filter_schema}. The first example is an instance 
of the Watts-Strogatz model (Figure~\ref{subfig:sa_filter_ws60-10-1}), 
corresponding to a noisy $k$-ring lattice (\textbf{WS60-10-.1}). The five lowest 
frequencies have been retained, and this leads to strengthening the diagonal, 
characterizing $k$-ring lattices. Besides, off-diagonal edges have been removed. 
The second example (Figure \ref{subfig:sa_filter_sbm60-2-7-10}) is an instance 
of the stochastic block model with communities (\textbf{SBM60-2}). This can be 
viewed in this case as two cliques whose random edges inside communities have 
been removed while random edges between communities have been added. After 
denoising, where the ten lowest frequencies of the matrix $\bm{M}$ have been 
retained, the missing edges inside communities are retrieved, while edges 
between them are removed: the communities are better-defined, appearing as 
actual cliques. These two examples highlight that denoising signals representing 
the graph is tantamount to denoising the graph itself.

\section{Conclusion}
\label{sec:conclusion}

In this article, we proposed a framework to study graphs in the classical domain 
of signals, enabling the analysis to take advantage of basic signal processing 
tools. From the method of transformation from graph to signals proposed by 
Shimada et al.~\cite{Shimada2012}, we discussed extensions to study the obtained 
signals and indirectly, the graph itself, using a robust inverse transformation. 
Denoising of graphs has been proposed to evaluate the method as a first 
potential application. The results we obtained suggest that the framework 
developed provides a connection between signal processing operations and 
modifications of graphs. This paves the road to new approach for the analysis of 
real-world networks, using more sophisticated signal processing tools, such as 
adaptive filtering.

\section*{Acknowledgments}
This work is supported by the programs ARC 5 and ARC 6 of the r\'egion 
Rh\^one-Alpes.

\appendix

\section{Upper bound for the parameter $w$}
\label{apx:choice_w}

We propose here a rationale for the choice of $w$, based on the study of 
matrix eigenvalues: As mentioned in the work of Gower 
\cite{Gower1985}, $\bm{\Delta}$ is exactly retrieved from $\bm{X}$ if and only 
if $\bm{B}$ is positive definite:
\begin{align}
  \langle \bm{z}, \bm{Bz} \rangle \geq 0 \text{ for all vectors } \bm{z} \in 
  \mathbb{R}^n
\end{align}
or equivalently, if and only if $\bm{\Delta}^{(2)}$ is conditionally negative 
definite:
\begin{align}
  \langle \bm{z}, \bm{\Delta}^{(2)}\bm{z} \rangle \leq 0 \quad{} \forall \bm{z} 
  \in \mathbb{R}^n \text{ such that } \sum_{i=1}^n z_i = 0
\end{align}
From the definition of $\bm{\Delta}$ in Eq.~\ref{eq:delta}, we have:
\begin{align}
  \langle \bm{z}, \bm{\Delta}^{(2)}\bm{z} \rangle &= \langle \bm{z}, 
  \bm{A}\bm{z} \rangle \\ \nonumber
  &+ w^2 ( \langle \bm{z}, \bm{1}_n\bm{1}_n^T \bm{z} \rangle 
  - \langle \bm{z},  \bm{I}_n \bm{z} \rangle - \langle \bm{z}, \bm{A}\bm{z} 
  \rangle) \\ \nonumber
  & = \langle \bm{z}, \bm{A}\bm{z} \rangle - w^2 ( \langle \bm{z}, \bm{z} 
  \rangle + \langle \bm{z}, \bm{A}\bm{z} \rangle)
\end{align}
Two cases can be then distinguished: 
\begin{enumerate}
\item If $\langle \bm{z}, \bm{A}\bm{z} \rangle > 
  - \langle \bm{z}, \bm{z} \rangle$, then 
  \begin{align}
    w^2 \geq \frac{\langle \bm{z}, \bm{A}\bm{z} \rangle}{\langle \bm{z}, 
    \bm{A}\bm{z} \rangle + \langle \bm{z}, \bm{z} \rangle}
  \end{align}
  which is hold as $w>1$.
\item If $\langle \bm{z}, \bm{A}\bm{z} \rangle < - \langle \bm{z}, \bm{z} 
  \rangle$, then 
  \begin{align}
    w^2 \leq \frac{\langle \bm{z}, \bm{A}\bm{z} \rangle}{\langle 
    \bm{z},\bm{A}\bm{z} \rangle + \langle \bm{z}, \bm{z} \rangle}
  \end{align}
\end{enumerate}

The upper bound depends on the adjacency matrix $\bm{A}$, i.e., on the 
structure of the graph. To have an idea of a suitable value of $w$, let us 
define $\bm{A}$ and $\bm{z}$ such that $\langle \bm{z},\bm{A}\bm{z} \rangle$ is 
minimal. $\bm{A}$ is defined as the adjacency matrix of a graph with $n$ 
vertices, with $n$ even, such that $a_{ij}=1$ if and only if $i$ and $j$ do not 
belong in the same subset among $\{1, \hdots, \frac{n}{2} \}$ and 
$\{\frac{n}{2}+1, \hdots, n \}$. $\bm{A}$ is then a $4$-block matrix, with the 
bottom-left block and the top-right block equal to $1$. As for $z$, it is equal 
to $-1$ for the first half of the vector and $1$ for the last half of the 
vector: $z = [-1, -1,  \hdots, 1, 1]$. $\langle \bm{z},\bm{A}\bm{z} \rangle$ is 
then equal to $-\frac{n^2}{2}$, while $\langle \bm{z},\bm{z} \rangle = n $. 
Hence, we obtain an approximation for the upper bound of $w$: $w \leq 
\sqrt{\frac{n}{n-2}}$.

\section{Theoretical results for the $k$-ring lattices}
\label{anx:kring}

A $k$-ring lattice is a graph where each vertex $i$ is connected to the vertices 
$\{i-\frac{k}{2}, i-\frac{k}{2} + 1, \hdots, i-1, i+1, \hdots, i+\frac{k}{2}-1, 
i+\frac{k}{2}\}$, for $k \in \{2, 4, \hdots, n\}$. As Shimada et al. discussed 
in \cite{Shimada2012}, it is immediate to find expected eigenvalues and 
eigenvectors in this case using circulant matrix theory \cite{Gray2005}. We 
explicit additionally here the connection between the parameters $w$ and $k$ and 
the resulting signals. Any circulant matrix $\bm{C}$ has its eigenvalues 
$\bm{\lambda}$ given $\forall q \in \{0,n-1\}$ by $\lambda_q = \sum_{j=0}^{n-1} 
{c_j \zeta^{kj}}$ where $\bm{c}$ is the circulant vector of $\bm{C}$ and $\zeta 
= e^{\frac{2i\pi}{n}}$ is the $n$th root of the unity. As for eigenvectors, 
they are given $\forall q \in [0,n-1]$ by $\bm{v}_q = \sqrt{n}[1, \zeta^q, 
\zeta^{2q}, \hdots, \zeta^{(n-1)q}]$, corresponding to the columns of the 
Fourier matrix noted $\bm{F}^{(q)}$. These eigenvalues and eigenvectors appear 
as complex conjugate pairs, namely $\bar{\lambda}_q = {\lambda}_{n-q}$ and 
$\bar{\bm{v}}_q = \bm{v}_{n-q}$ for $q \neq 0$. As we consider symmetric 
matrices, the eigenvalues are real and double ($\lambda_q = {\lambda}_{n-q}$ for 
$q>0$) and the corresponding eigenvectors are the real and imaginary parts of 
$\bm{v}_q$, normalized by $\sqrt{2}$ to obtain an orthonormal matrix: $\bm{v}_q 
= \sqrt{2}\Re(\bm{F}^{(q)}) = \sqrt{2} \cos(\frac{2\pi q}{n})$ and $\bm{v}_{n-q} 
= \sqrt{2}\Im(\bm{F}^{(q)}) = \sqrt{2} \sin(\frac{2\pi q}{n})$, corresponding to 
harmonic oscillations. If $n$ is even, $\lambda_{\frac{n}{2}}$ is single and the 
corresponding eigenvector is not normalized by $\sqrt{2}$.

Starting from the circulant vector $\bm{\delta}$ with three values $0$, $1$ and 
$w$, the circulant vector $\bm{b}$ is defined by $b_i = -\frac{1}{2} [ 
\delta_i^2 -\frac{\alpha}{n}]$, with $\alpha = k + (n-1-k)w^2$. The vector 
$\bm{b}$ is then completely defined by three values: $\frac{\alpha}{2n}$ when 
$i=0$, $-\frac{\alpha}{2}(1-\frac{\alpha}{n})$ when $i\in \{1, \hdots, 
\frac{k}{2} \} \cup\{n - \frac{k}{2}, \hdots, n-1\} $  and 
$-\frac{\alpha}{2}(w^2-\frac{\alpha}{n})$ when $i \in \{ \frac{k}{2}+1, \hdots, 
n-\frac{k}{2} - 1 \}$. The computation of eigenvalues of $\bm{B}$ leads to $ 
\lambda_q = \frac{\alpha}{2n} \sum_{j=0}^{n-1} {\zeta^{jq}} - \frac{1}{2} \left( 
  \sum_{j=1}^{\frac{k}{2}} \zeta^{jq} + \sum_{j=n - \frac{k}{2}}^{n-1} \zeta^{jq} 
  + w^2\sum_{j=\frac{k}{2}+1}^{n-\frac{k}{2}-1} \zeta^{jq} \right)$. From these 
eigenvalues, the obtained signals are similar to those obtained by standard 
methods of diagonalization of matrices, up to rotation, reflection and 
translation.

\bibliographystyle{plain}
\bibliography{bibliography}

\end{document}